\documentclass[useAMS,usenatbib]{mn2e}

\usepackage{graphicx}
\usepackage{captcont}
\usepackage{times}

\def\aj{\,{AJ}}
\def\apj{\,{\rm ApJ}}

\def\apjs{\,{\rm ApJS}}

\def\pasj{\,{\rm PASJ}}
\def\araa{\,{\rm ARAA}}
\def\aap{\,{\rm A\&A}}

\def\mnras{\,{\rm MNRAS}}

\def\araa{\,{\rm ARA\&A}}

\title[Radial and 2D Colour Properties of E+A Galaxies]
{Radial and 2D Colour Properties of E+A Galaxies}
\author[Yamauchi \& Goto]
{
  Chisato Yamauchi$^{1,2}$\thanks{E-mail:cyamauch@a.phys.nagoya-u.ac.jp}
  and Tomotsugu Goto$^{3}$\thanks{E-mail:tomo@jhu.edu}
  \\
  $^{1}$Department of Physics and Astrophysics, Nagoya University,
  Chikusa-ku, Nagoya 464-8602, Japan\\
  $^{2}$National Astronomical Observatory, 2-21-1 Osawa, Mitaka, Tokyo
  181-8588, Japan\\
  $^{3}$Department of Physics and Astronomy, The Johns Hopkins
  University, 3400 North Charles Street, Baltimore, MD 21218-2686, USA
}
\begin{document}

\pagerange{\pageref{firstpage}--\pageref{lastpage}} \pubyear{2004}

\maketitle

\label{firstpage}

\begin{abstract}

We investigate the radial colour gradient and
 two-dimensional (2D) colour properties of 22 E+A galaxies with 
$5.5{\rm \AA} < {\rm H}\delta~{\rm EW} < 8.5{\rm \AA}$
and 49 normal early-type galaxies as a control sample 
at a redshift of $<0.2$ 
in the Second Data Release of the Sloan Digital Sky Survey. 
We found that a substantial number of E+A galaxies exhibit positive slopes
of radial colour gradient (bluer gradients toward the centre)
which are seldom seen in normal early-type galaxies.
We found irregular `Colour Morphologies' 
-- asymmetrical and clumpy patterns --
at the centre of $g{\rm -}r$ and $r{\rm -}i$ 2D colourmaps
of E+A galaxies with positive slopes of colour gradient.
 Kolomogorov-Smirnov two-sample tests
show that $g{\rm -}r$ and $r{\rm -}i$ colour gradient distributions of E+A galaxies
differ from those of early-type galaxies
with a more than 99.99\% significance level.
We also found 
 a tight correlation between radial colour gradients and colours,
and 
between radial colour gradients and 4000\AA~break
in the E+A sample; E+A galaxies which exhibit bluer colour or
weaker $D_{4000}$ tend to have
positive slopes of radial colour gradient.
We compared the GISSEL model
and E+A's
observational quantities, ${\rm H}\delta~{\rm EW}$, $D_{4000}$ and
 $u{\rm -}g$ 
colour, 
and found that almost all our E+A galaxies are 
situated along
a single evolution track.
Therefore, these results are interpreted as E+A galaxies evolving from  
${\rm H}\delta~{\rm EW} \sim 8{\rm \AA}$ 
to ${\rm H}\delta~{\rm EW} \sim 5{\rm \AA}$,
with colour gradients changing from positive to negative,
and with the irregular 2D colourmap becoming smoother, 
during a time scale of $\sim$300 Myr.
Our results favor the hypothesis that E+A galaxies are post-starburst galaxies caused by 
merger/interaction, having undergone a centralized violent starburst.
\end{abstract}


\section{Introduction}
\label{section:intro}

\citet{dre83,dre92} discovered galaxies with mysterious spectra while
investigating high redshift cluster galaxies.  These galaxies had strong
Balmer absorption lines with no emission in [OII].  They were named
``E+A'' galaxies, since their spectra resembled a superposition of those
of elliptical galaxies (Mg$_{5175}$, Fe$_{5270}$ and Ca$_{3934,3468}$
absorption lines) and those of A-type stars (Strong Balmer absorption)%
\footnote{Because the spectra of elliptical galaxies are characterized
by K stars, these galaxies are sometimes called
``K+A'' galaxies \citep[e.g.,][]{fra93,dre99,bar01}.
Following the first discovery, we refer to them as ``E+A'' throughout this work.
}%
.
Since the lifetime of an A-type star is about 1 Gyr, the existence of
strong Balmer absorption lines shows that these galaxies have
experienced starburst within the last Gyr.  However, they show no
sign of on-going star formation as non-detection in the [OII] emission
line indicates.  Therefore E+A galaxies are interpreted as 
post-starburst galaxies, that is, galaxies which have undergone
truncated starburst
\citep{dre83,dre92,cou87,mac88,new90,fab91,abr96}.

At first, ``E+A'' galaxies were found in cluster regions, both in low
redshift clusters \citep{fra93,cal93,cal97,cas01,ros01} and high
redshift clusters
\citep{sha85,lav86,cou87,bro88,fab91,bel95,bar96,fis98,mor98,cou98,dre99}.
Therefore, a cluster specific phenomenon 
(e.g., Goto et al. 2003b) was thought to be responsible
for the violent star formation history of E+A galaxies.  A ram-pressure
stripping model
\citep{spi51,gun72,far80,ken81,aba99,fuj99,qui00,fuj04,fg04} 
as well as tides from the cluster potential \citep[e.g.,][]{fuj04}
may first
accelerate star formation of cluster galaxies and later turn it off.
However, recent large surveys of the nearby universe found many E+A
galaxies in the field regions \citep{got03,got05,got03c,qui04}.  At the very
least, it is clear that these E+A galaxies in the field regions cannot
be explained
by a physical mechanism that works in the cluster region.  E+A galaxies
have often been thought to be transitional objects during cluster
galaxy evolution, evolving phenomena such as the Butcher-Oemler effect
\citep[e.g.,][]{got03a}, 
the morphology-density relation \citep[e.g.,][]{got03e},
and the correlation between various properties of the galaxies with the
environment \citep[e.g.,][]{tan04}.  However, explaining cluster
galaxy evolution using E+A galaxies may not be realistic anymore.

One possible explanation for E+A phenomena is dust-enshrouded star
formation, where E+A galaxies are actually star-forming, but emission
lines are invisible in optical wavelengths due to heavy obscuration
by dust.
As a variant, \citet{pog00} presented the selective dust extinction
hypothesis, where dust extinction is dependent on stellar age,
since the youngest stars inhabit very dusty star-forming HII regions while
older stars have had time to migrate out of such dusty regions.
If O, B-type stars in E+A galaxies are embedded in dusty regions and
only A-type stars have long enough lifetimes ($\sim$1 Gyr) to move out
from such regions, this scenario can naturally explain the E+A
phenomena.  A straightforward test for these scenarios is to observe in
radio wavelengths where the dust obscuration is negligible.
At 20cm radio wavelengths, the synchrotron radiation from electrons accelerated by supernovae can be observed.
Therefore, in the absence of a radio-loud active nucleus, the radio flux
of a star-forming galaxy can be used to estimate its current massive
star formation rate (SFR) \citep{con92,ken98,hop03}. \citet{sma99}
performed such a radio observation and found that among 8 galaxies
detected in radio, 5 
have strong Balmer absorption with no
detection in [OII].
They concluded that massive stars are currently forming in these 5 galaxies.  \citet{owe99} investigated the radio properties of galaxies in a rich cluster at $z \sim 0.25$ (A2125) and found that optical line luminosities (e.g., ${\rm H}\alpha$+[NII]) were often weaker than one would expect for the SFRs implied by the radio emission.  \citet{mil01} observed radio continua of 15 E+A galaxies and detected moderate levels of star formation in only 2 of them.  
\citet{got04a} performed 20cm radio continuum observation of 36 E+A
galaxies, and none of them are detected in 20cm, suggesting that E+A galaxies are not
dusty-starburst galaxies.


Alternatively, galaxy-galaxy interaction has been known to trigger star
formation in the pair of galaxies
\citep{sch82,lav88,liu95a,liu95b,sch96,nik04}.  \citet{oeg91} found a
nearby E+A galaxy with a tidal tail feature.  High resolution Hubble
Space Telescope imaging
supported the galaxy-galaxy interaction scenario
by
showing
some 
post-starburst (E+A) galaxies in high
redshift clusters 
as having
disturbed or interacting signatures
\citep{cou94,cou98,dre94,oem97}.  \citet{liu95a,liu95b} observed 40
merging/interacting systems and found that some of their spectra
resemble E+A galaxies.  \citet{bek01} modeled galaxy-galaxy mergers with
dust extinction, confirming that such systems can produce spectra which
evolve into E+A spectra.  Recently, \citet{got03d,got05} found that young E+A
galaxies have 
more companion galaxies within 100 kpc, providing 
strong support for the merger/interaction origin of E+A galaxies.

As is mentioned above, most of the previous work is focused on the
global properties of E+A galaxies. Needless to say, 
investigation of the 
internal properties of E+A galaxies such as
age and metallicity gradients
is also indispensable in solving
the E+A mystery.  Do E+A
galaxies leave traces of merger/interaction?  Where do the traces
remain, in a centralized or decentralized starburst?  
The elucidation of
these questions is essential in testing E+A evolution scenarios. 
\citet{nor01} performed long-slit spectroscopic observation of 21 E+A
galaxies, and found young stellar populations of E+A galaxies are more
centrally concentrated than older populations, and old components of E+A
galaxies conform to the Faber-Jackson relation.  \citet{bar01} also
reported that E+A galaxies on average tend to have slightly bluer radial
gradients toward the centre than do normal early-type galaxies.
\citet{yan04} presented HST observations of the five bluest E+A galaxies
with $z \sim 0.1$ and reported details of disturbed morphologies and
detected compact sources associated with E+A which are consistent with
the brightest clusters in nearby starburst galaxies.  
Although these are important results, previous work often lacked
statistical significance, since E+A galaxies are extremely rare. In
addition, it is important to understand the spatial properties of E+A
galaxies along an evolutionary sequence.  Furthermore, there is a
contamination problem in the E+A sample selected without using  
${\rm H}\alpha$.  
\citet{ken92a,ken92b} shows, ${\rm H}\alpha$ is the best
star formation indicator in the optical wavelength since it is a strong line
and it has fewer uncertainties (e.g., dust extinction, self-absorption,
metallicity dependence) than the other lines. Indeed, \citet{got03}
reported that an E+A sample without ${\rm H}\alpha$ cut-off contains
$\sim$50\% contamination.  \citet{bla04} also found that the
criterion using ${\rm H}\delta$ and [OII] leads to a significant
sub-population of disk systems with detectable ${\rm H}\alpha$ emissions.  
We stress that a clean sample selection is essential in order  to study a rare population of galaxies such as E+As.



\begin{figure}
\includegraphics[scale=1.1]{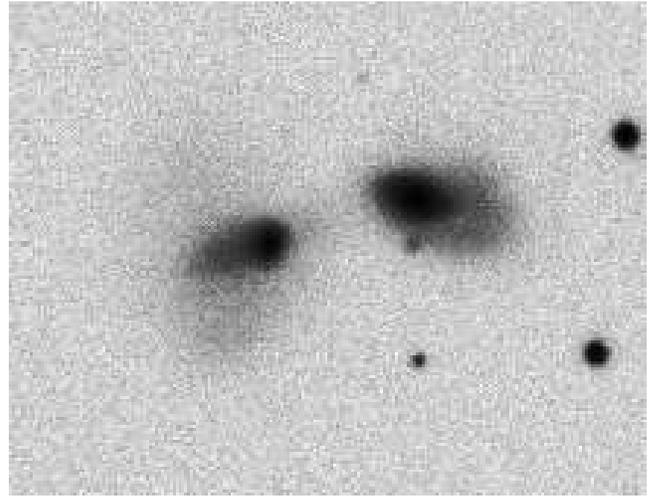}
\caption{SDSS $r$-band image of one of the nearest E+A galaxies 
(left;$z$=0.034) captured from the SDSS DR2 sample.
 The image size is 86.3 arcsec $\times$ 66.9 arcsec.
Note the dramatic tidal tails.
}\label{fig:nearest_e+a}
\end{figure}

In this paper, we use publicly available 
{\it true} E+A galaxies (without ${\rm H}\alpha$ nor [OII] emission)
selected from the Sloan Digital Sky Survey (SDSS, York et al. 2000;
Early Data Release, Stoughton et al. 2002, hereafter EDR; First Data
Release, Abazajian et al. 2003, hereafter DR1; Second Data Release,
Abazajian et al. 2004, hereafter DR2) by \citet{got05}.  Both
broad-band imaging and the spectroscopic survey of 10,000 deg$^2$ of
SDSS provides us with the first opportunity to study E+A galaxies in a
very large number.  \citet{got05} analyzed $\sim$250,000 galaxy spectra
in the DR2, and the number of E+A homogeneous galaxies reached 266.
The SDSS imaging is somewhat poor (typical seeing size is $\sim$1.5$''$)
compared with 8m-class telescopes or HST, but this sample includes
very nearby or large E+A galaxies like Figure \ref{fig:nearest_e+a},
which exhibits dramatic tidal tails or conspicuous disturbed
morphologies.  We then select 22 E+A galaxies with $z < 0.2$, large
apparent size, and more strict criteria 
$5.5 {\rm \AA} < H\delta~{\rm EW}$, 
$-1.0 {\rm \AA} < {\rm H}\alpha~{\rm EW}$ 
and $-2.5 {\rm \AA} < {\rm OII}~{\rm EW}$ 
(a positive sign is absorption), and investigate spatial properties; 2D
colourmaps and $g{\rm -}r$ and $r{\rm -}i$ radial colour gradients.  The
radial colour gradients are compared with spectroscopic properties, and
we compare these properties with  evolution scenarios using SED models
in order to understand the evolutionary sequence of E+A galaxies.

This paper is organized as follows: In section \ref{section:samples},
the definitions of two galaxy samples are summarized, as well as the
spectroscopic and imaging data.  In section \ref{section:reductions}, we
explain the reduction of the imaging data, e.g. an image convolution,
position matching between passbands, a review of the concentration index and
a definition of radial colour gradients.  In section
\ref{section:results}, we briefly describe our E+A morphologies, and
study 2D $g{\rm -}r$ and $r{\rm -}i$ colour properties, 2D colourmaps
and radial colour gradients, with relations to other
photometric/spectroscopic properties.  We then investigate the evolution
scenario for E+A radial colour gradients using a comparison between SED
models and our observational data.  In section \ref{section:discussion},
we discuss our results with other E+A-related studies.  Our conclusions
are presented in section \ref{section:conclusions}.  The cosmological
parameters adopted throughout this paper are 
$H_0=75 {\rm km s}^{-1} {\rm Mpc}^{-1}$, and 
$(\Omega_m,\Omega_{\Lambda},\Omega_k)=(0.3,0.7,0.0)$.

\section{Data Samples}
\label{section:samples}

Our E+A galaxies are selected from the publicly available catalog described in \cite{got05}.  The mother sample of this catalog contains $\sim$250,000 objects classified as galaxies with spectroscopic information in SDSS DR2 \citep{aba04}.
\citet{got05} selected 266 E+A galaxies 
as those with 
$5.0{\rm \AA}<{\rm H}\delta~{\rm EW}$,
$-3.0{\rm \AA}<{\rm H}\alpha~{\rm EW}$ and
$-2.5{\rm \AA}<{\rm [OII]}~{\rm EW}$ (absorption lines have a positive sign)
after redshift ($>0.01$) and S/N ($>$10 per pixel) cut-off.

We have restricted our targets to those galaxies which have a redshift $z$
of less than 0.2 and an apparent size of $a_{60}$ (Petrosian 60\% flux semi-major
axis; described in section \ref{section:reductions}) more than 2.82$''$
to maintain enough sampling points for radial colour gradients.  This
selection leaves 
50 E+A galaxies.  To minimize contamination of our
E+A galaxies with remaining star formation activity, we set more
strict spectroscopic criteria 
$5.5{\rm \AA} < {\rm H}\delta~{\rm EW}$ and 
$-1.0 {\rm \AA}<{\rm H}\alpha~{\rm EW}$, 
and finally, we are left with 22 E+A galaxies of greater purity.


We also randomly selected 50 normal early-type galaxies within $z<0.2$
as a control sample from objects classified as galaxies in the SDSS DR2
catalog.  The normal early-type galaxies are selected using the
elliptical-based concentration index, $C_e$ \citep{yam04s}, a ratio of
Petrosian 50\% flux semi-major axis to 90\% flux semi-major axis (See
section \ref{section:reductions}); $C_e<0.33$ for early-type galaxies
\citep{shi01}.  Although the standard (inverse) concentration index
$C=r_{50}/r_{90}$ defined with the Petrosian flux in the circular
apertures is significantly affected by inclination, the effect of
inclination is removed with the use of elliptical apertures
\citep{yam04s}.  The cut-off size of $a_{60}$ is the same as that of
E+A galaxies.  But this subsample contains 1 galaxy with a deblending
problem, which we eliminated.  Finally, we use 49 normal early-type
galaxies as a control sample.

The spectroscopic data, ${\rm H}\delta$, ${\rm H}\alpha$ and [OII]
equivalent width (${\rm EWs}$) and their errors are measured by the 
flux-summing method described in \citet{got05} (See also Goto et al. 2003c).
The values of 4000\AA~break($D_{4000}$) in our paper are the reciprocal
numbers of values in the SDSS catalog.  We use DR2 atlas images
\citep{sto02} as imaging data to calculate 2D colourmaps and radial
colour gradients of $g{\rm -}r$ and $r{\rm -}i$, and concentration index
$C_e$.  The photometric system, imaging hardware and astrometric
calibration of SDSS are described in detail elsewhere
\citep{fuk96,gun98,hog01,smi02,pie03}.

\section{Data Reductions}
\label{section:reductions}

From the SDSS database, we can easily download the deblended galaxy
images, called `atlas images', by fpAtlas file in which images are
archived using Rice compression.  In the SDSS images, however, the
seeing sizes differ from passband to passband, and the relation between
physical and imaging positions of a passband on an Atlas image is not equal
to others, but slightly differs.  Therefore we cannot directly compose
$g$- and $r$-band images, or $r$- and $i$-band images using the
coordinates given by the fpAtlas file. We adjust the positions of the
multi-band images in the following way:

First, we smear all images except those with the worst seeing
$\sigma_{\rm max}$ among the passbands to equalize the seeing sizes of all
passbands.  Assuming that the point-spread function is Gaussian,
images are convolved with a Gaussian function of $\sigma_{\rm c}$,
\begin{equation}
 \sigma_{\rm c} = \sqrt{\sigma_{\rm max}^2 - \sigma^2} ,
\end{equation}
where $\sigma$ is 
the actual seeing of each image derived from 
the ${\tt psf\_ width}$ parameter in tsField files:
\begin{equation}
 \sigma = ({\tt psf\_ width}/1.06)/2.35,
\end{equation}
where ${\tt psf\_ width}$ is equivalent to
1.06 FWHM for a Gaussian profile \citep{fan03}.

Next, each passband image is resampled using 
positional information, 
``${\tt colc}$'' and ``${\tt rowc}$''(real) 
in tsObj file, and
``offsets wrt%
\footnote{with reference to}
 reference colour''(${\tt dcol}$ and ${\tt drow}$; integer)
and ``bounding box''(${\tt cmin}$ and ${\tt rmin}$; integer) in 
``master mask'' in fpAtlas file (see source code of readAtlasImages).
Where ``${\tt colc}$'' and ``${\tt rowc}$'' are at the exact centre
of the object in fpC coordinates, and
$({\tt cmin}+{\tt dcol},{\tt rmin}+{\tt drow})$
is the fiducial point of atlas images in fpC coordinates.
The typical errors of ``${\tt colc}$'' and ``${\tt rowc}$'' are
$0.01\sim0.1$ pixel($0.004\sim0.04$ arcsec).

In the case of making $g{\rm -}r$ 2D colourmaps
without resampling $r$-band images,
the preliminary position offset ($x0_{p{g}},y0_{p{g}}$) of
$g$-band pixels 
needed for $1\times$ $g$-band resampling in an atlas image coordinate
is
\begin{equation}
\begin{array}{ccl}
 x0_{p{g}} & = & ({\tt colc}_g-{\tt col0}_g) - ({\tt colc}_r-{\tt col0}_r) \\
 y0_{p{g}} & = & ({\tt rowc}_g-{\tt row0}_g) - ({\tt rowc}_r-{\tt row0}_r)~,
\end{array}
\label{eq:x00g}
\end{equation}
where ${\tt col0}={\tt cmin}+{\tt dcol}$ and 
${\tt row0}={\tt rmin}+{\tt drow}$.
And we consider the values $(d0_x,d0_y)$:
\begin{equation}
\begin{array}{ccl}
 d0_x & = & x0_{p{g}} - [x0_{p{g}}] \\
 d0_y & = & y0_{p{g}} - [y0_{p{g}}] \\
\end{array}
\label{eq:d0x}
\end{equation}
where $[x]$ is the floor function and
$(d0_x,d0_y)$ indicates the distance from pixel grid and 
the resampled position of $g$-band pixels
(That is, resampling is not required when $(d0_x,d0_y)=(0,0)$).

Resampling only $g$-band images is not reasonable.
Both $r$- and $g$-bands should be resampled and 
optimized.
Then we determine the position offsets by $d0_x$,
\begin{equation}
\begin{array}{ccl}
 x0_{og} & = & x0_{p{g}} + (-d0_x/2) \\
 x0_{or} & = &  (-d0_x/2) \\
\end{array}
  ~~~~~~~~~~~{\rm for }~~~d0_x<=0.5
\label{eq:under0.5x}
\end{equation}
or
\begin{equation}
\begin{array}{ccl}
 x0_{og} & = & x0_{p{g}} + (1-d0_x)/2 \\
 x0_{or} & = & (1-d0_x)/2 \\
\end{array}
  ~~~~~~{\rm for }~~~0.5<d0_x~,
\label{eq:over0.5x}
\end{equation}
and by $d0_y$,
\begin{equation}
\begin{array}{ccl}
 y0_{og} & = & y0_{p{g}} + (-d0_y/2) \\
 y0_{or} & = &  (-d0_y/2) \\
\end{array}
  ~~~~~~~~~~~{\rm for }~~~d0_y<=0.5
\label{eq:under0.5y}
\end{equation}
or
\begin{equation}
\begin{array}{ccl}
 y0_{og} & = & y0_{p{g}} + (1-d0_y)/2 \\
 y0_{or} & = & (1-d0_y)/2 \\
\end{array}
  ~~~~~~{\rm for }~~~0.5<d0_y~.
\label{eq:over0.5y}
\end{equation}
These position offsets minimize 
declines in the resolution of $r$- and $g$-band images.
The position offsets of $u$-, $i$- and $z$-bands,
$x0_{ou}$, $x0_{oi}$ and $x0_{oz}$,
are also derived by equation 
(\ref{eq:under0.5x}) or (\ref{eq:over0.5x}), and
(\ref{eq:under0.5y}) or (\ref{eq:over0.5y}),
replacing $g$ with $u$, $i$ or $z$.
The $x0_{pu}$, $x0_{pi}$ and $x0_{pz}$ are calculated by
equation (\ref{eq:x00g}),
replacing $g$ with $u$, $i$ or $z$
(Note that equation (\ref{eq:d0x}) is applied for the $g$-band only).
In the case of $r{\rm -}i$ 2D colourmaps,
$x1_{oi}$, $y1_{oi}$, $x1_{or}$ 
and $y1_{or}$ are calculated using the same convention.
When performing 1$\times$ resampling, 
bilinear filtering is adopted.

Then we compute the surface brightness $SB$ of each pixel on
a 1$\times$ resampled image and correct the reddening
due to dust extinction in our Galaxy, using \citet{sch98}.
The SDSS magnitudes $m$ 
are derived by the following equations used in 
the SDSS photometric pipeline (PHOTO; Lupton et al. 2001),
\begin{equation}
f/f0 = N_{\rm cnt}/T_{\rm exp} \cdot 
 10^{0.4(v_{\rm aa} + v_{\rm kk} \cdot v_{\rm air})}
\end{equation}
\begin{equation}
 m_{\rm asinh} = - \frac{2.5}{\ln 10} \cdot \left\{ {\rm asinh} \left( \frac{f/f0}{2b}
					    \right)+\ln b \right\}
\end{equation}
\begin{equation}
 m_{\rm Pogson} = - 2.5 \cdot \log(f/f0)
\end{equation}
where $N_{\rm cnt}$ is
counts 
within an aperture on the atlas image,
$T_{\rm exp}$ is exposure time in the fpC file,
$v_{\rm aa}$ , $v_{\rm kk}$ and $v_{\rm air}$
are zeropoint, extinction coefficient and airmass
in the tsField file,
and
$b$ is the asinh softening parameters.
See the DR2 Photometric Flux Calibration page
(http://www.sdss.org/dr2/algorithms/fluxcal.html)
for details of computing magnitude from SDSS CCD images.
Finally, we calculate `pixel by pixel' $K$-corrected
surface brightness using the ${\tt kcorrect.v3\_ 2}$ library
\citep{bla03}.
The same $K$-correction procedure is applied to each pixel in our galaxy
based on the $u$,$g$,$r$,$i$ and $z$ colours of the pixel. The $K$-correction
code uses several templates obtained as eigen vectors from the real SDSS
spectra. During the fitting procedure, the code finds the bestfit out
of the liner combinations of these template spectra.
Therefore, the code does not try to break the 
age-metallicity degeneracy, but uses typical spectra with typical ages and 
metallicities 
to find the restframe colour of 
galaxies. 
Details of the $K$-correction code itself are written in \citet{bla03}.

The concentration index $C_e$ is computed using an $r$-band image
using elliptical apertures which eliminates the effect of the apparent
axis ratio of the galaxy.
The details of the effect and method of
deriving the axis ratio and position angle are described in \citet{yam04s}.
To calculate the concentration index $C_e$ for the ellipse,
we consider area $A_e(a)$ of the ellipse of the semi-major axis
$a$ and axis ratio $\alpha$, and the integrated flux $F_e(a)$ within $A_e(a)$.
The Petrosian semi-major axis $a_{\rm P}$ for a given $\eta$ 
is defined by 
\begin{equation}
\eta{\rm =}
 \frac{ \{ F_e(1.25a_{\rm P}){\rm -}F_e(0.8a_{\rm P}) \} / \{ A_e(1.25a_{\rm P})\rm{-}A_e(0.8a_{\rm P}) \} }
      { F_e(a_{\rm P}) / A_e(a_{\rm P}) },
\end{equation}
where we take $\eta = 0.2$, and
the elliptical Petrosian flux $F_{\rm P}$ as
\begin{equation}
F_{\rm P} = F_e(ka_{\rm P})
\end{equation}
with $k$ set equal to 2, following the SDSS definition
\citep{str02}.
The Petrosian half-, 60\%- and 90\%-light
semi-major axes $a_{\rm 50}$, $a_{\rm 60}$ and $a_{\rm 90}$ are defined 
in such a way that the flux in the elliptical apertures of these
semi-major axes are
50\%, 60\% and 90\% of the elliptical Petrosian flux:
\begin{equation}
 F_e(a_{50}) = 0.5F_{\rm P},~~
 F_e(a_{60})=0.6F_{\rm P},~~
 F_e(a_{90})=0.9F_{\rm P}~.
\label{eq:F_p}
\end{equation}
We define our concentration index $C_{e}$ 
by 
\begin{equation}
 C_{e} = a_{\rm 50} / a_{\rm 90}~.
\end{equation}
The axis ratio and the position angle used above
are also adopted for the calculation of
the radial colour gradient.

The radial colour gradients are based on our pixel-to-pixel 
$K$-corrected $g{\rm -}r$ and $r{\rm -}i$ 2D colourmap
with Pogson magnitude.
The regression lines to radial colour $(SB_g - SB_r)$
and $(SB_r - SB_i)$,
\begin{equation}
\begin{array}{ccl}
 (SB_g - SB_r) & = & t_{g\mbox{-}r} + {\it CI_{g\mbox{-}r}} \cdot \log (a/a_{60}) \\
 (SB_r - SB_i) & = & t_{r\mbox{-}i} + {\it CI_{r\mbox{-}i}} \cdot \log (a/a_{60}),
\end{array}
\end{equation}
are calculated by linear least-squares fitting, 
and we define ${\it CI_{g\mbox{-}r}}$ and ${\it CI_{r\mbox{-}i}}$ 
as the radial colour gradient.  
The
$a$ is the equivalent distance from galaxy centre 
derived
by the axis ratio and position angle.  
The Petrosian 60\% flux radius corresponds to an effective radius 
with a pure de Vaucouleurs' profile.  
Generally the effective radius is used for the normalization 
of the radial colour gradient, but profile fitting is needed 
to calculate the effective radius.  
However, some E+A morphologies are somewhat disturbed 
and it is difficult to obtain a stable profile fit. 
Thus, we use the Petrosian 60\% flux semi-major axis 
for stability of calculation.  
When this linear least-squares fitting is applied, 
the $a/a_{60}$ of data points are restricted 
to $0.35 < a/a_{60} < 1.0$ to avoid a cusp 
in the galaxy centre and lower S/N regions in the outskirts.  
The minimum size of $a_{60}$ in our samples applied 
to the colour gradient analysis is $2.83''$($7.14$ pixels), 
and $a/a_{60}=0.35$ corresponds to the diameter of $2.0''$, 
since the typical FWHM of the SDSS point-spread function is $\sim 1.5''$ \citep{shi01}.
We confirmed by our tests using a control sample that the effects of
seeing on the derived colour gradients
are not significant.
The $({\rm FWHM}/2)/a_{60}$ v.s. $|CI_{g{\rm -}r}|$  
shows neither correlation (only 0.17 of correlation coefficient)
nor systematic effect of seeing.

\section{Results}
\label{section:results}

\subsection{E+A Morphologies}

Since E+A galaxies have experienced starburst truncation fairly recently ($<$ 1Gyr), E+A galaxies might still hold some traces
in their morphology (e.g., dynamically disturbed signs).  Therefore we might obtain some hint of the origins of E+A galaxies by examining their morphology.

\begin{figure}
\includegraphics[scale=0.68]{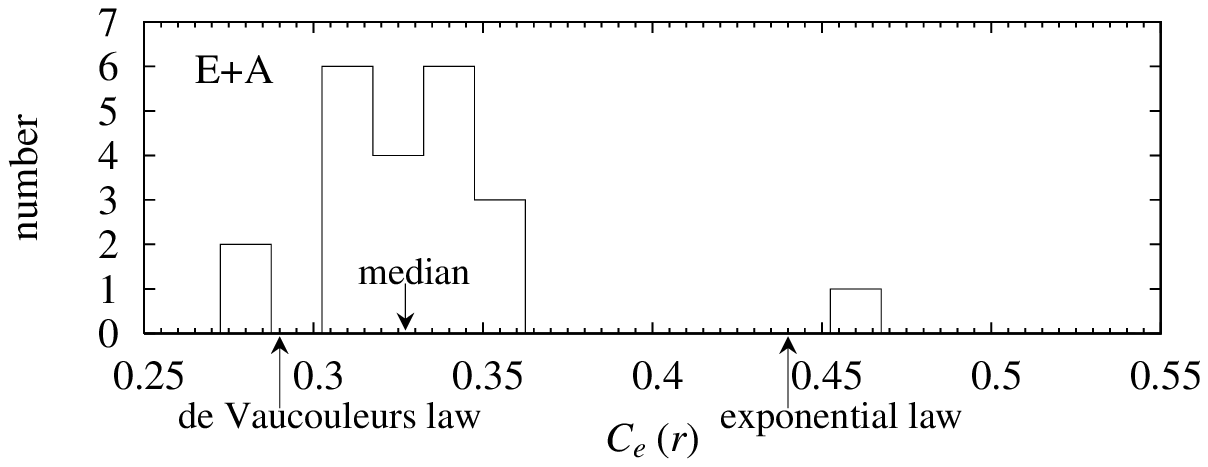}
 \caption{The distributions of concentration index $C_e$ of 22 E+A
 galaxies.  Galaxies that follow de Vaucouleurs' law give $C_e=0.29$
 and those with the exponential profile give 0.44.  All our E+A galaxies
 except one are centrally concentrated galaxies.  }\label{fig:ce_hist}
\end{figure}

\begin{figure*}
 \begin{tabular}{cc}
\includegraphics[scale=0.80]{e+a/atlas-002206-3-0155-0080.fit_colors.eps3} &
\includegraphics[scale=0.80]{e+a/atlas-000752-3-0251-0008.fit_colors.eps3} \\
\includegraphics[scale=0.64]{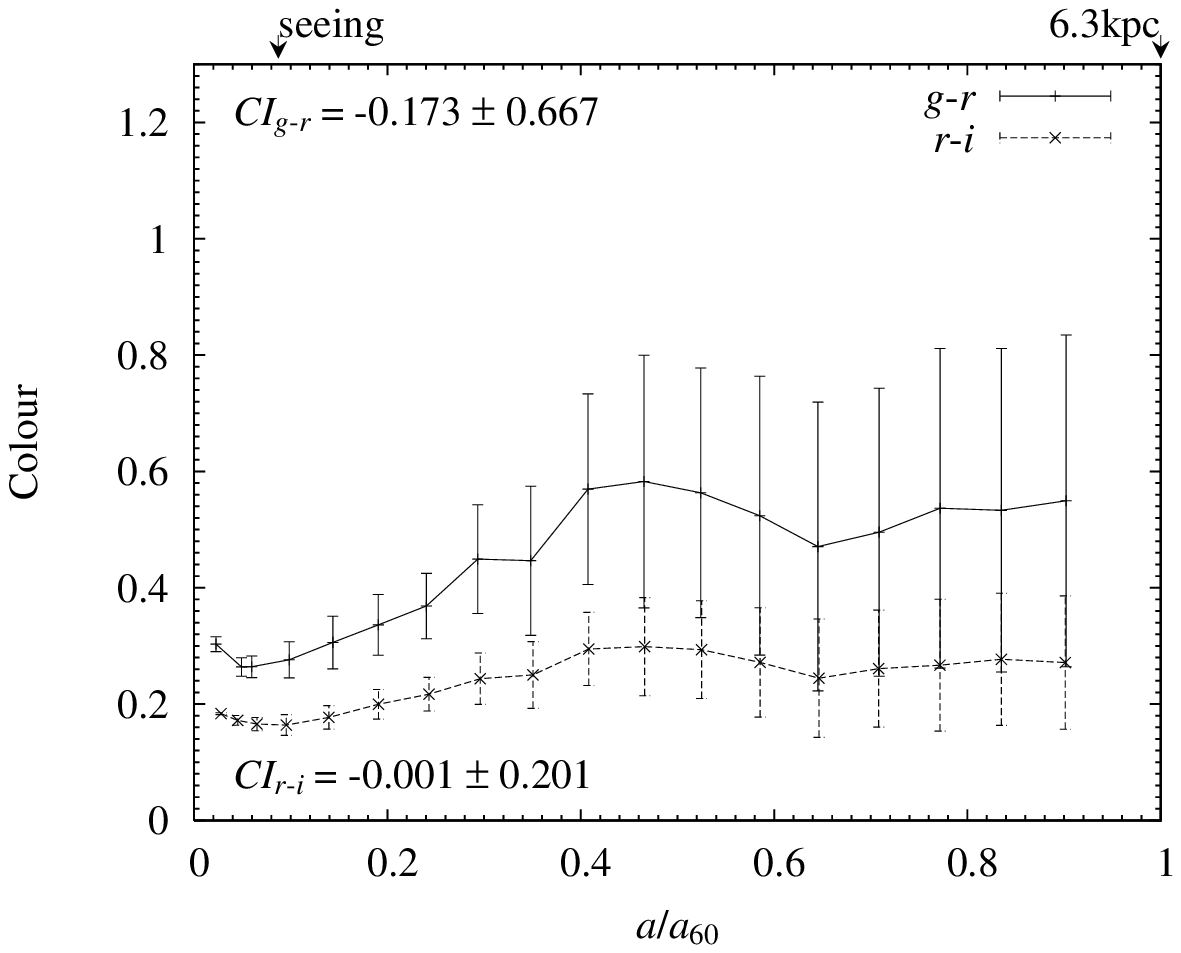} &
\includegraphics[scale=0.64]{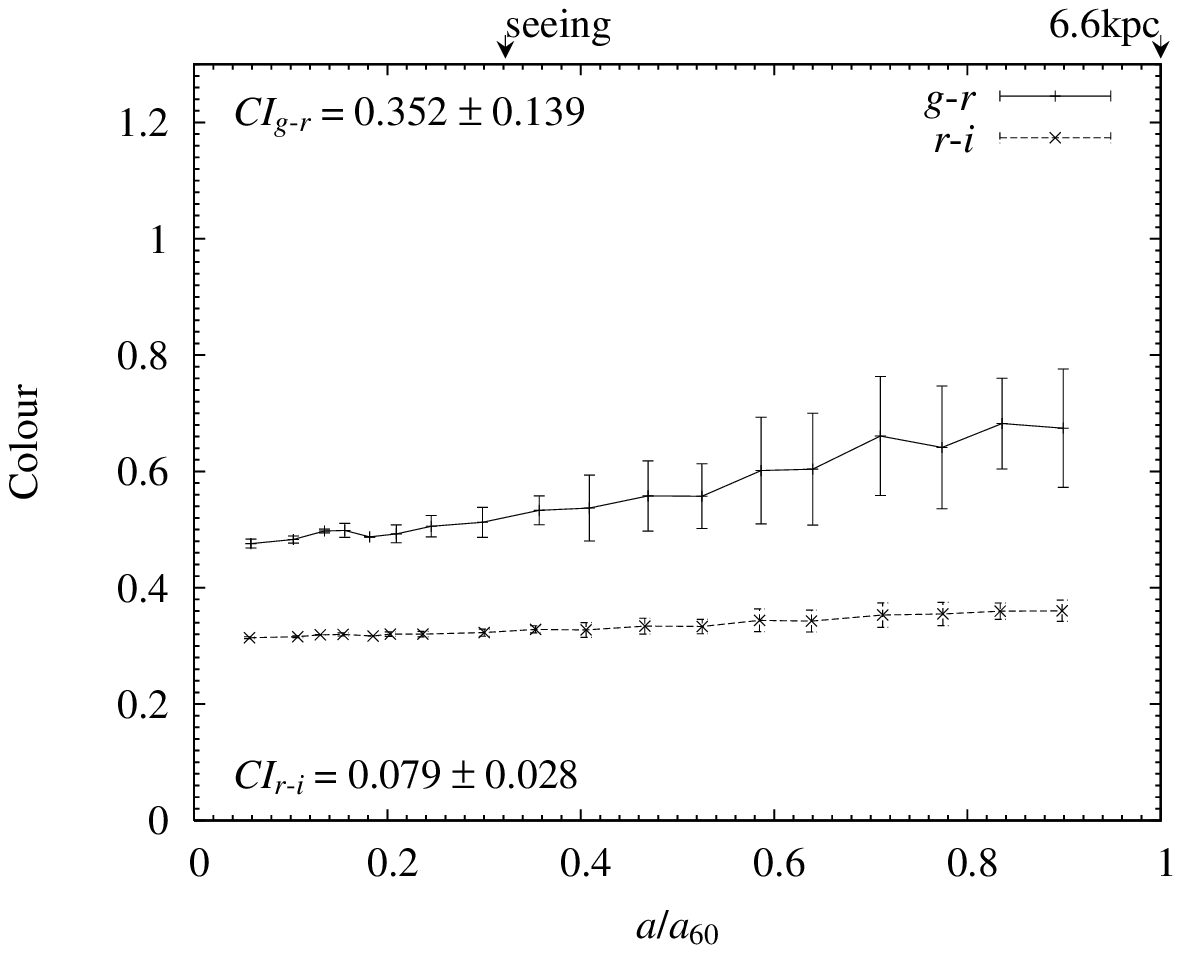} \\
\includegraphics[scale=0.80]{e+a/atlas-002137-3-0220-0074.fit_colors.eps3} &
\includegraphics[scale=0.80]{e+a/atlas-000756-5-0347-0219.fit_colors.eps3} \\
\includegraphics[scale=0.64]{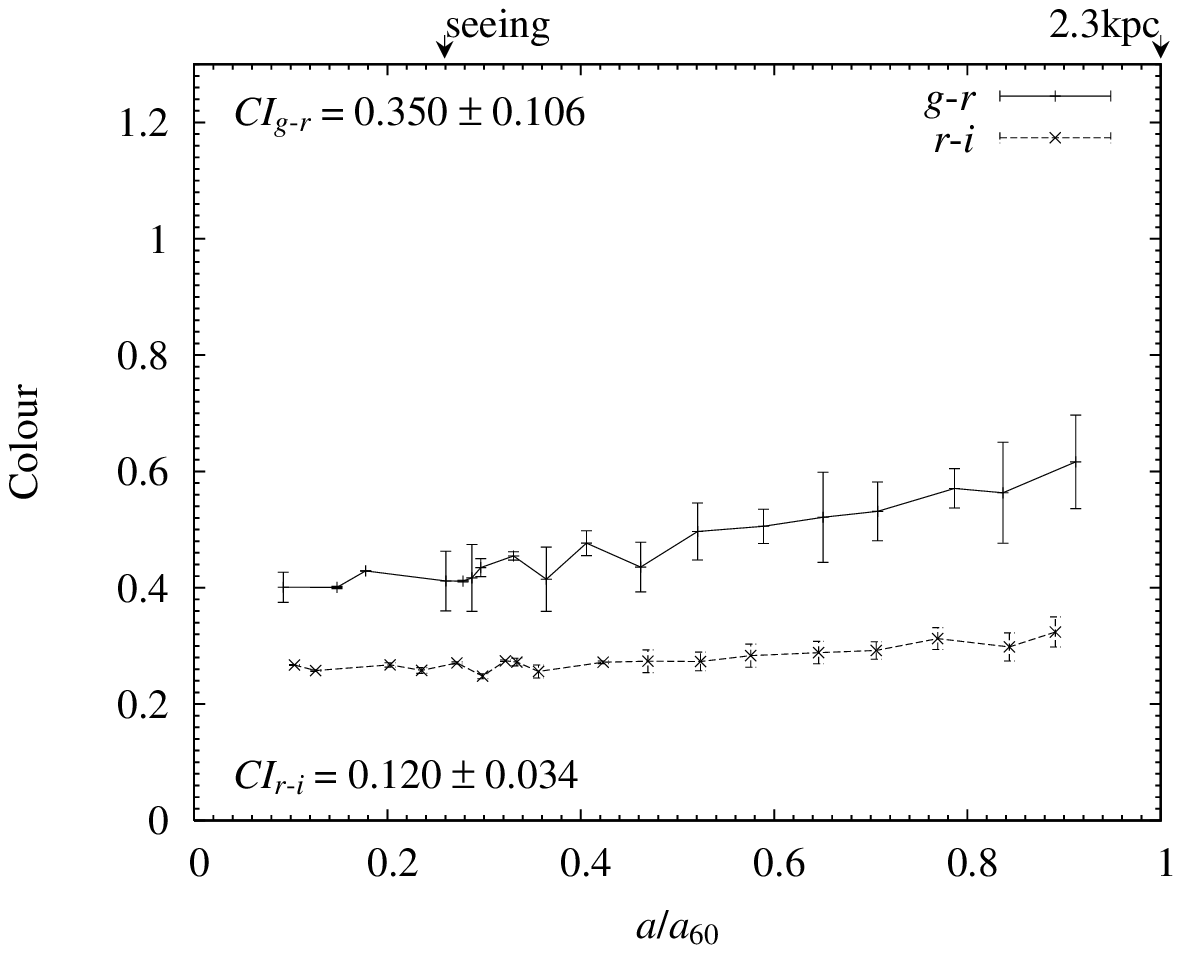} &
\includegraphics[scale=0.64]{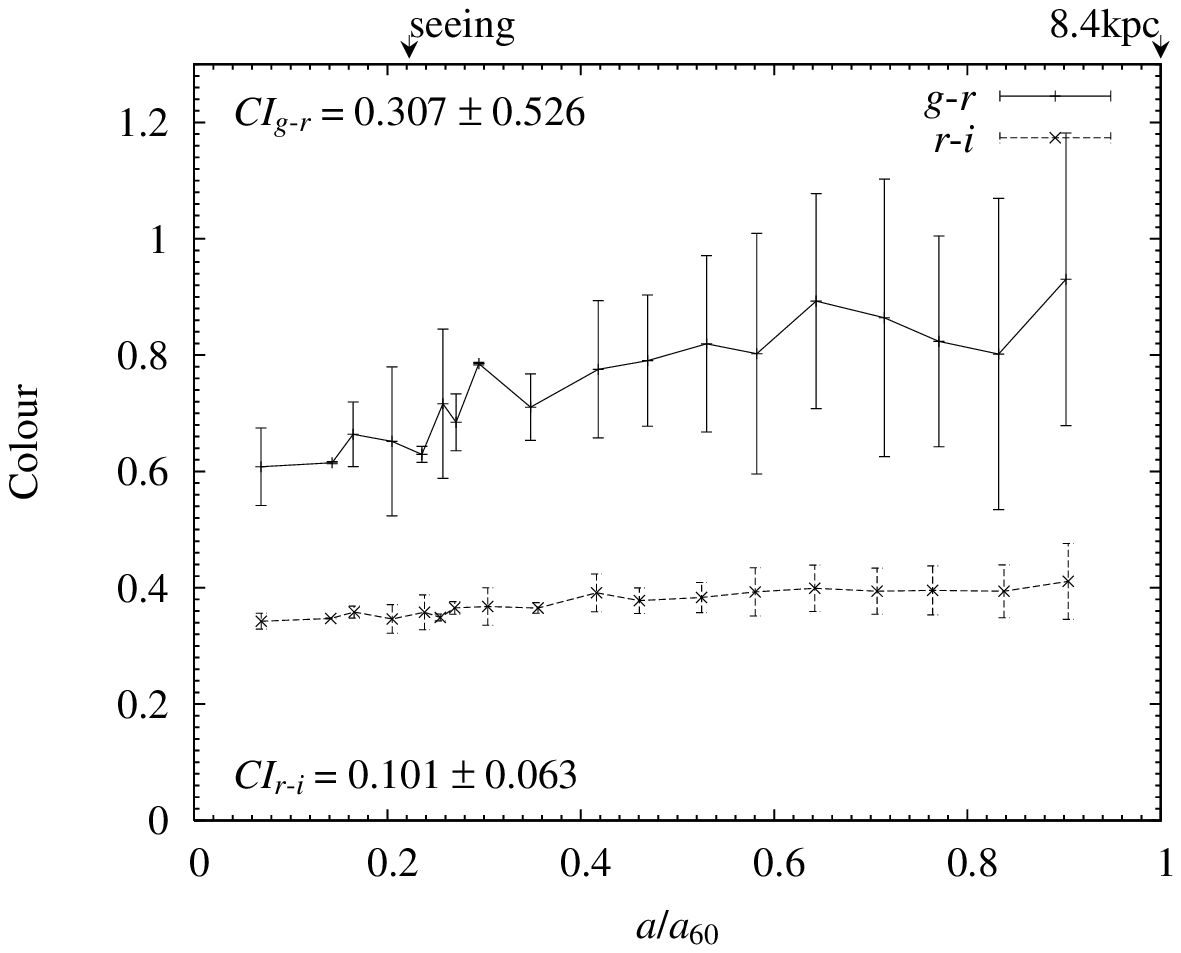} \\
 \end{tabular}
\caption{ $g$-, $r$- and $i$-band images, 
$g{\rm -}r$ and $r{\rm -}i$ 2D colourmaps ({\it each top}) 
and the rest colour profiles ({\it each bottom}) of our all E+A galaxies.  
The error bars represent standard deviation
of $g{\rm -}r$ or $r{\rm -}i$ values on sampling points.
The details of explanation of image panels are inlaid in Figure \ref{fig:example_profiles_early}.  The panels are placed in ascending order with respect to $D_{4000}$.  We numbered our E+As in this order, so this Figure includes No.1({\it top left}), No.2({\it top right}), No.3({\it bottom left}) and No.4({\it bottom right}) E+A galaxies.  E+As with smaller $D_{4000}$ tend to show positive slopes of radial colour gradients (bluer gradients toward the centre) and blue irregular patterns in 2D colourmaps.  Meanwhile, those with larger $D_{4000}$ show negative or flat slope and moderate property in 2D colourmap.  }\label{fig:e+a_all}
\end{figure*}

\begin{figure*}
 \begin{tabular}{cc}
\includegraphics[scale=0.80]{e+a/atlas-002326-6-0153-0137.fit_colors.eps3} &
\includegraphics[scale=0.80]{e+a/atlas-000752-6-0161-0162.fit_colors.eps3} \\
\includegraphics[scale=0.64]{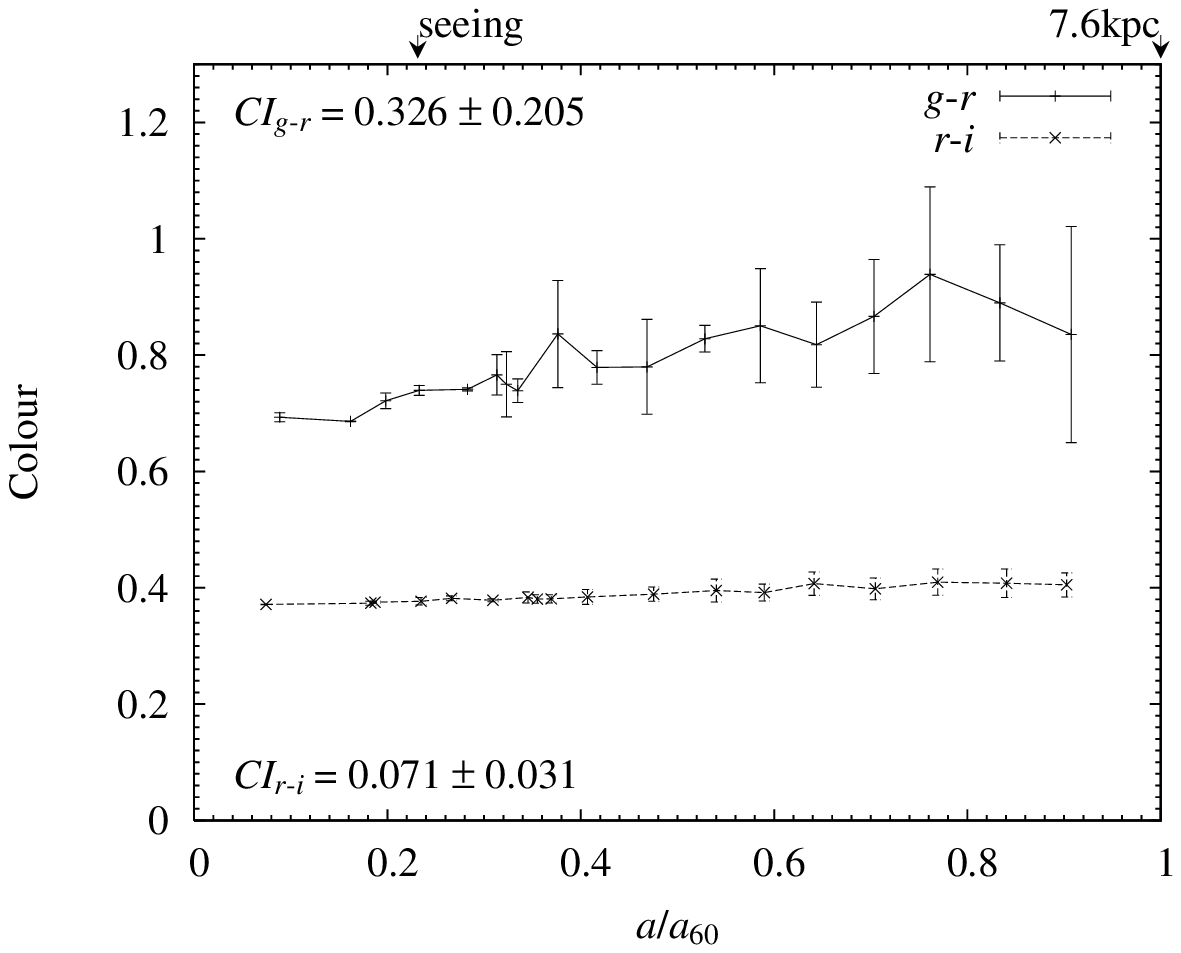} &
\includegraphics[scale=0.64]{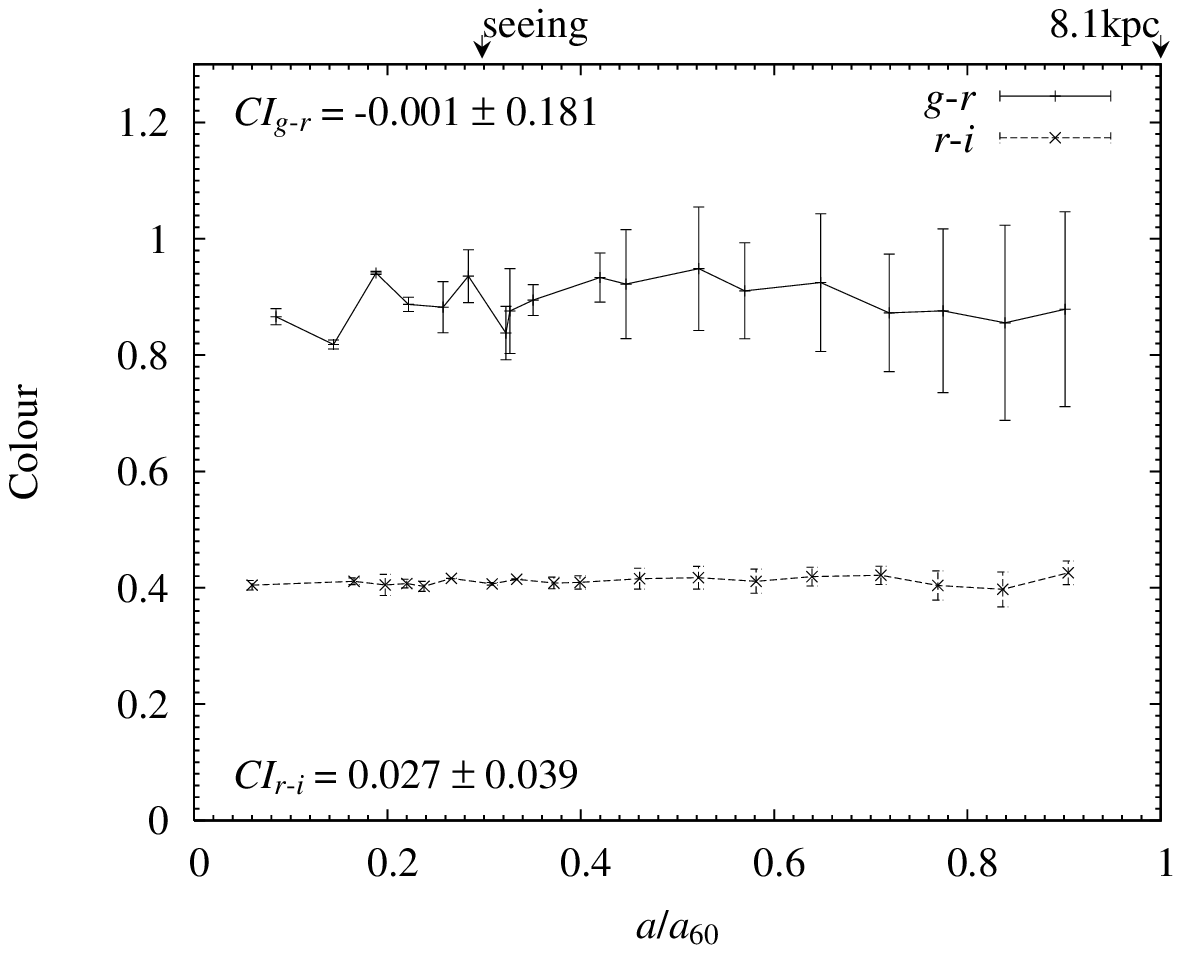} \\
\includegraphics[scale=0.80]{e+a/atlas-001729-3-0475-0027.fit_colors.eps3} &
\includegraphics[scale=0.80]{e+a/atlas-002335-3-0085-0502.fit_colors.eps3} \\
\includegraphics[scale=0.64]{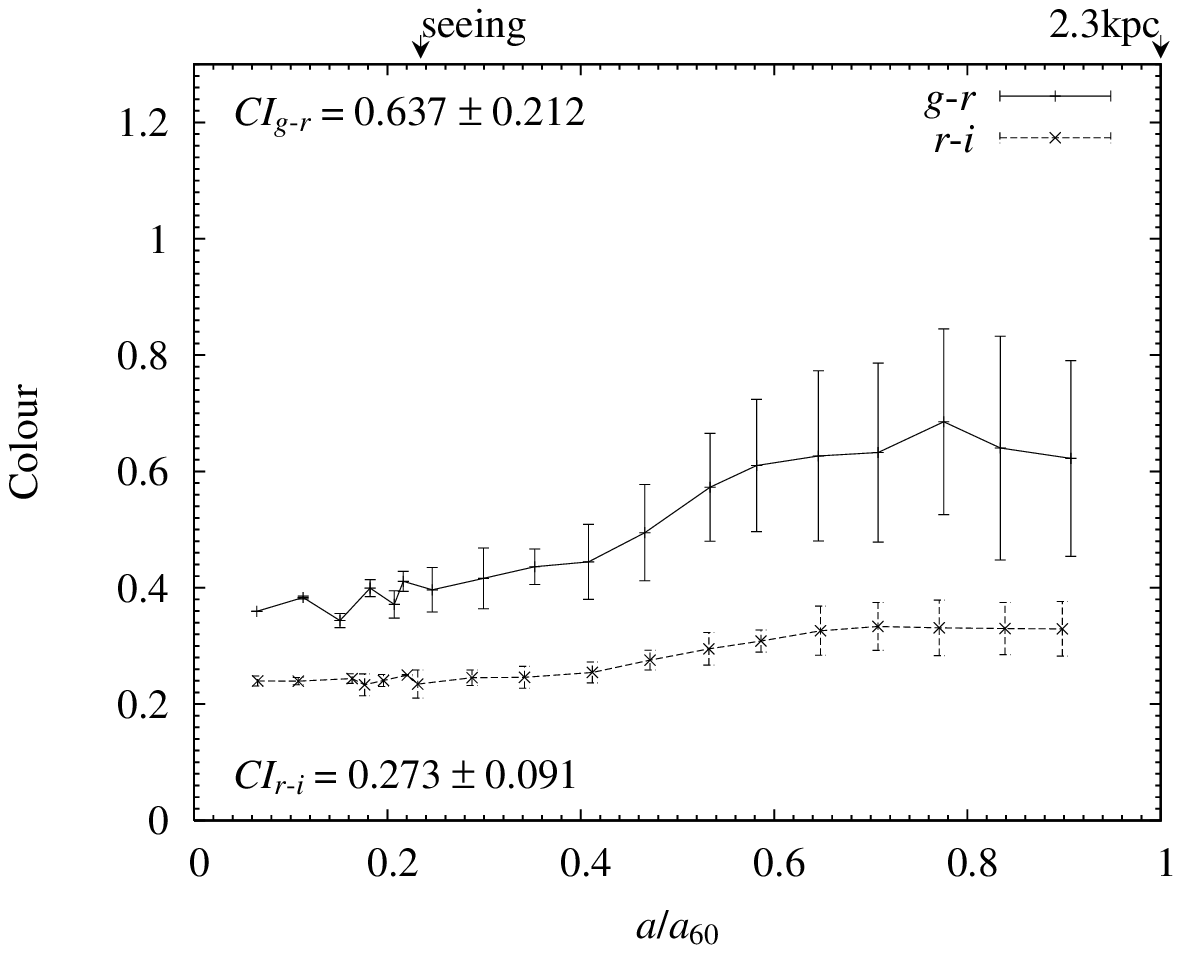} &
\includegraphics[scale=0.64]{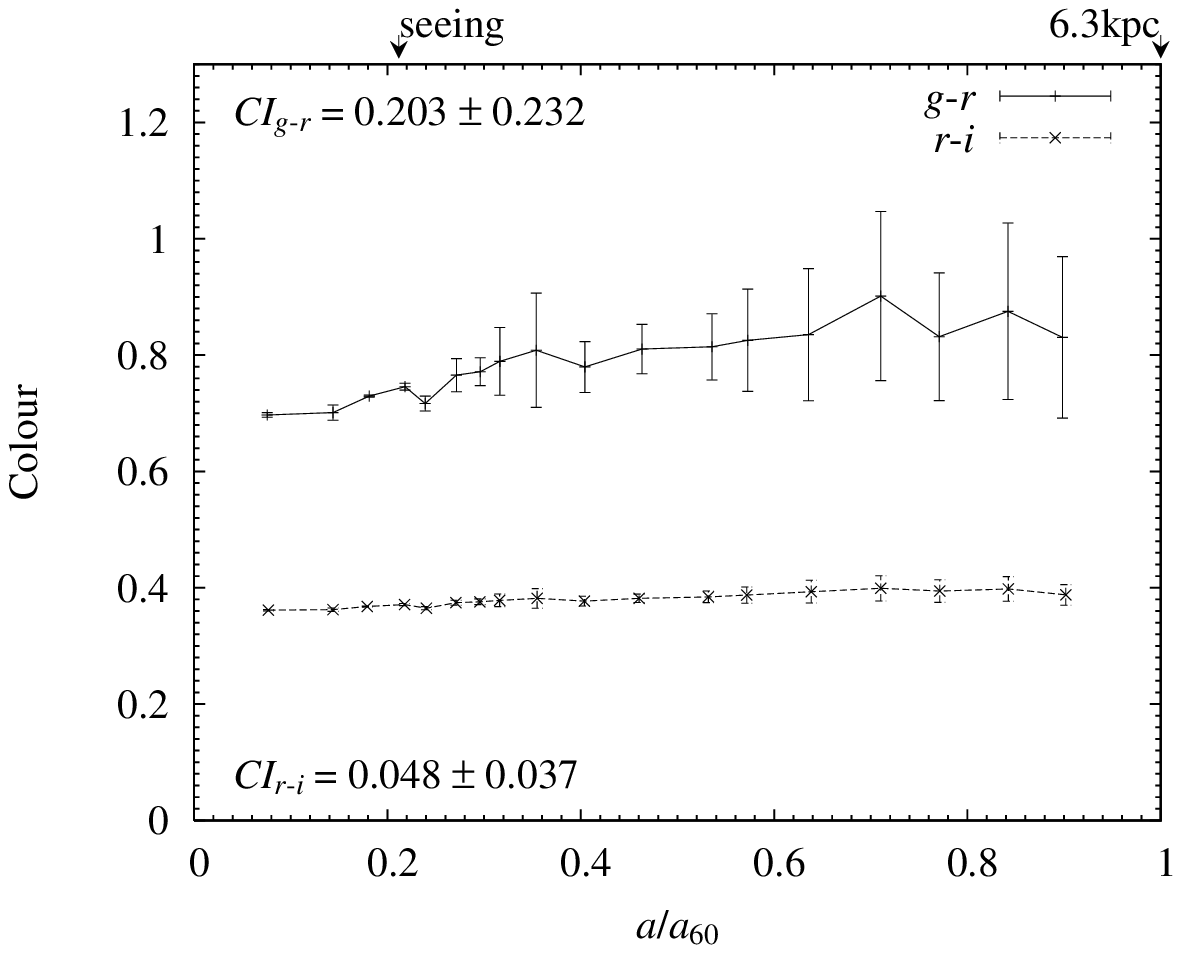} \\
 \end{tabular}
\contcaption{No.5, No.6, No.7 and No.8 E+A galaxies}
\end{figure*}

\begin{figure*}
 \begin{tabular}{cc}
\includegraphics[scale=0.80]{e+a/atlas-003180-3-0060-0125.fit_colors.eps3} &
\includegraphics[scale=0.80]{e+a/atlas-002305-2-0080-0363.fit_colors.eps3} \\
\includegraphics[scale=0.64]{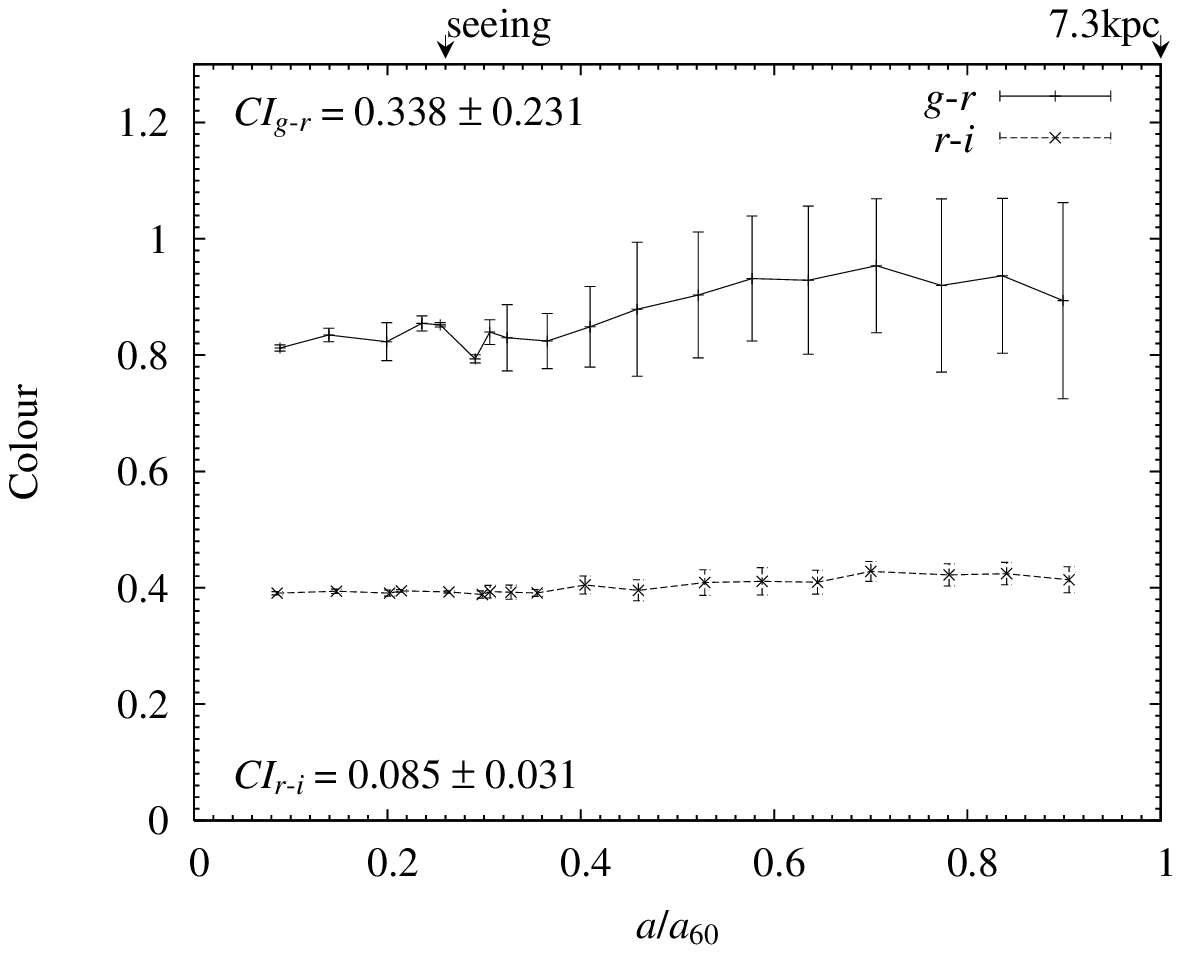} &
\includegraphics[scale=0.64]{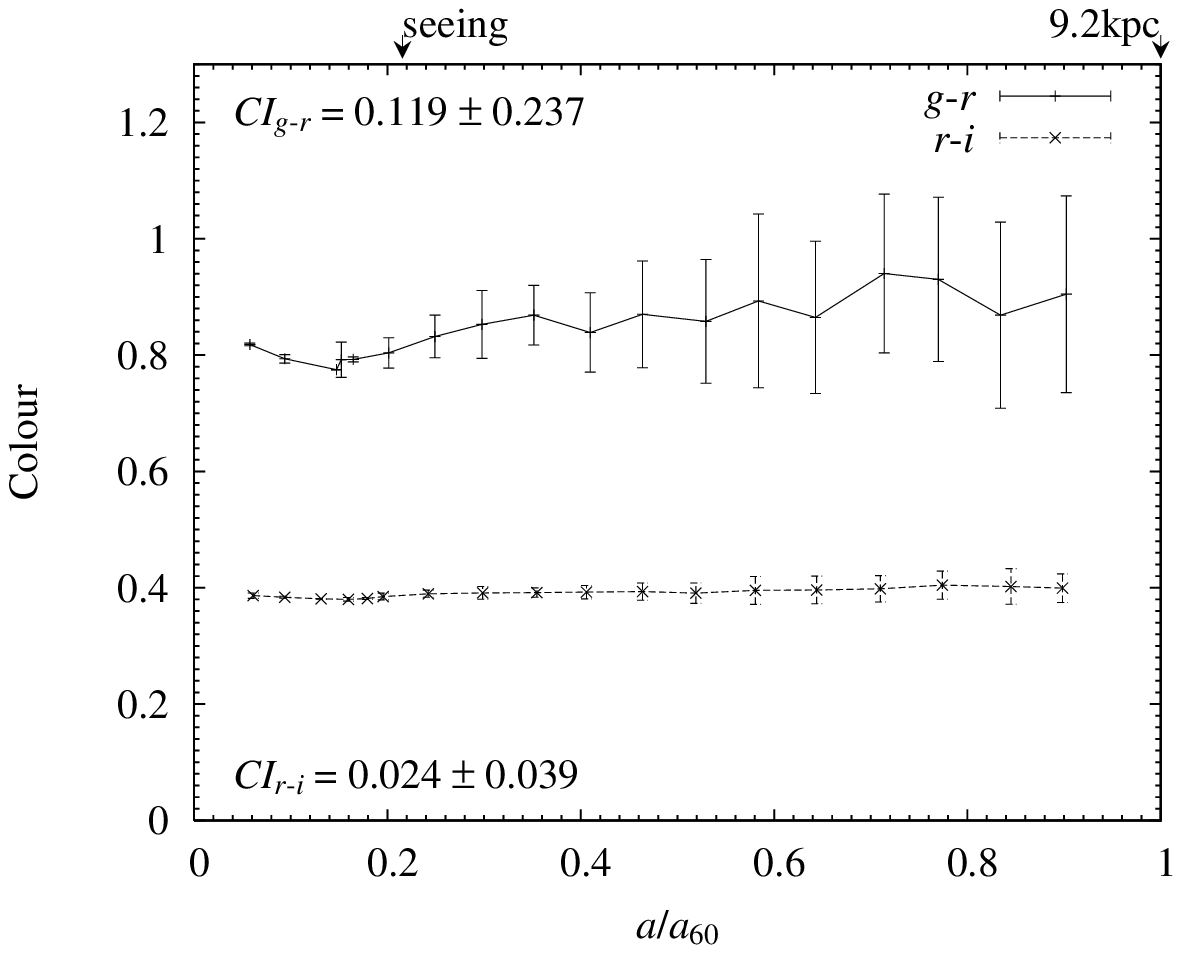} \\
\includegraphics[scale=0.80]{e+a/atlas-001231-4-0069-0110.fit_colors.eps3} &
\includegraphics[scale=0.80]{e+a/atlas-002987-4-0033-0037.fit_colors.eps3} \\
\includegraphics[scale=0.64]{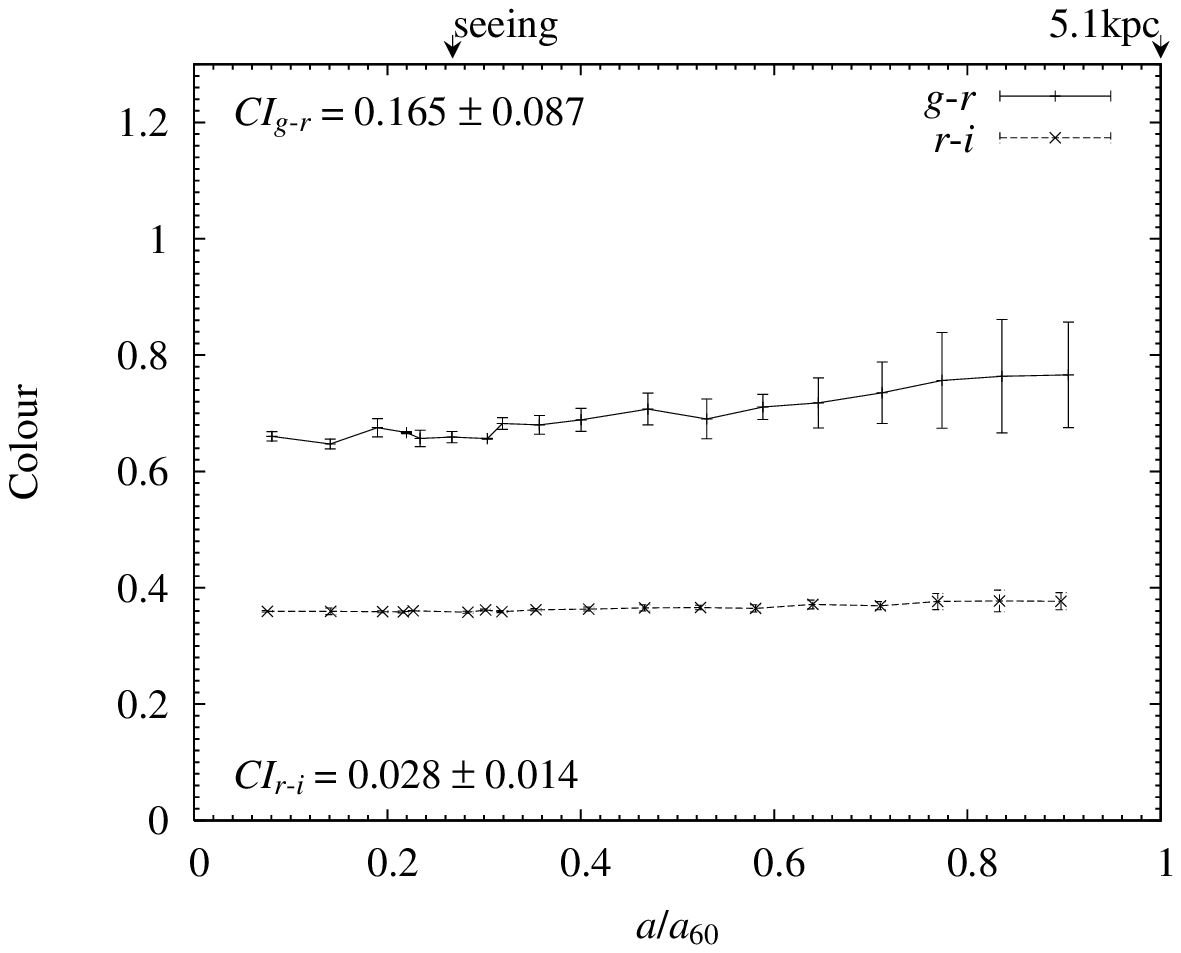} &
\includegraphics[scale=0.64]{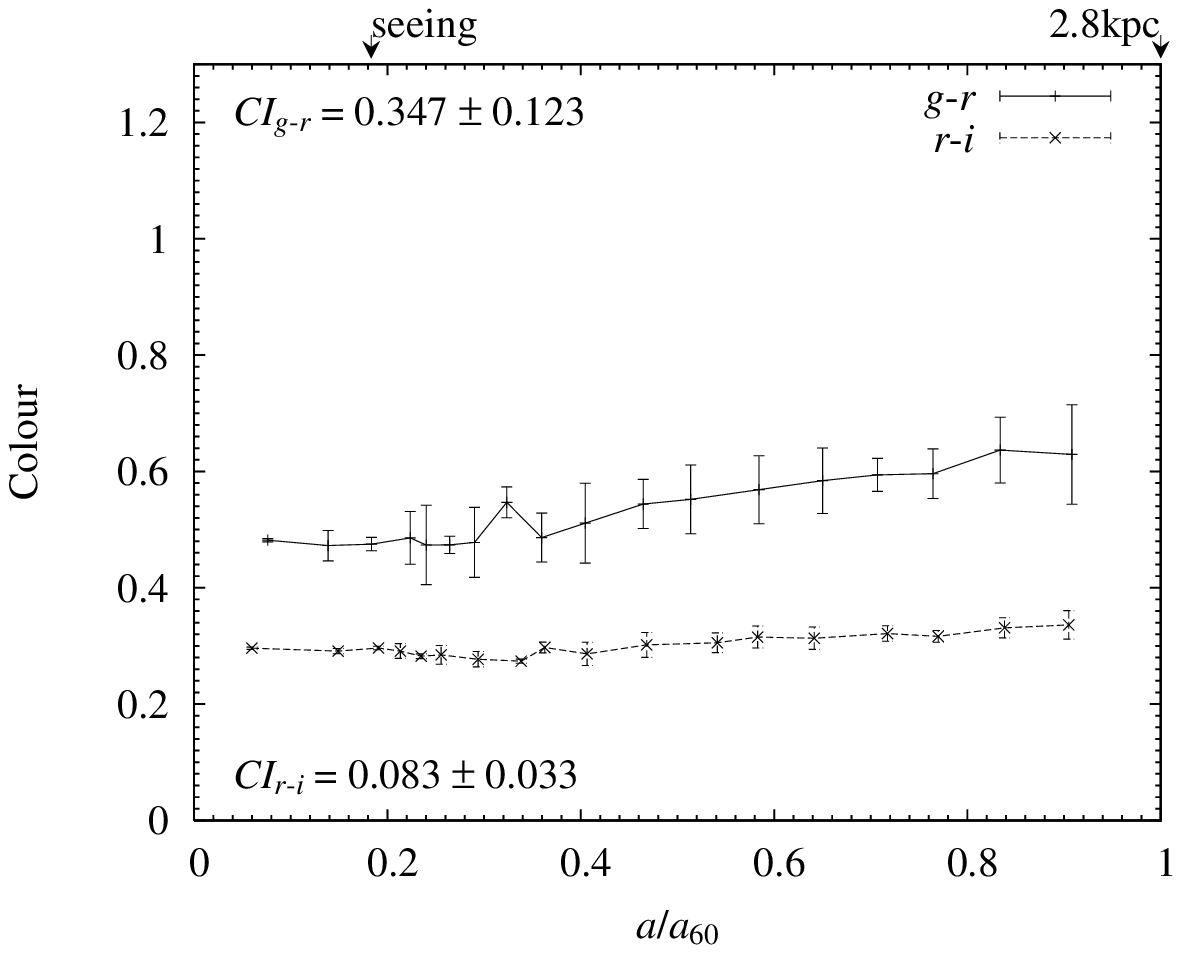} \\
 \end{tabular}
\contcaption{No.9, No.10, No.11 and No.12 E+A galaxies}
\end{figure*}

\begin{figure*}
 \begin{tabular}{cc}
\includegraphics[scale=0.80]{e+a/atlas-002189-2-0088-0019.fit_colors.eps3} &
\includegraphics[scale=0.80]{e+a/atlas-001659-4-0089-0037.fit_colors.eps3} \\
\includegraphics[scale=0.64]{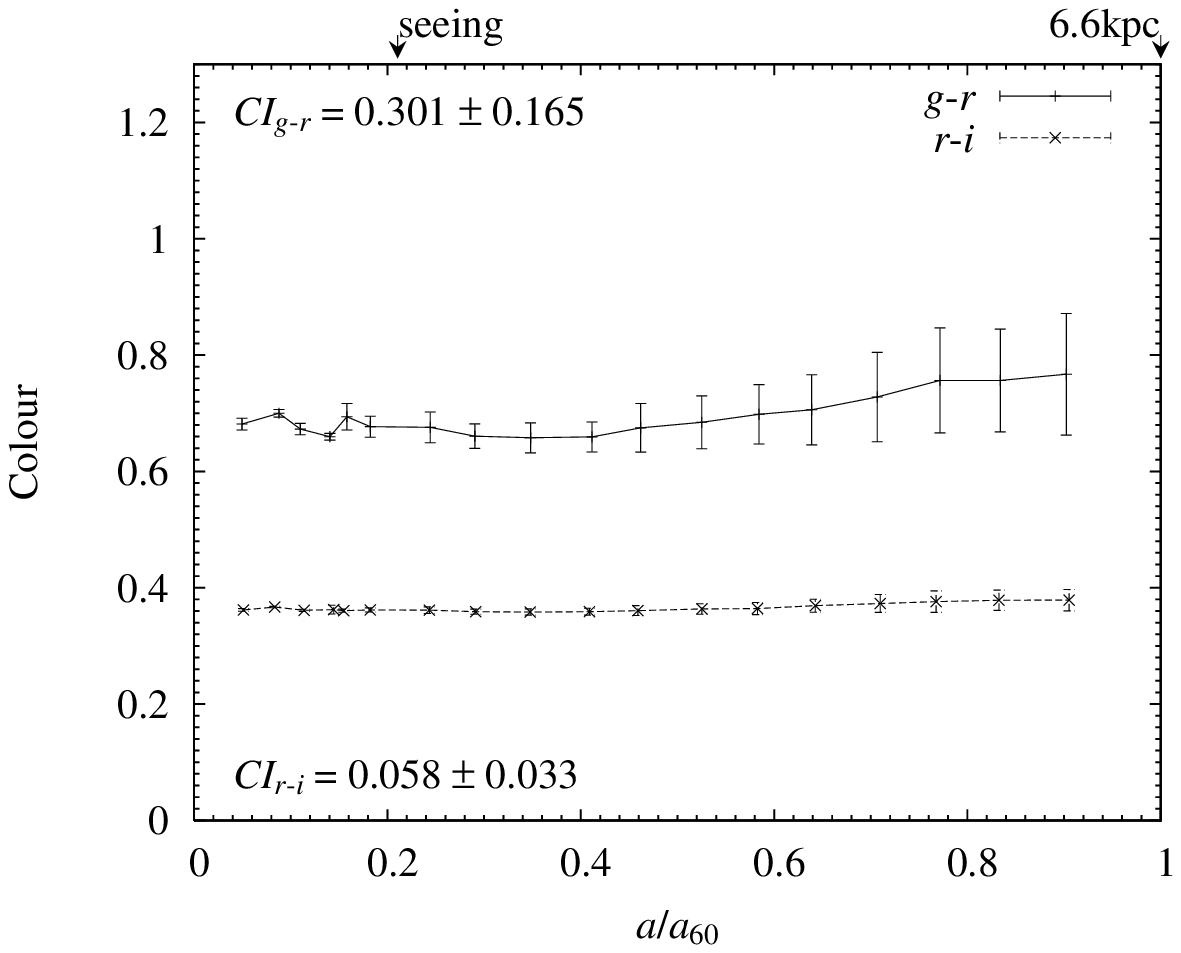} &
\includegraphics[scale=0.64]{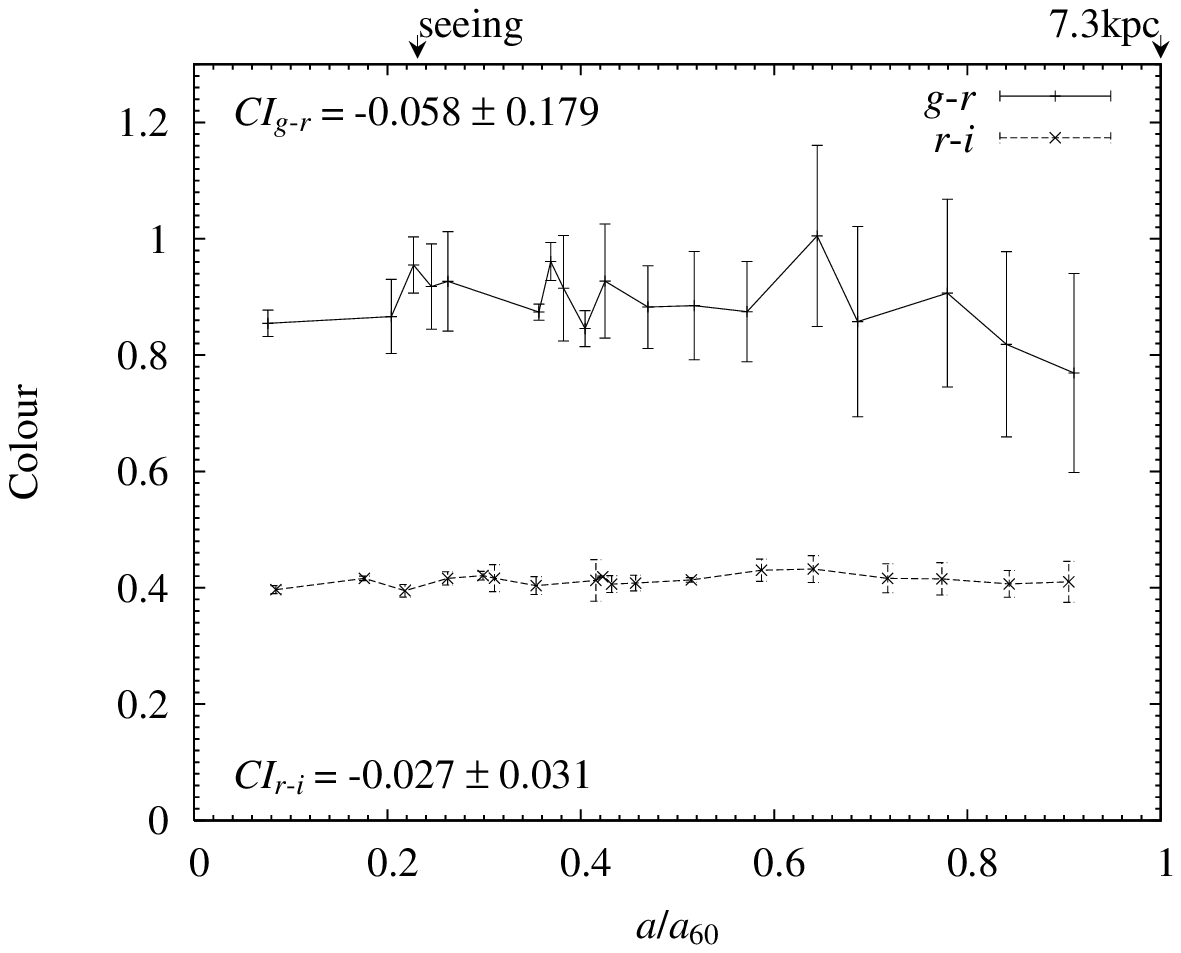} \\
\includegraphics[scale=0.80]{e+a/atlas-002305-5-0029-0126.fit_colors.eps3} &
\includegraphics[scale=0.80]{e+a/atlas-001035-2-0073-0009.fit_colors.eps3} \\
\includegraphics[scale=0.64]{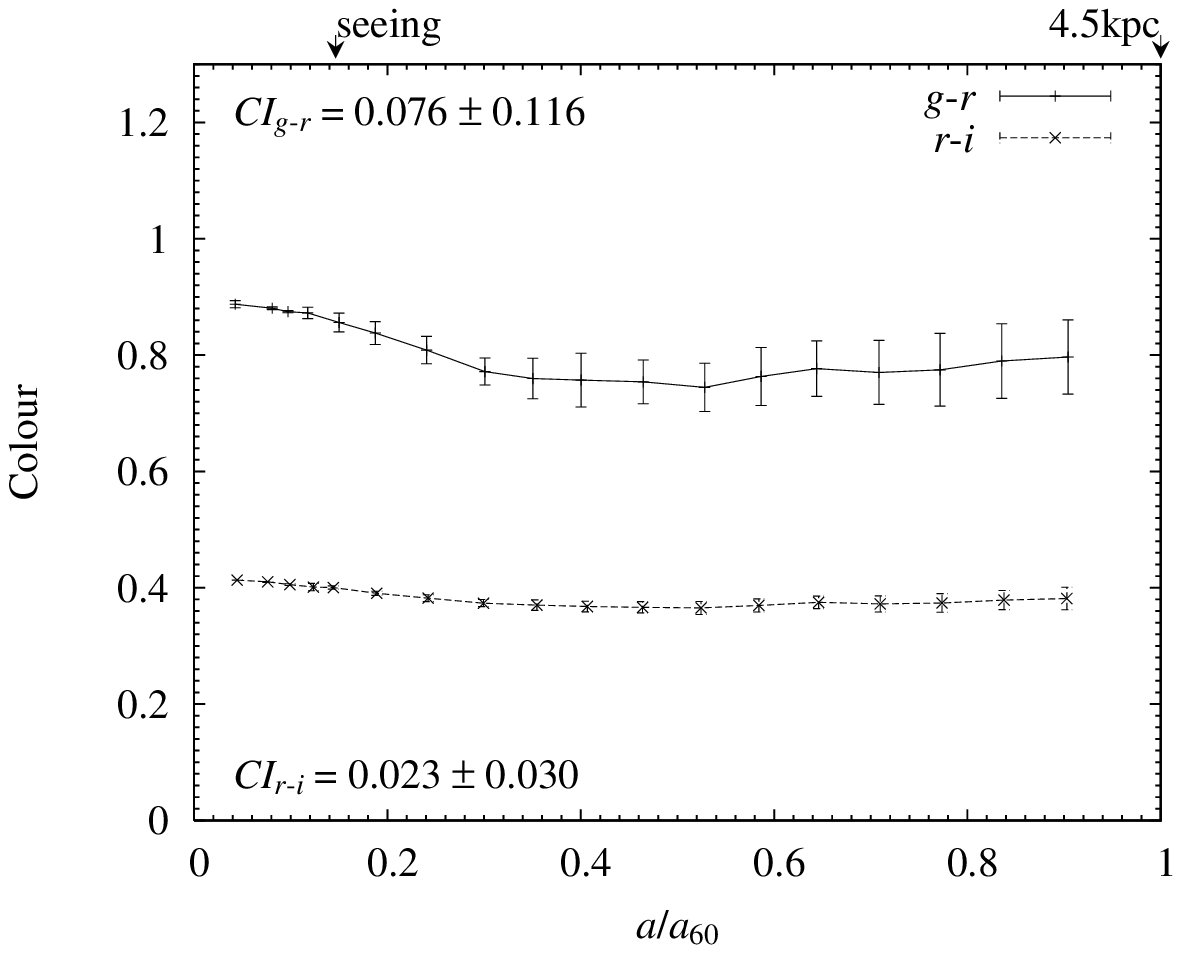} &
\includegraphics[scale=0.64]{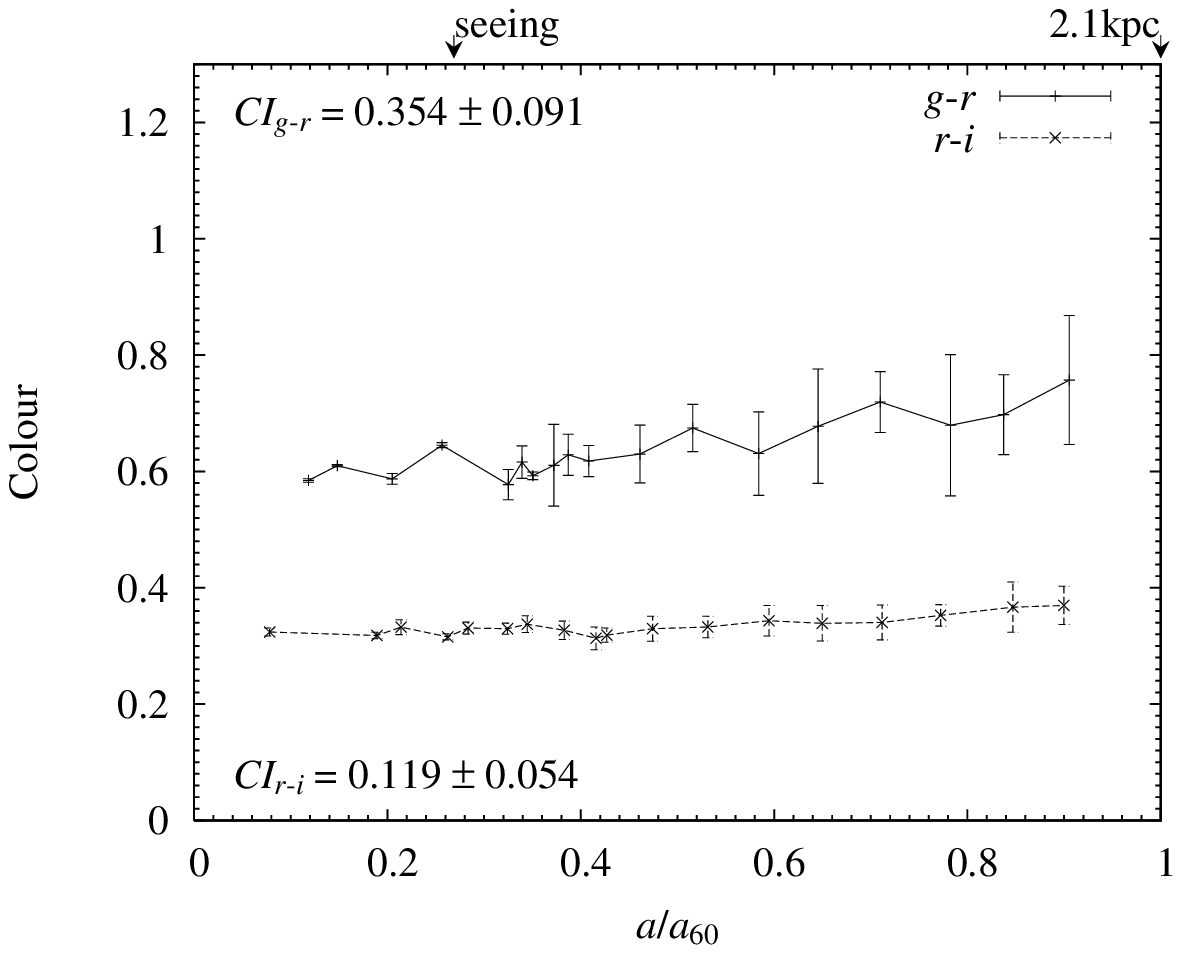} \\
 \end{tabular}
\contcaption{No.13, No.14, No.15 and No.16 E+A galaxies}
\end{figure*}

\begin{figure*}
 \begin{tabular}{cc}
\includegraphics[scale=0.80]{e+a/atlas-000752-3-0400-0183.fit_colors.eps3} &
\includegraphics[scale=0.80]{e+a/atlas-002740-3-0163-0118.fit_colors.eps3} \\
\includegraphics[scale=0.64]{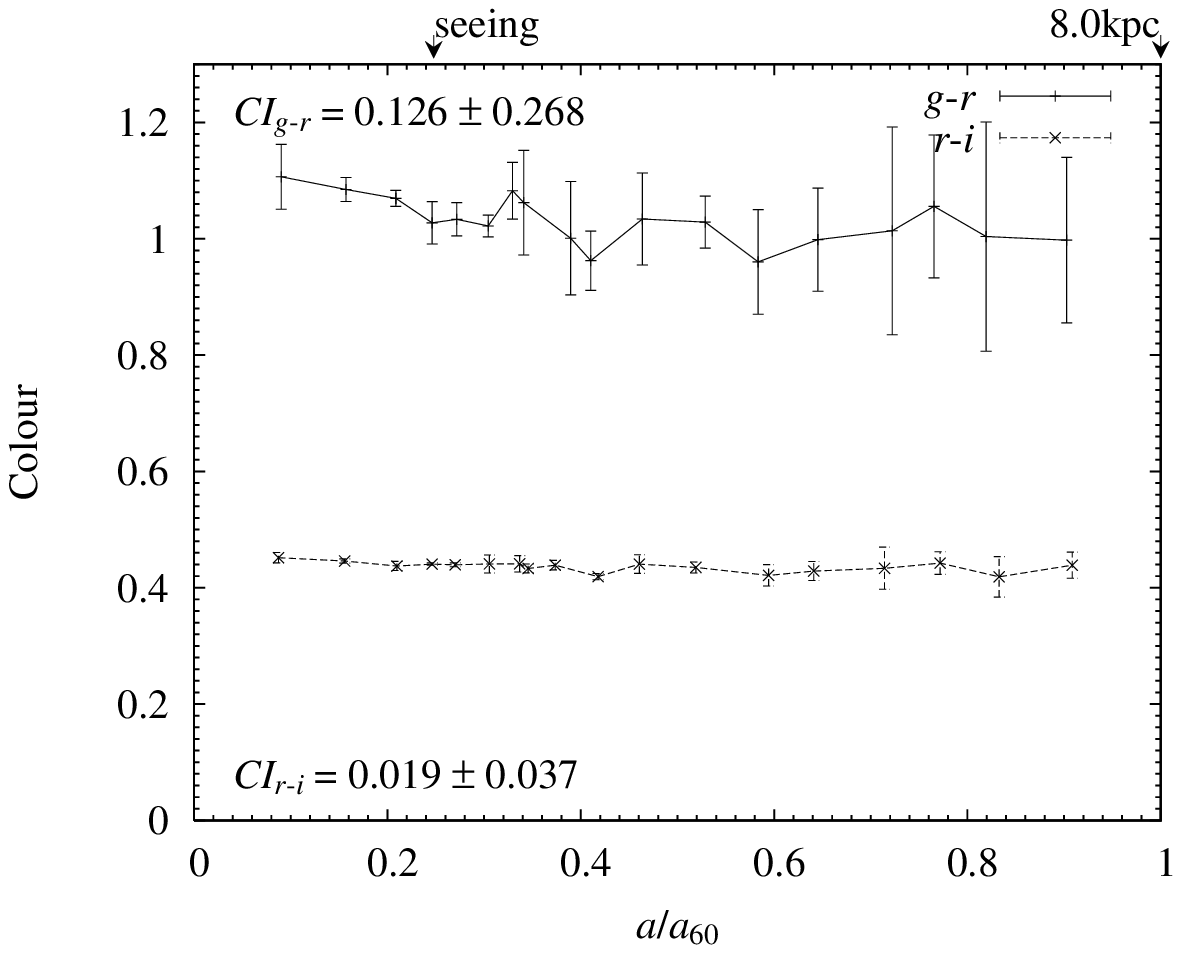} &
\includegraphics[scale=0.64]{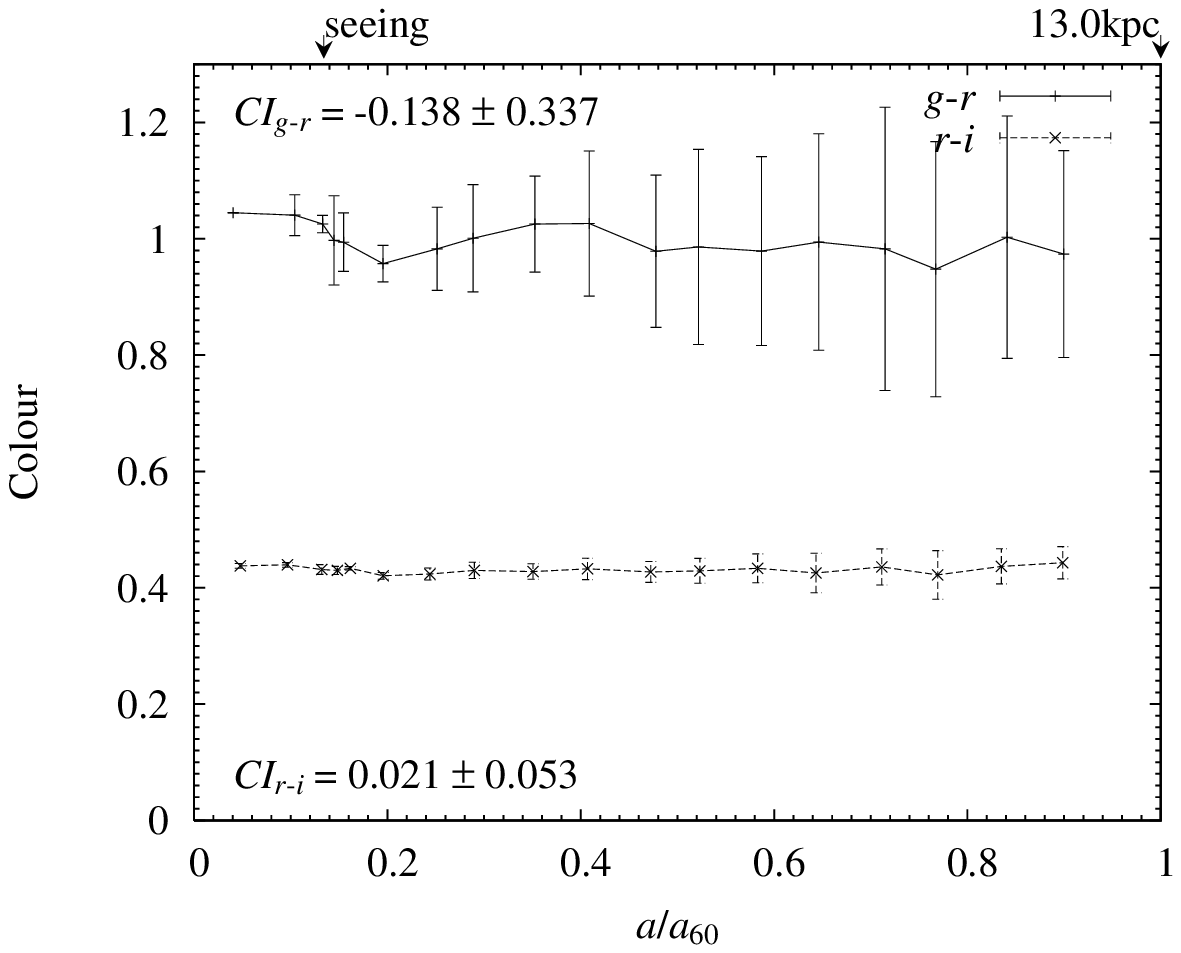} \\
\includegraphics[scale=0.80]{e+a/atlas-001239-4-0162-0256.fit_colors.eps3} &
\includegraphics[scale=0.80]{e+a/atlas-001345-1-0409-0130.fit_colors.eps3} \\
\includegraphics[scale=0.64]{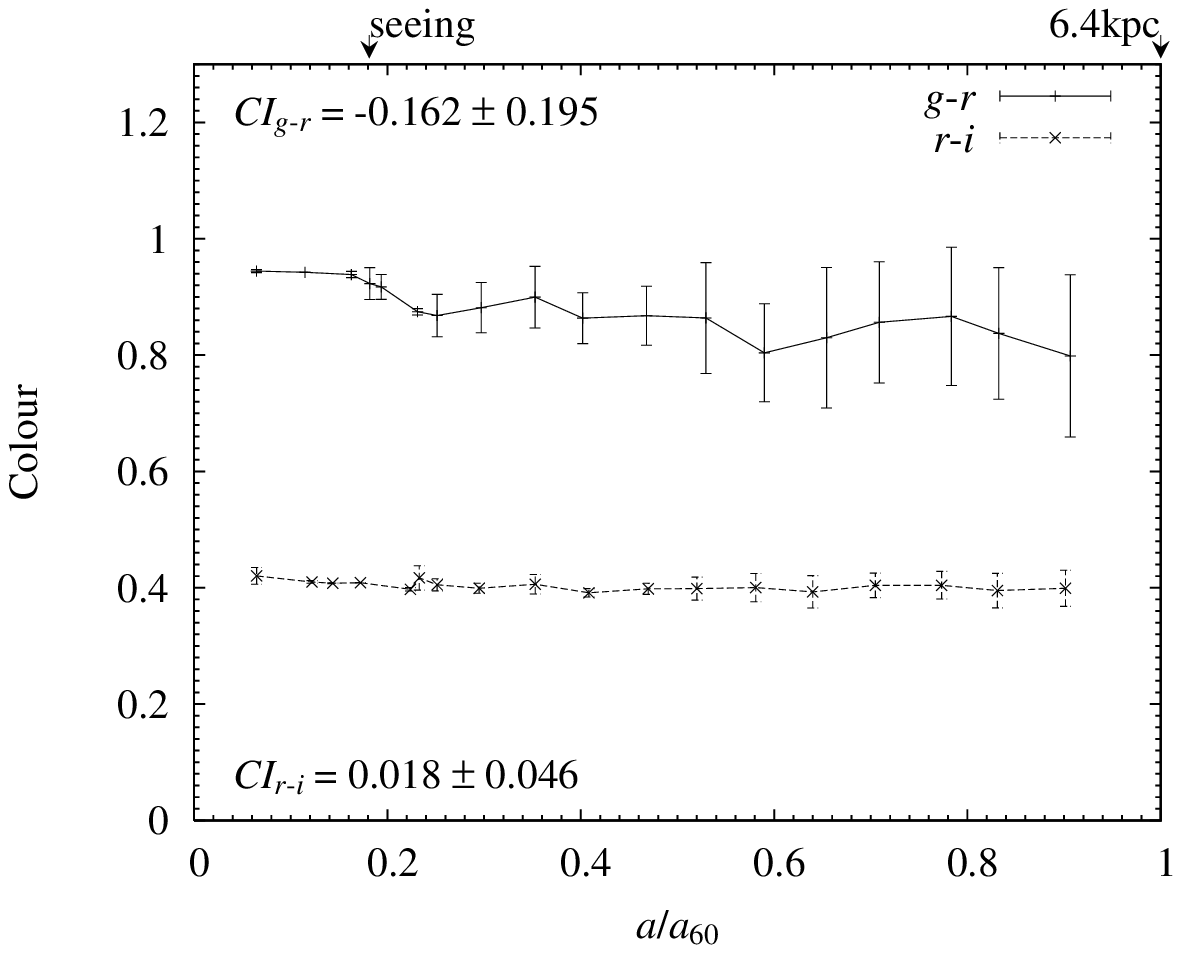} &
\includegraphics[scale=0.64]{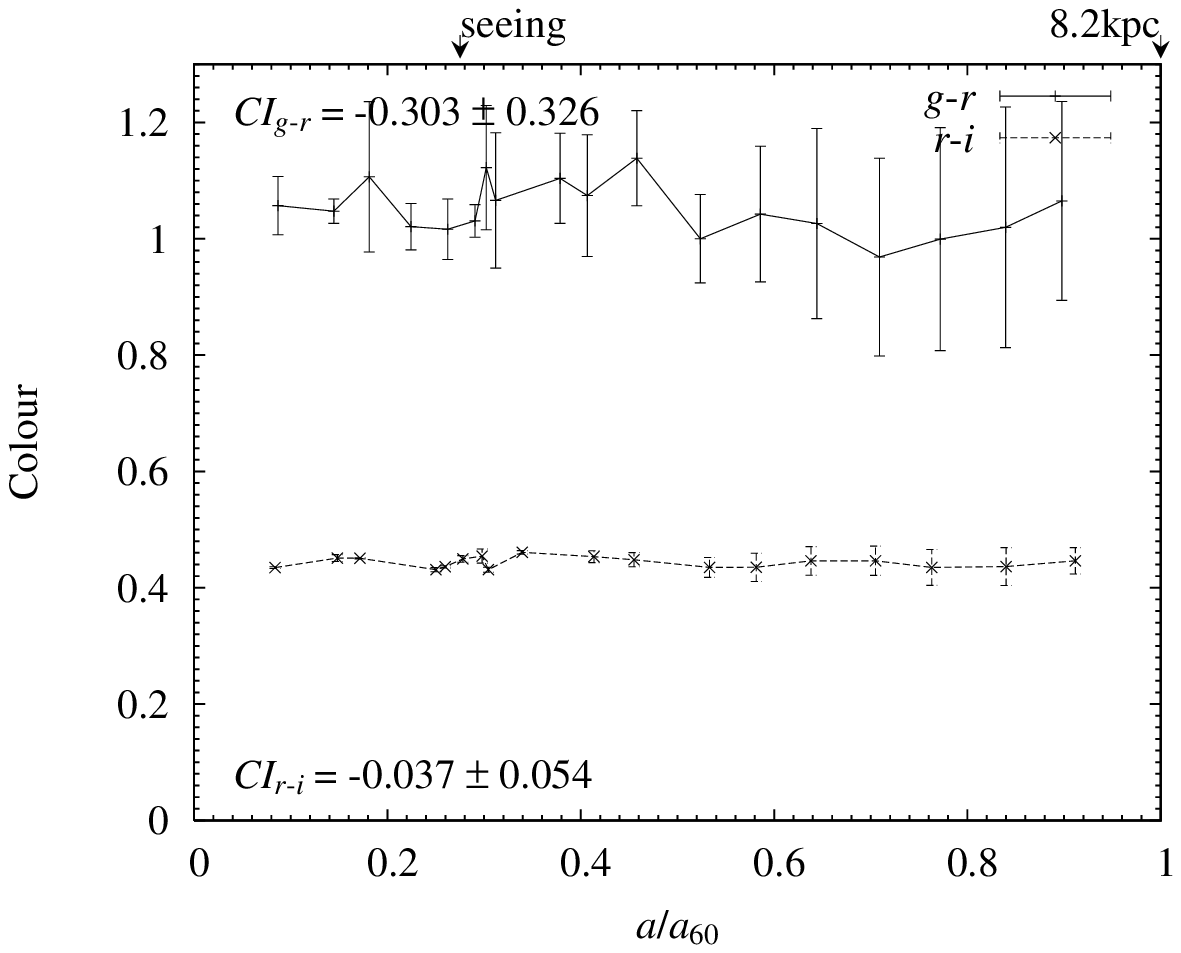} \\
 \end{tabular}
\contcaption{No.17, No.18, No.19 and No.20 E+A galaxies}
\end{figure*}

\begin{figure*}
 \begin{tabular}{cc}
\includegraphics[scale=0.80]{e+a/atlas-002334-1-0118-0218.fit_colors.eps3} &
\includegraphics[scale=0.80]{e+a/atlas-000752-1-0282-0090.fit_colors.eps3} \\
\includegraphics[scale=0.64]{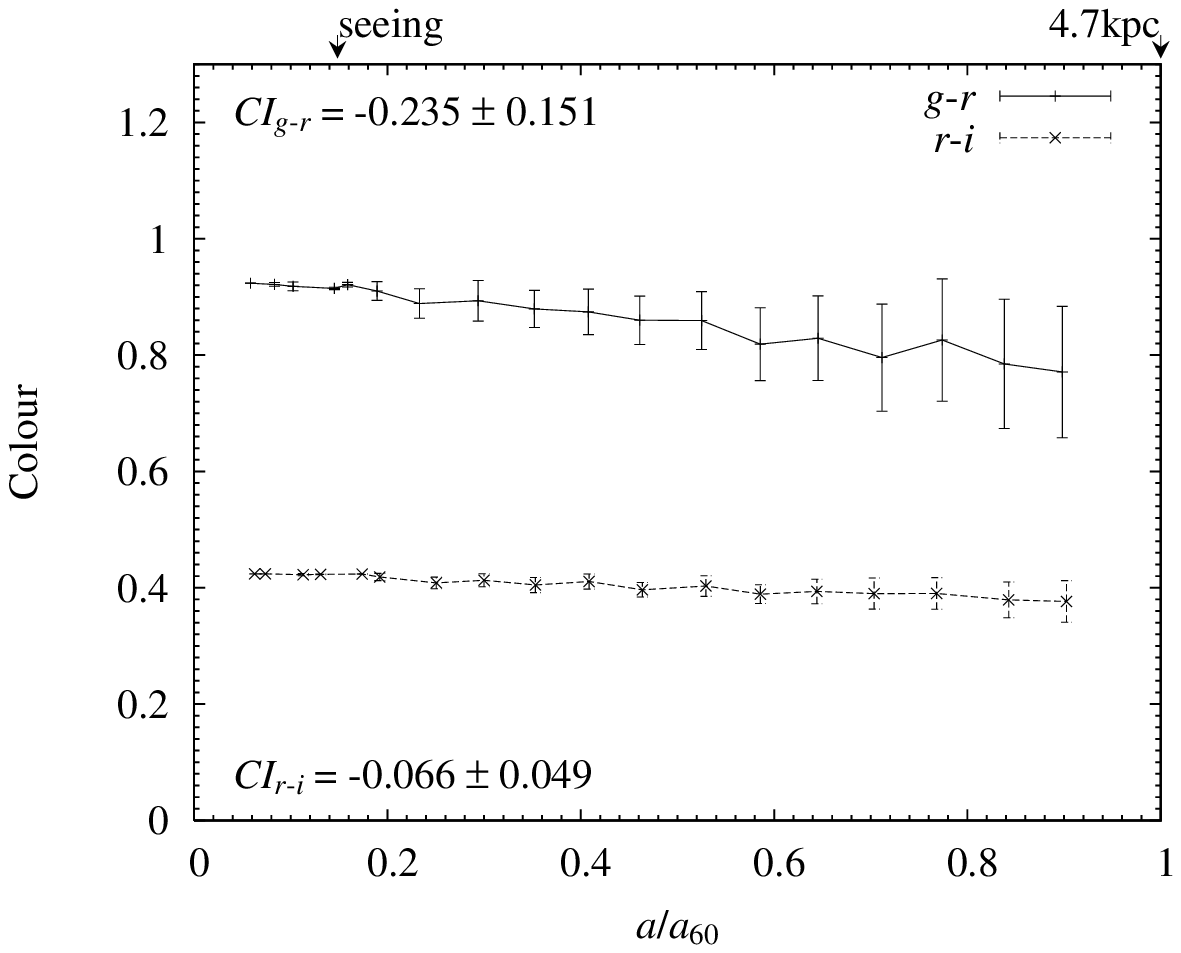} &
\includegraphics[scale=0.64]{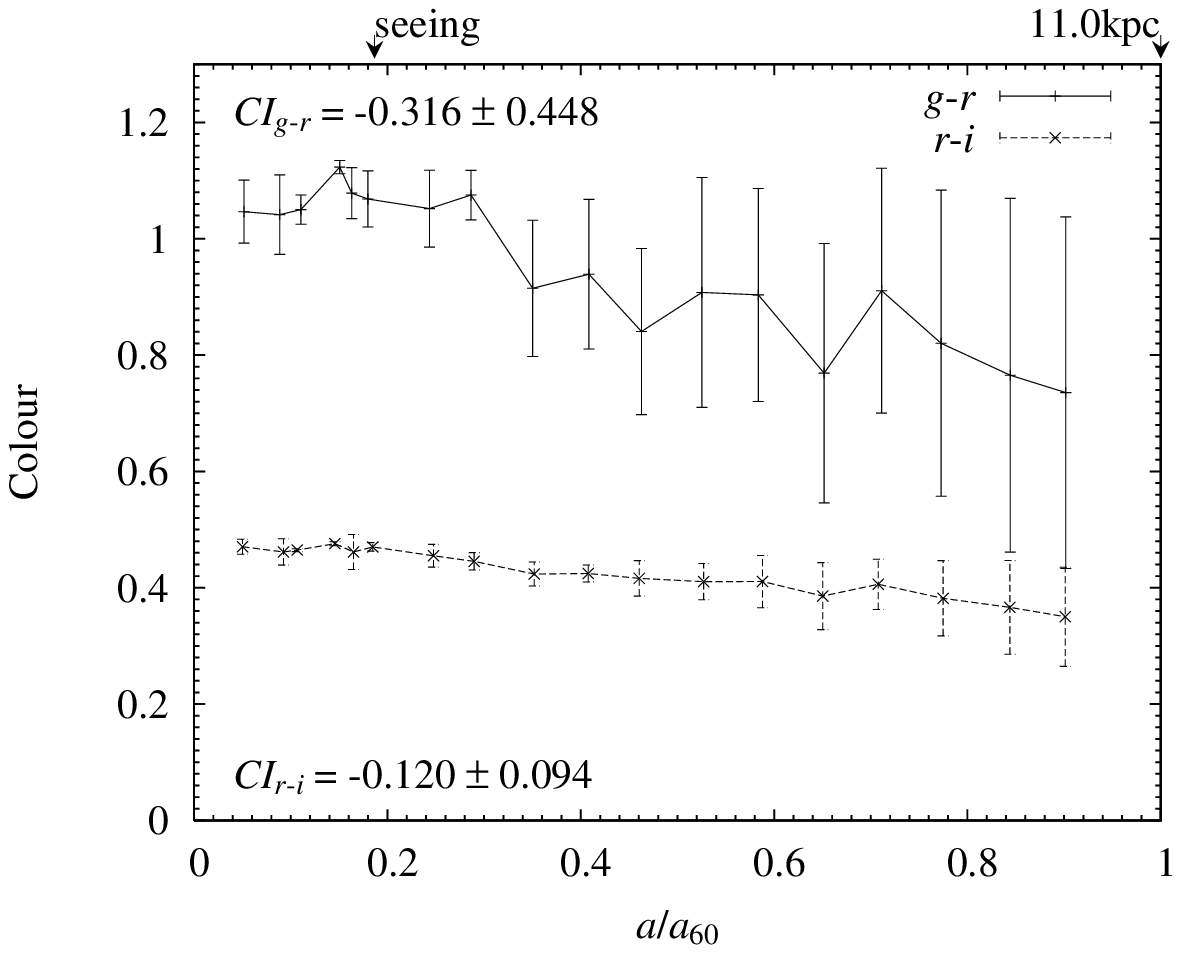} \\
 \end{tabular}
\contcaption{No.21 and No.22 E+A galaxies}
\end{figure*}

\begin{figure}
\includegraphics[scale=0.39]{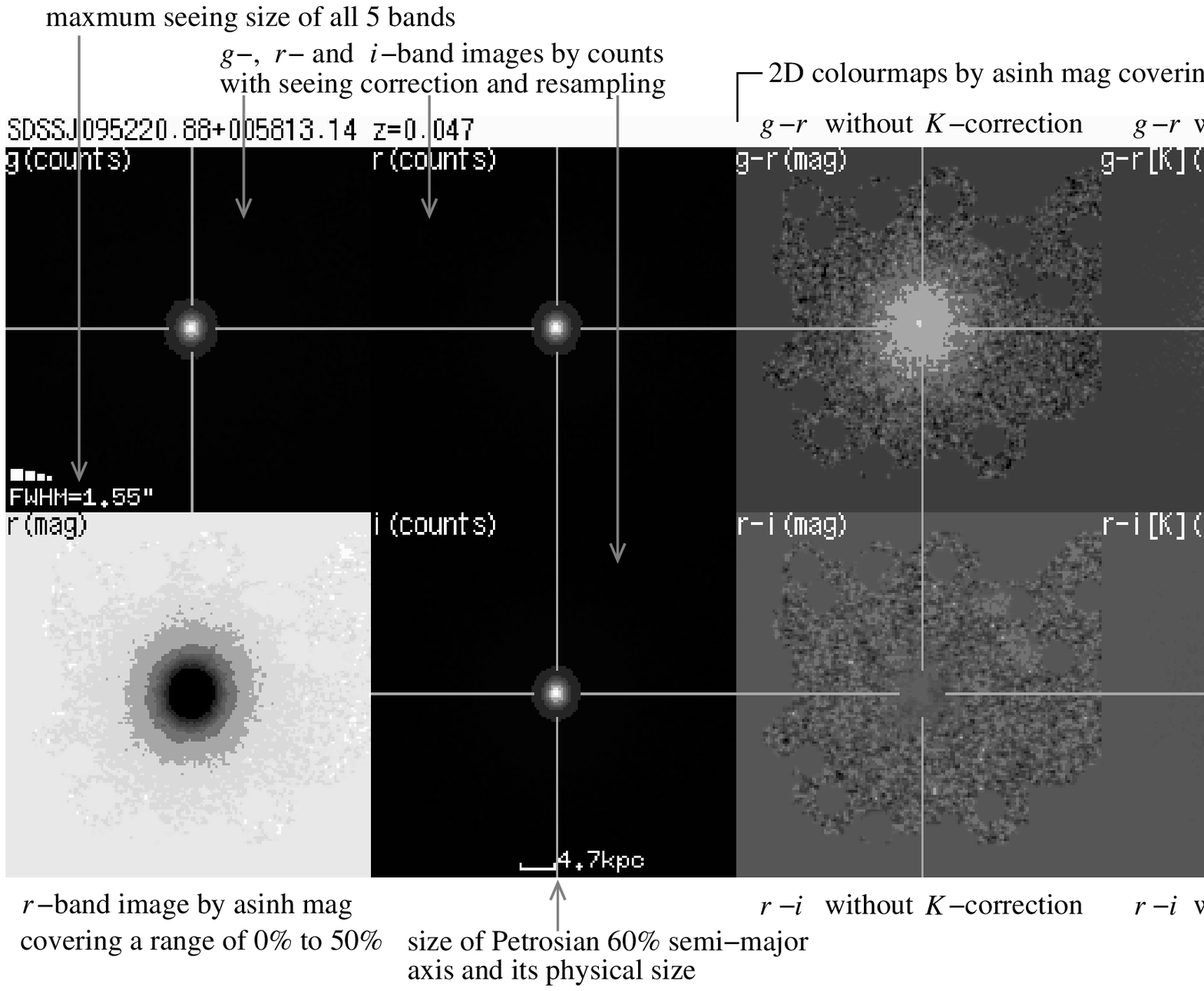}
\includegraphics[scale=0.64]{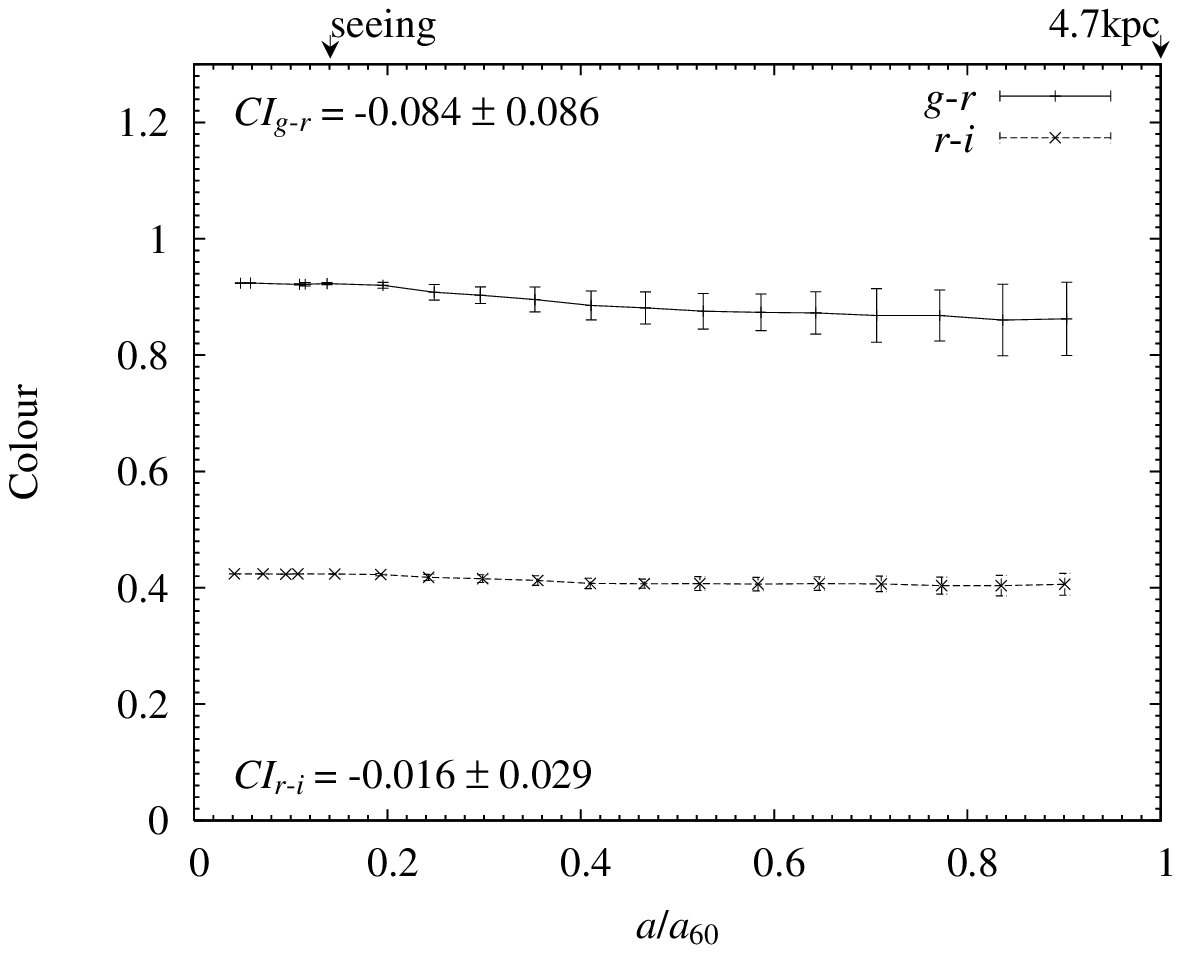}
\caption{Example of $g$-, $r$- and $i$-band image, 
$g{\rm -}r$ and $r{\rm -}i$ 2D colourmaps ({\it top}) and the rest colour
 profiles ({\it bottom}) of a normal early-type galaxy.
 2D colourmaps are based on ${\rm asinh}$ magnitude, but
 Pogson magnitude is used for colour profiles.
 The $K$-corrected 2D colourmaps are on the right, another has no
 $K$-correction applied.
 The error bars represent standard deviation
 of $g{\rm -}r$ or $r{\rm -}i$ values on sampling points within an annulus.
 The details of explanation are inlaid in the top panel.
 The 2D colourmaps of Elliptical galaxy are featureless, and
 many of them have moderate negative slope 
 (redder gradients toward the centre)
 at $g{\rm -}r$ and
 quite flat $r{\rm -}i$ colour gradient.
}\label{fig:example_profiles_early}
\end{figure}

First, we can confirm that E+A galaxies are centrally concentrated using
concentration index $C_e$.  In Figure \ref{fig:ce_hist}, we show the
distribution of $C_e$ in 22 E+A galaxies.  Galaxies that follow de
Vaucouleurs' law give $C_e=0.29$ and those with the exponential profile
give 0.44, so $C_e$ of early-type and late-type galaxies are clustered
on each point, respectively (See Yamauchi et al. 2005).  $C_e$ of 21 E+A
galaxies are distributed in the same way as early-type galaxies, 
and the median $C_e=0.33$ is a typical value of an E or S0 galaxy \citep{shi01,str01}.  
Our visual inspection%
\footnote{%
One of us (C.Y.) has performed a visual classification of the E+A galaxies in 
reference to the Third Reference Catalogue of Bright Galaxies 
\citep[][RC3]{vau91}.
}%
of the SDSS images
in Table \ref{tab:e+a_list}
also shows that our E+A galaxies are predominantly early-type galaxies%
.
We can interpret these results as a suggestion that 
all except one of our E+A galaxies
are bulge-dominated systems.

We showed one of the most splendid SDSS E+A galaxies with a tidal
feature in Figure \ref{fig:nearest_e+a}.  The SDSS imaging does not have
high resolution (typical seeing size is $\sim$1.5$''$) compared with the
HST or 8m-class telescopes, but our 22 nearby or large E+A galaxies make
a show of dramatic tidal tails or conspicuous disturbed morphologies.
We present all $r$-band negative images of our E+A galaxies in Figure
\ref{fig:e+a_all} on a logarithmic flux scale with 0\% to 50\%.  The
images in Figure \ref{fig:e+a_all} are placed in ascending order with
respect to 4000\AA~break($D_{4000}$), and we numbered our E+A galaxies
according to this order.  The physical size of $a_{60}$ is displayed on
the $i$-band image (right of the negative $r$-band image).  To examine E+A
morphologies, we present an image of a normal elliptical galaxy in Figure
\ref{fig:example_profiles_early}.  Comparison between Figures
\ref{fig:example_profiles_early} and \ref{fig:e+a_all} helps our
examination of E+A morphologies.  These 22 E+A galaxies exhibit a
variety of disturbed features, ranging from what could visually be
classified as a normal elliptical galaxy without disturbance (E+A No.7)
to one with impressive tidal features (E+A No.1).  We notice that at least half of
our E+A galaxies have traces of merger/interaction, and exhibit tidal
features or disturbed morphologies.  If we observe our E+A galaxies with
a higher resolution using an 8m-class telescope, we should be able to
examine the details of these features and the fraction of E+A galaxies
which leave traces of dynamically disturbed signs might increase.


\subsection{2D Colour Properties: Radial Colour Gradients and Colour Morphologies}

An example of $g{\rm -}r$ and $r{\rm -}i$ 2D colourmaps and their radial
colour profiles of normal early-type galaxy are presented in Figure
\ref{fig:example_profiles_early}.  
The 3 positive images in the top left panel show $g$-, 
$r$- and $i$-band images by counts with seeing correction and
resampling, and the 4 images on the right are $g{\rm -}r$ and 
$r{\rm -}i$ 2D colourmaps in asinh magnitude.  The two rightmost panels of
2D colourmaps are pixel-to-pixel $K$-corrected, while the others are not.  The
seeing sizes of all images are indicated in the $g$-band image.
The bottom panel shows $g{\rm -}r$ and $r{\rm -}i$ radial colour profiles using
Pogson magnitude, and seeing size(FWHM$/2$) and physical scale of
$a_{60}$ are displayed on the upper abscissa.  In this figure, we inlay
the 
explanation of images and 2D colourmap, and the same form
is used in Figure \ref{fig:e+a_all}, where we present all of our E+A
galaxies.

The 2D colourmaps of elliptical galaxies are basically featureless, and
the major difference between $K$-corrected and uncorrected is small.  
The radial colour profiles derived from 2D rest colour information also
show moderate properties.  Many elliptical galaxies have quite flat
$r{\rm -}i$ colour gradients and moderate negative slopes at $g{\rm -}r$.
Although there are many kinds of $g{\rm -}r$ radial gradients in
early-type galaxies,
the positive slope (bluer gradients toward the centre)
at $g{\rm -}r$ profile is seldom seen.

\begin{figure}
\includegraphics[scale=0.68]{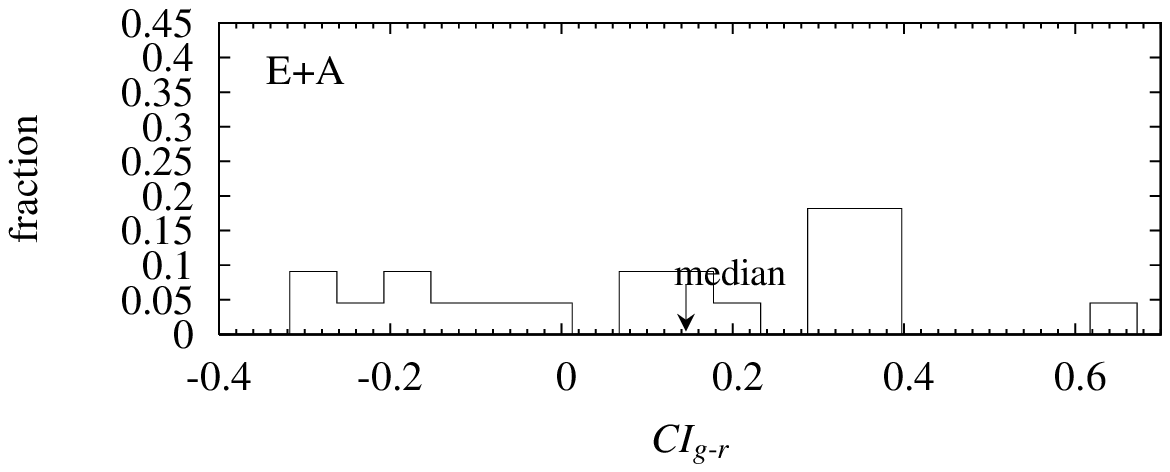}
\includegraphics[scale=0.68]{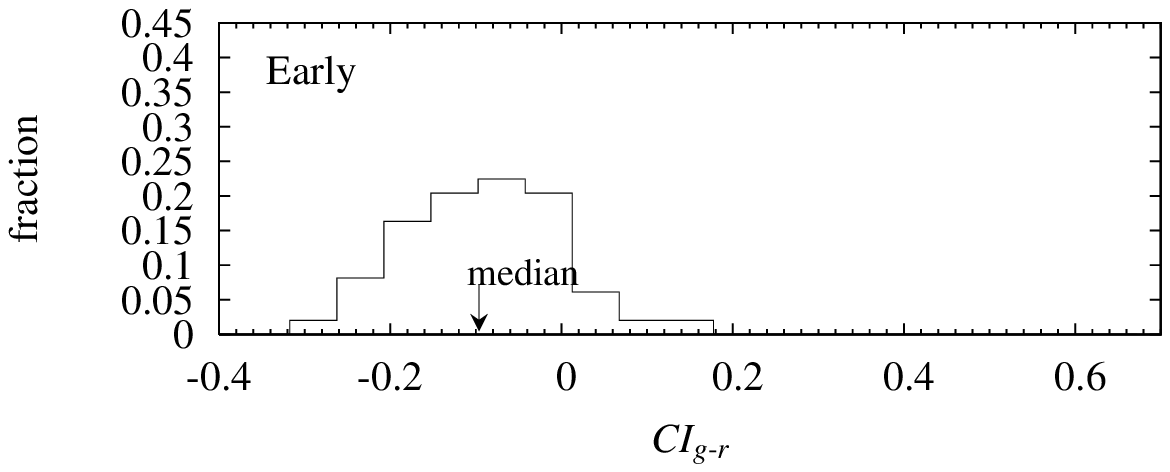}
\caption{The histograms of $g{\rm -}r$ radial colour gradient,
showing E+A ({\it top}) and early-type ({\it bottom}) galaxies.
}\label{fig:histo_g-r} 
\end{figure}

\begin{figure}
\includegraphics[scale=0.68]{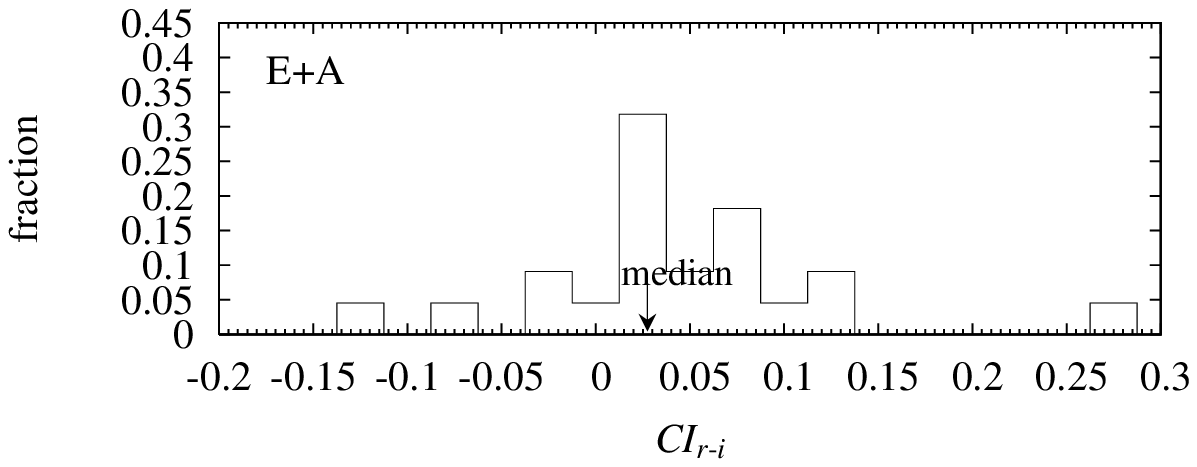}
\includegraphics[scale=0.68]{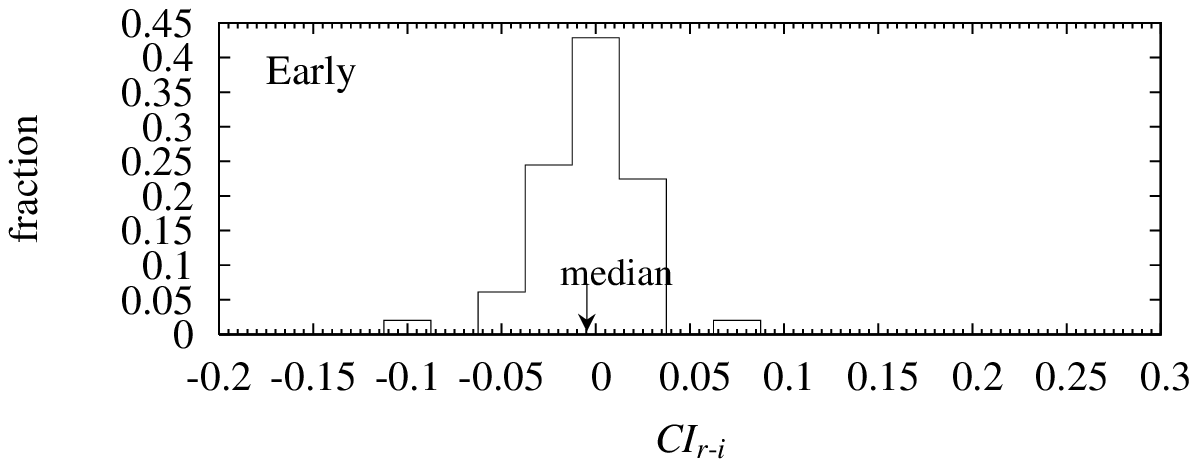}
\caption{The histograms of $r{\rm -}i$ radial colour gradient,
showing E+A ({\it top}) and early-type ({\it bottom}) galaxies.
}\label{fig:histo_r-i} 
\end{figure}

Meanwhile, many of our E+A galaxies show mysterious 2D colourmaps and
radial profiles.  We present 2D colourmaps and radial profiles of all our
E+A galaxies in Figure \ref{fig:e+a_all}.  The images and colour
profiles in Figure \ref{fig:e+a_all} are placed in ascending order with
respect to the 4000\AA~break($D_{4000}$), and we have numbered our E+A galaxies
according to this order.  The slopes of $g{\rm -}r$ radial profile of a number of
E+A galaxies are positive (bluer gradients toward the centre) which is
seldom seen in normal early-type galaxies.  We list our 22
E+A galaxies in Table \ref{tab:e+a_list} with ${\it CI_{g\mbox{-}r}}$
and ${\it CI_{r\mbox{-}i}}$ results.  
It turns out that $\sim$63\% of our E+A galaxies have a positive slope
of radial $g{\rm -}r$ colour gradient.  In the case of the $r{\rm -}i$
colour gradient, $\sim$77\% of E+As have a positive slope.  In addition,
some E+A galaxies show an irregular pattern in their 
$g{\rm -}r$ and $r{\rm -}i$ 2D colourmap without $K$-correction ---
`Colour Morphology' --- asymmetrical and clumpy bluer region.  The No.1,
No.2 and No.3 E+A galaxies are especially conspicuous; the bluer regions
are shown near the galaxy centre, but some of the bluest regions are 
placed slightly away from nucleus.  However, the $K$-corrected 2D colourmaps tend
to weaken this asymmetrical feature.  We expect that such a region has a
particular SED, and this may be a limit of $K$-correction by fitting
template SEDs.  These `irregularities' on 2D colourmaps are observed in
E+A galaxies with `positive slopes'.  But not all of the E+A galaxies with positive
slopes necessarily have conspicuous irregularities on the 2D
colourmap.  That is, the E+A galaxies with `positive slopes' of radial
colour gradient tend to show irregular patterns on the 2D colourmap.

To compare the radial colour gradients of the E+A galaxies with those of
normal early-type galaxies, the distributions of radial colour gradients
${\it CI_{g\mbox{-}r}}$ and ${\it CI_{r\mbox{-}i}}$ are displayed in
Figures \ref{fig:histo_g-r} and \ref{fig:histo_r-i}.  It is clear
that normal early-type galaxies have negative $g{\rm -}r$ colour
gradients, and almost all are placed within 
$-0.2 < {\it CI_{g\mbox{-}r}} < 0$, and $r{\rm -}i$ colour gradients are
concentrated at ${\it CI_{r\mbox{-}i}} \sim 0$.  Each median is $-0.10$
and $0.005$, respectively.  On the other hand, ${\it CI_{g\mbox{-}r}}$
and ${\it CI_{r\mbox{-}i}}$ of E+A galaxies are dispersed, and galaxies
with large positive slopes are obviously increased.  Each median is also
shifted to positive, $0.14$ and $0.03$, respectively.
%
%
We then applied a Kolomogorov-Smirnov (K-S) two-sample test to find the
probability that the two samples are drawn from the same parent
distribution.  The results show that $g{\rm -}r$ and $r{\rm -}i$ colour
gradient distributions of E+A are different from those of early-type
galaxies with a more than 99.99\% significance level.

\subsection{Comparison between Radial Colour Gradient and Other Photometric/Spectroscopic Properties}

\begin{figure}
\includegraphics[scale=0.55]{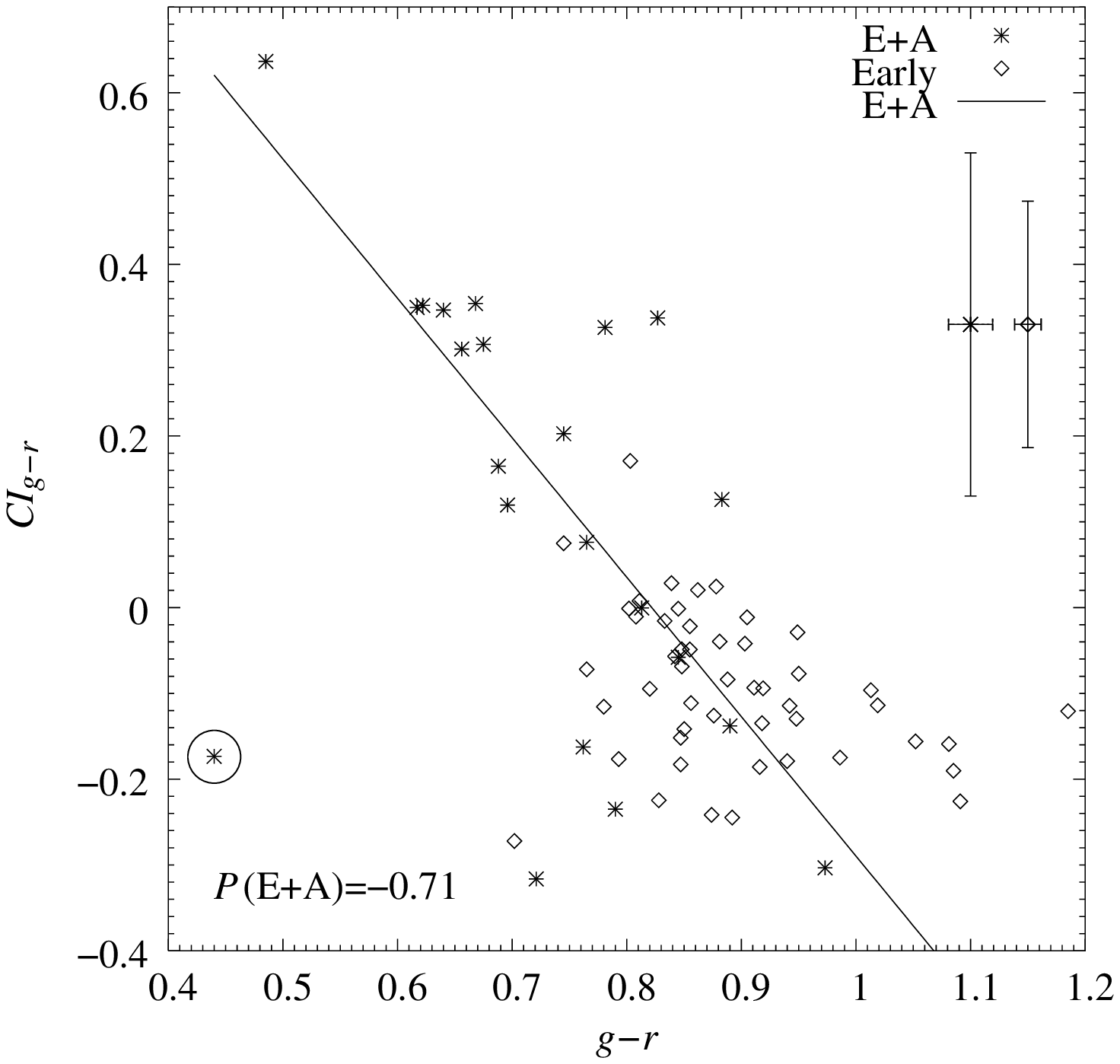}
\includegraphics[scale=0.55]{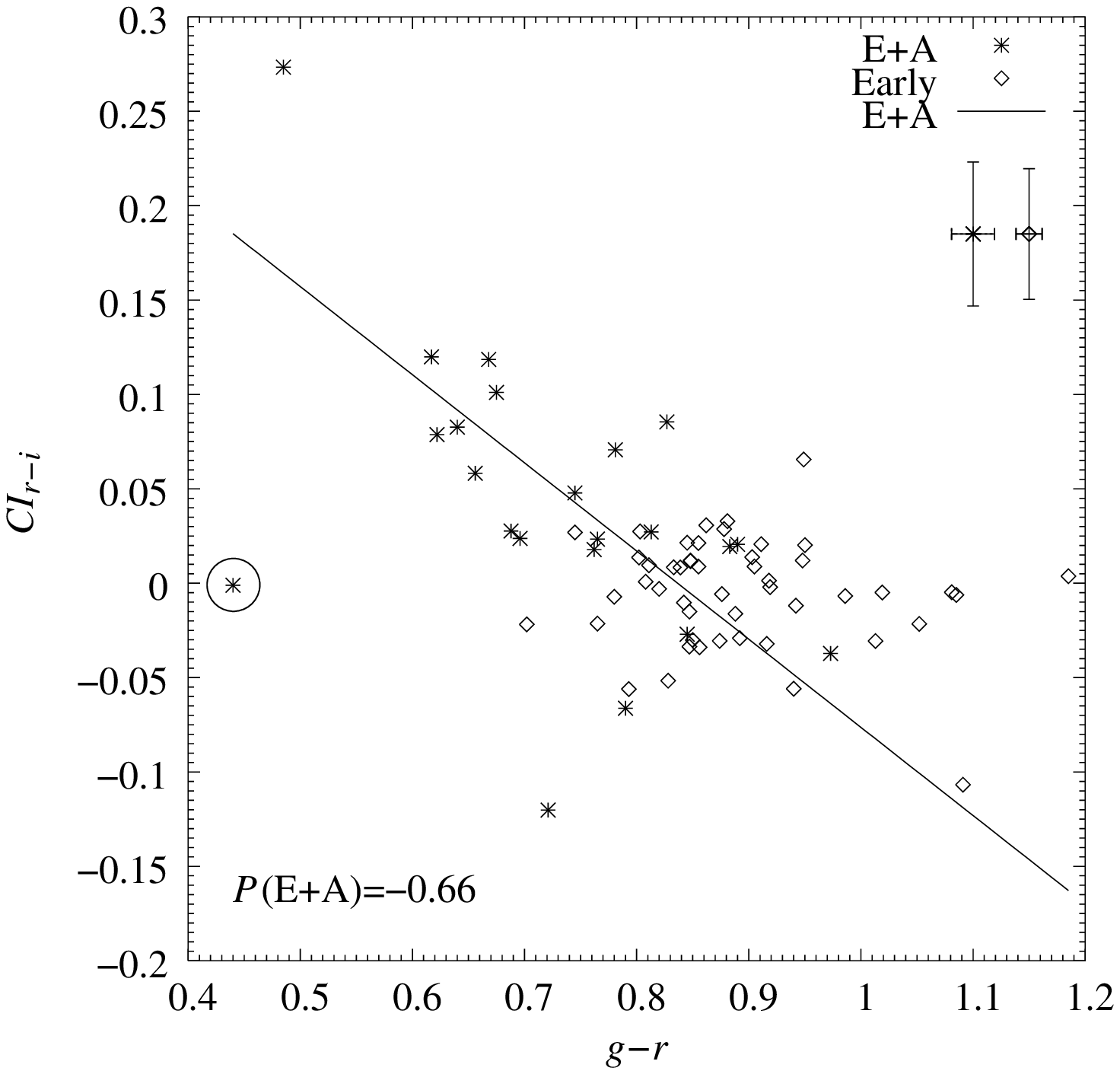}
\caption{The $g{\rm -}r$ radial colour gradient plotted against
$g{\rm -}r$ rest colour ({\it top}) and 
$r{\rm -}i$ radial colour gradient plotted against
$g{\rm -}r$ rest colour ({\it bottom}),
showing 22 E+A and 49 early-type galaxies.
E+A and early-type galaxies are plotted using asterisks and
open lozenges, respectively.
The coefficient $P$ on the bottom left is the Spearman linear correlation
coefficient, and the solid line represents a linear least-squares fit
using E+A data points.
The circled asterisks are No.1 E+A galaxy rejected
for $P$ and linear least-squares fit.
Error bars at the top right are typical errors of the 
observational data.
}\label{fig:ci_g-r} 
\end{figure}

\begin{figure}
\includegraphics[scale=0.55]{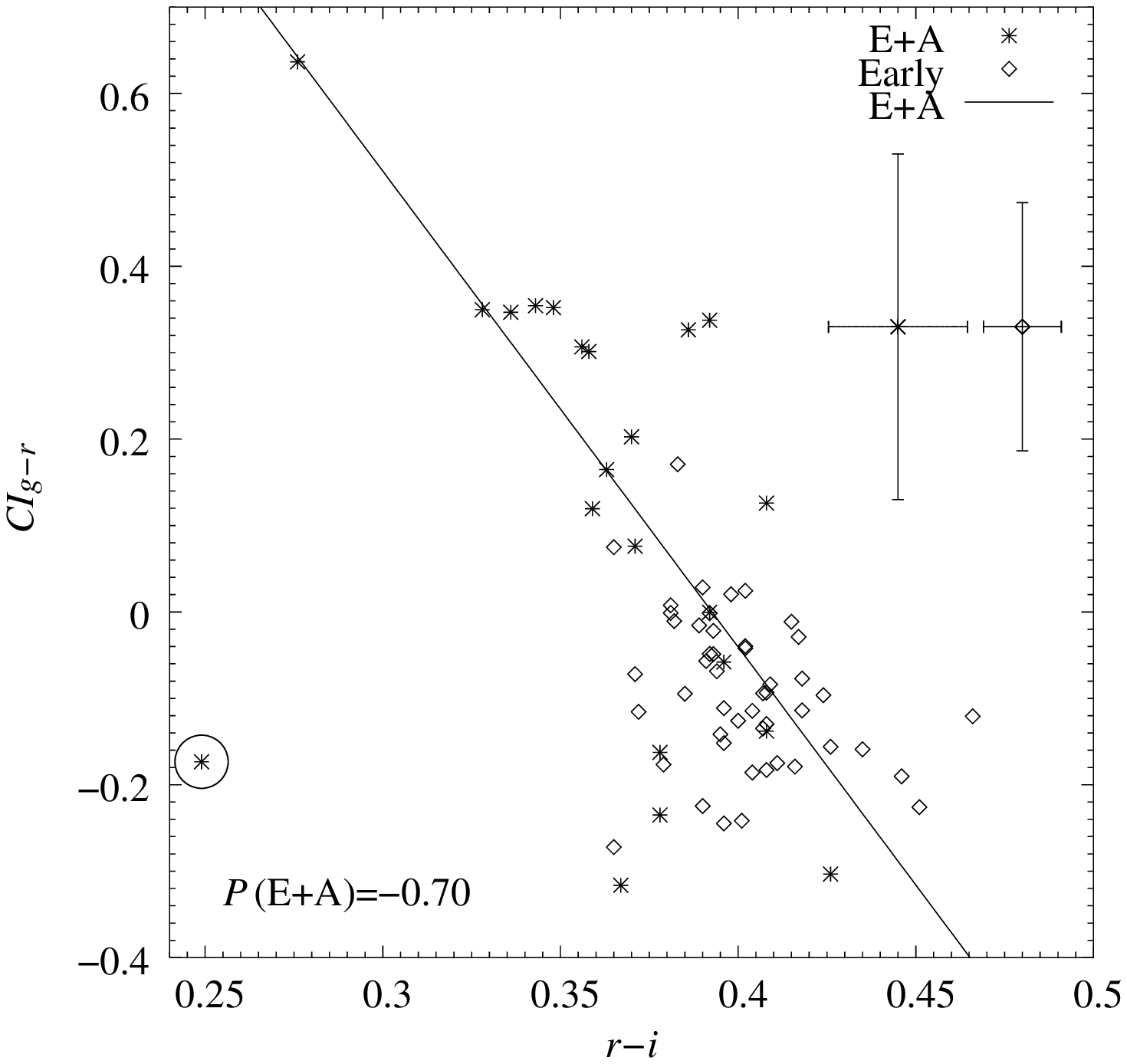}
\includegraphics[scale=0.55]{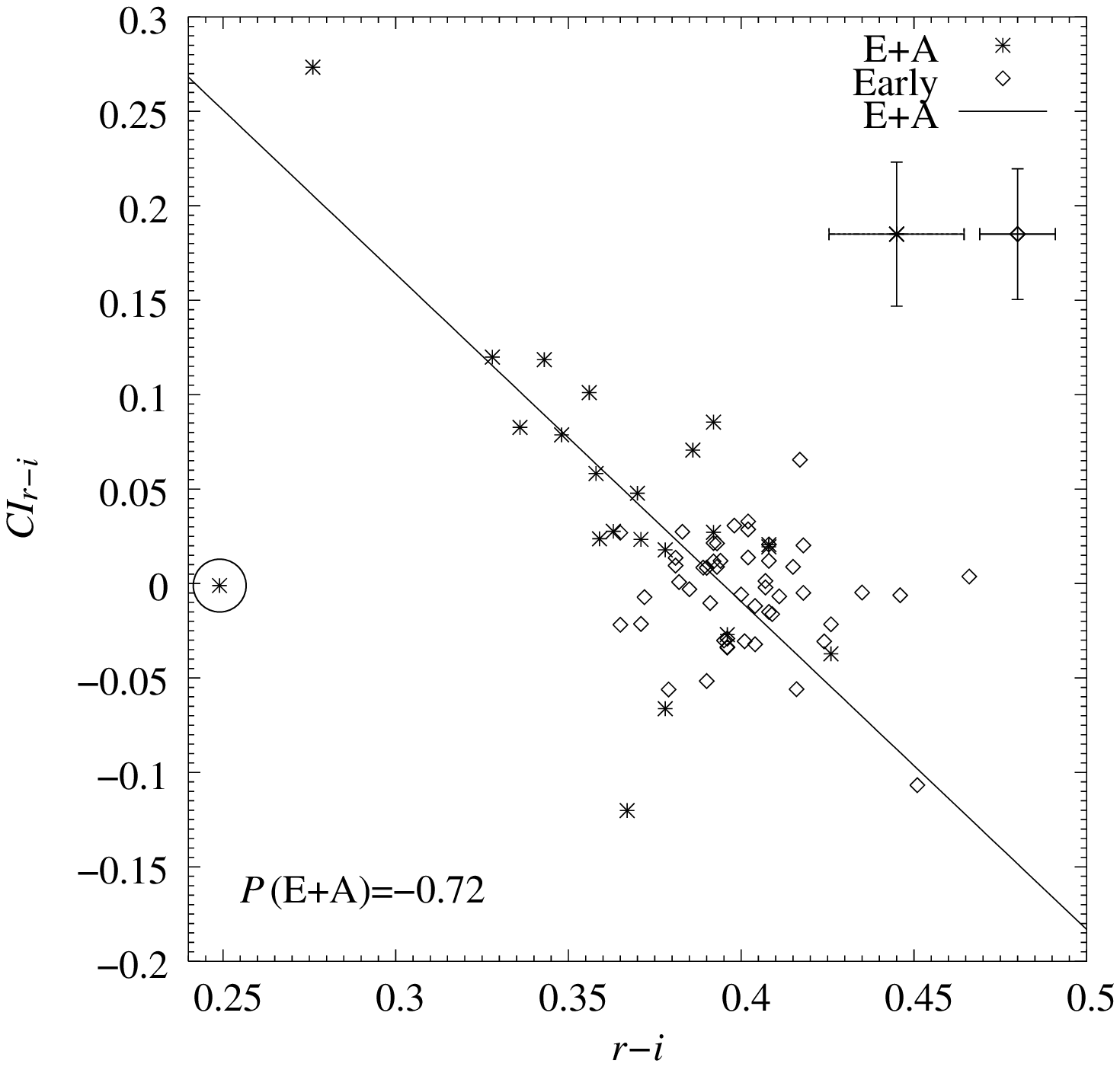}
\caption{The $g{\rm -}r$ radial colour gradient plotted against
$r{\rm -}i$ rest colour ({\it top}) and 
$r{\rm -}i$ radial colour gradient plotted against
$r{\rm -}i$ rest colour ({\it bottom}),
showing 22 E+A and 49 early-type galaxies.
E+A and early-type galaxies are plotted using asterisks and
open lozenges, respectively.
The coefficient $P$ at the bottom left is the Spearman linear correlation
coefficient, and the solid line represents a linear least-squares fit
using E+A data points.
The circled asterisks are No.1 E+A galaxy rejected
for $P$ and linear least-squares fit.
Error bars at the top right are typical errors of the 
observational data.
}\label{fig:ci_r-i} 
\end{figure}

We showed that E+A galaxies have interesting radial colour gradients.
Comparison between radial colour gradient and other
photometric/spectroscopic properties may provide a clue to the E+A evolution
scenario.

Figures \ref{fig:ci_g-r} and \ref{fig:ci_r-i} show the relation between
radial colour gradients ${\it CI}$ and rest $g{\rm -}r$ or $r{\rm -}i$
colours.  Observed colours in this paper are shifted to the restframe
using the $K$-correction software (${\tt v3\_ 2}$) by \citet{bla03}.
Our 22 E+A and 49 early-type galaxies are plotted using asterisks and
open lozenges, respectively.  The error bars on the top right indicate a
typical error.  Coefficient $P$ on the bottom left is the Spearman
linear correlation coefficient, and the solid line represents a linear
least-squares fit using E+A data points.  A data point with an open
circle is No.1 E+A galaxy rejected for $P$ and linear least-squares
fit, since the errors of ${\it CI_{g\mbox{-}r}}$ and 
${\it CI_{r\mbox{-}i}}$ are too large (0.67 and 0.20, respectively).  We
expect that E+A galaxy No.1 has too heavy a disturbance feature to
permit calculation of the 
appropriate Petrosian radius.  We can find obvious
correlations of $P\sim -0.7$ level
between radial colour gradients and colours.
But some of the derived regressions appear to rely heavily on the one
point, $g{\rm -}r=0.48$ E+A galaxy. So
we have examined the Spearman
linear correlation coefficient $P$ with $0.5<g{\rm -}r$ points only
in Figure \ref{fig:ci_g-r}, and
gotten $P=-0.62$ and $-0.59$, respectively.
On the other hand, the early-type galaxies show only
$P\sim -0.32$ at the maximum absolute value in these diagrams.
Therefore, derived $P$ of E+As which indicate correlations
between colour gradients and colours
have some dependence 
upon the one point; however, it is not a prime factor.
The results also support the idea that the photometric properties of E+A are
significantly different from those of early-type galaxies.

\begin{figure}
\includegraphics[scale=0.55]{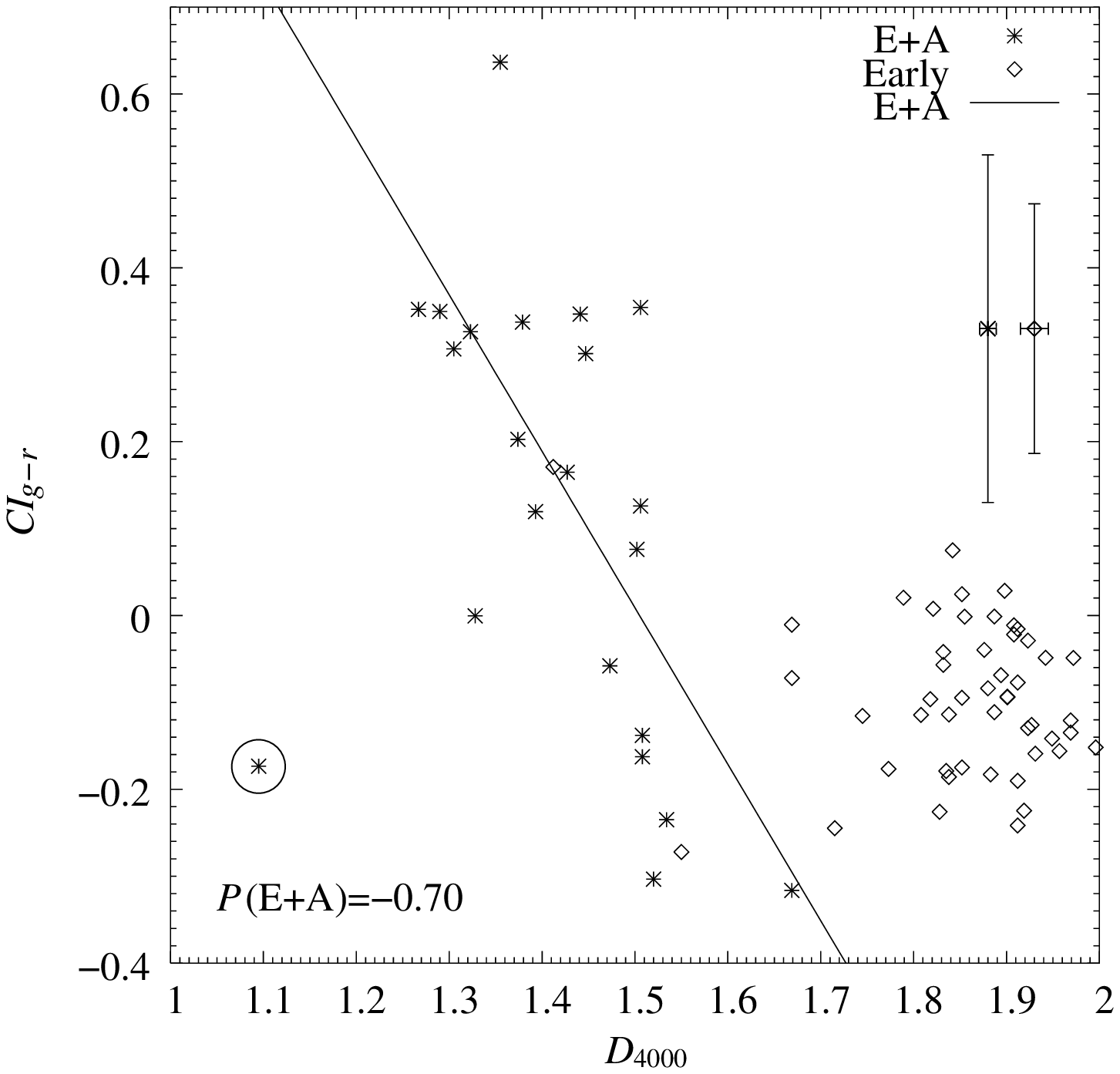}
\includegraphics[scale=0.55]{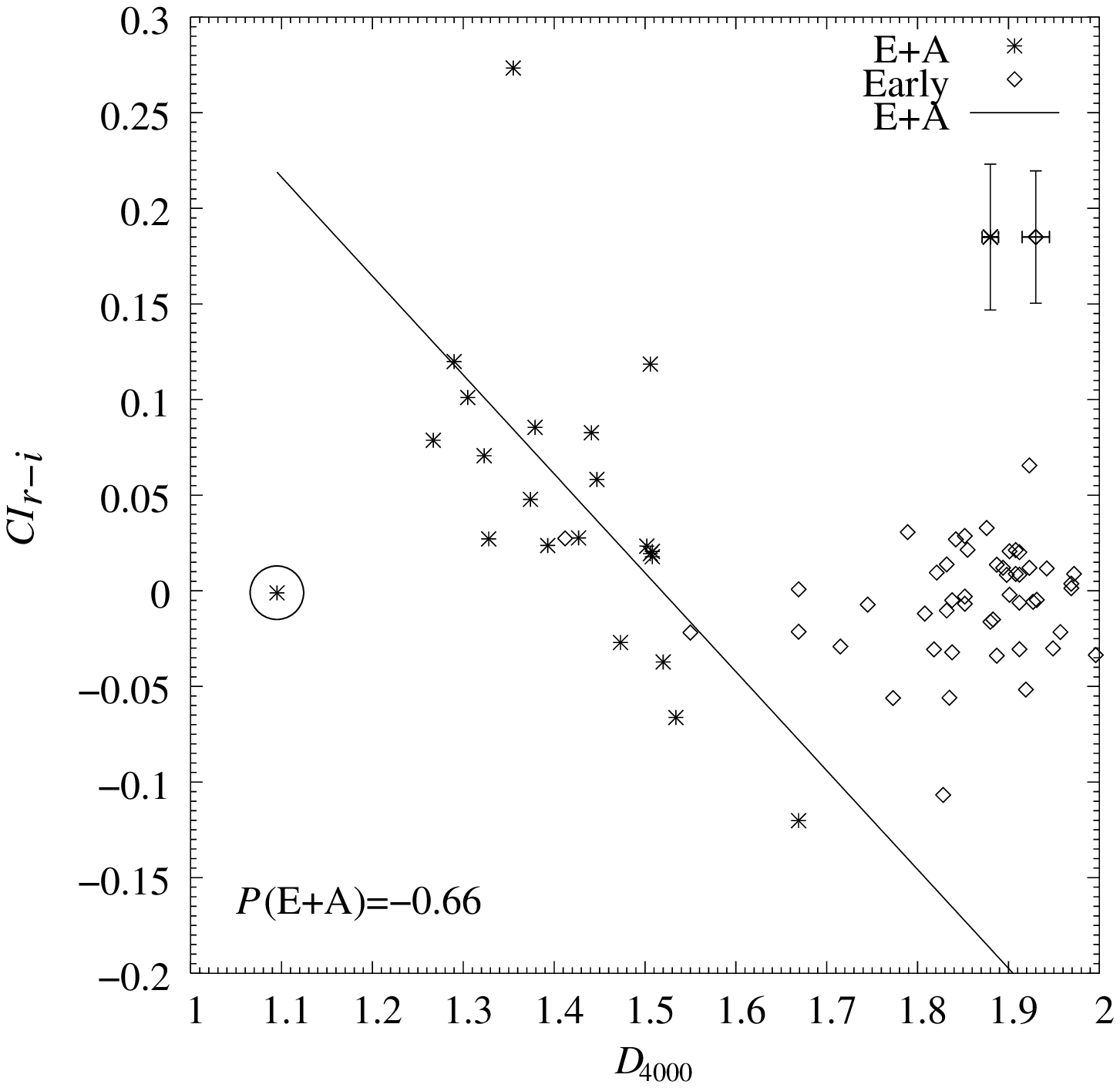}
\caption{The $g{\rm -}r$ radial colour gradient plotted against
$D_{4000}$ ({\it top}) and 
$r{\rm -}i$ radial colour gradient plotted against
$D_{4000}$ ({\it bottom}),
showing 22 E+A and 49 early-type galaxies.
E+A and early-type galaxies are plotted using asterisks and
open lozenges, respectively.
Coefficient $P$ on the bottoms left is the Spearman linear correlation
coefficient, and the solid line represents a linear least-squares fit
using E+A data points.
The circled asterisks are No.1 E+A galaxy rejected
for $P$ and linear least-squares fit.
Error bars at the top right are typical errors of the
observational data.
}\label{fig:ci_d4000} 
\end{figure}

Next, we compare the radial colour gradients with spectroscopic
properties.  
In Figure \ref{fig:ci_d4000}, we plot radial colour gradients against the
4000\AA~break ($D_{4000}$) which is sensitive to old stellar
populations.  Symbols and rejections are the same as in previous figures.
This diagram evidently indicates the correlation
between radial colour gradients and the 4000\AA~break.  Early-type galaxies
show less than $|P|\sim 0.2$ level correlations like those of colour
gradients and colours.  But normal early-type galaxies are not placed
under the regression line, different from plots of colour gradients
against colours. $D_{4000}$ is shown to separate the two
populations more clearly than the broad-band colours. 

\begin{figure}
\includegraphics[scale=0.55]{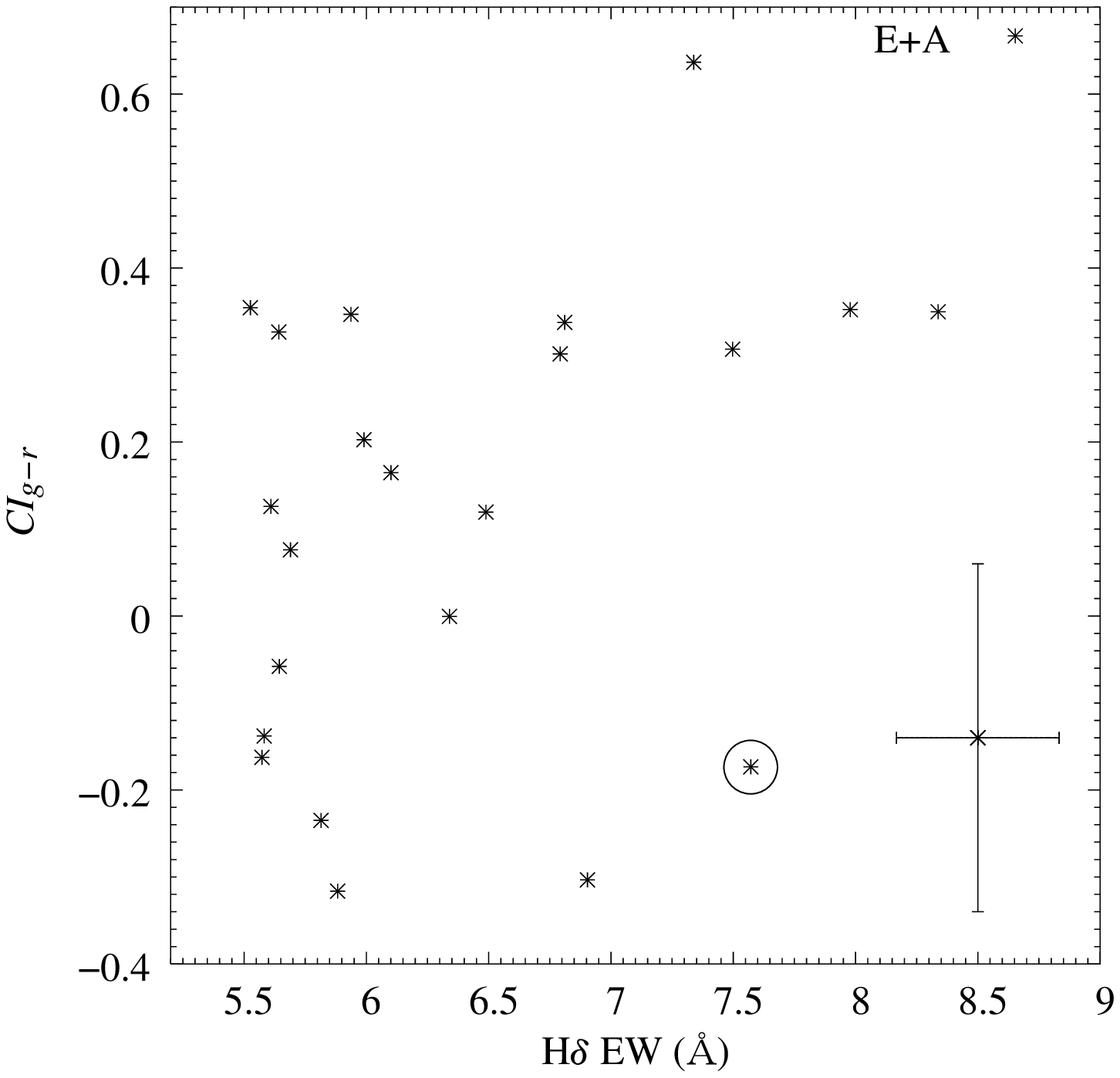}
\includegraphics[scale=0.55]{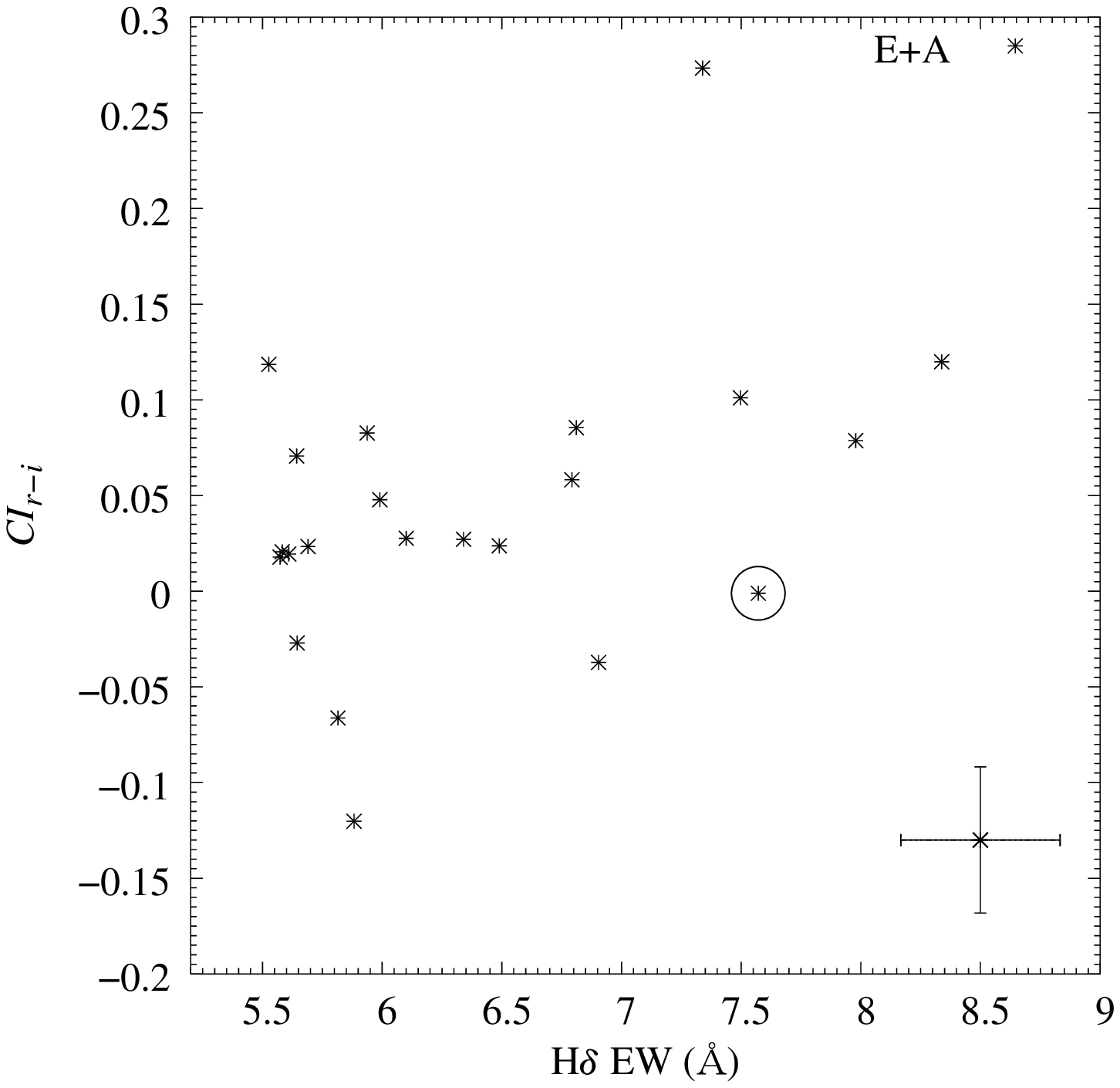}
\caption{The $g{\rm -}r$ radial colour gradient plotted against
${\rm H}\delta~{\rm EWs}$ ({\it top})
and 
$r{\rm -}i$ radial colour gradient plotted against
${\rm H}\delta~{\rm EWs}$ ({\it bottom}),
showing 22 E+A galaxies.
The circled asterisks are No.1 E+A galaxy.
Error bars at the bottom right are typical errors of the 
observational data.
}\label{fig:ci_hd} 
\end{figure}

To examine the relation between radial colour gradients and the amount of
young A-type stars, we plot radial colour gradients against 
${\rm H}\delta~{\rm EW}$ in Figure \ref{fig:ci_hd}.  Symbols are the
same as in previous figures.  Compared with the case of $D_{4000}$, we
cannot find a tight correlation.  However, these panels show a trend for E+A
galaxies with large positive slope of radial colour gradient to
show strong ${\rm H}\delta~{\rm EWs}$.  However, this result might be
affected by the cut-off of $a/a_{60} < 0.35$ for measuring radial colour
gradient.

\subsection{Evolution Scenario for E+A Radial Colour Gradients}

\begin{figure}
\includegraphics[scale=0.55]{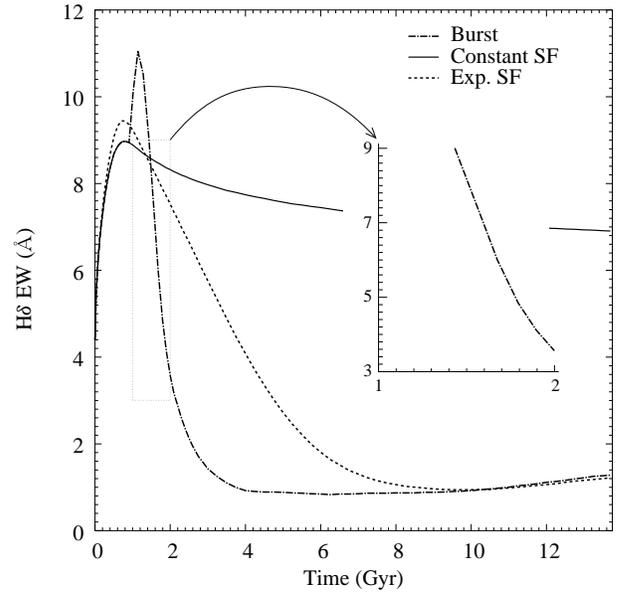}
\caption{%
${\rm H}\delta~{\rm EWs}$ are plotted against time (age) for three star formation
histories with the GISSEL model.
The dot-dashed, solid and dotted lines show the models with
instantaneous burst, constant star formation and exponentially decaying
star formation rate.
The models in this figure assume Salpeter IMF and solar metallicity.
The inlaid panel is an enlarged plot of the burst model with 
$3 < {\rm H}\delta~{\rm EW} < 9$.
}\label{fig:hd_time_gissel_model}
\end{figure}

\begin{figure}
\includegraphics[scale=0.55]{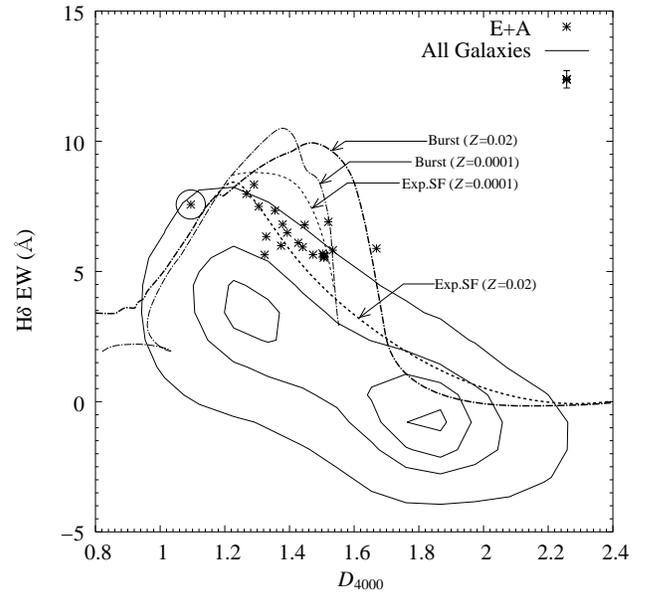}
\caption{%
${\rm H}\delta~{\rm EWs}$ are plotted against $D_{4000}$ for the models with
$Z=0.0001$($0.5$\% solar) and $Z=0.02$(solar).
Star formation histories are the burst and exponentially
decreasing, shown by the dot-dashed and dotted,
respectively.
Observational data are plotted using asterisks.
The circled asterisk is No.1 E+A galaxy.
Error bars at the top right are typical errors of the
observational data.
The contours show the distribution of all DR1 94770 galaxies in
\citet{got03}.
}\label{fig:hd_d4000}
\end{figure}

\begin{figure}
\includegraphics[scale=0.55]{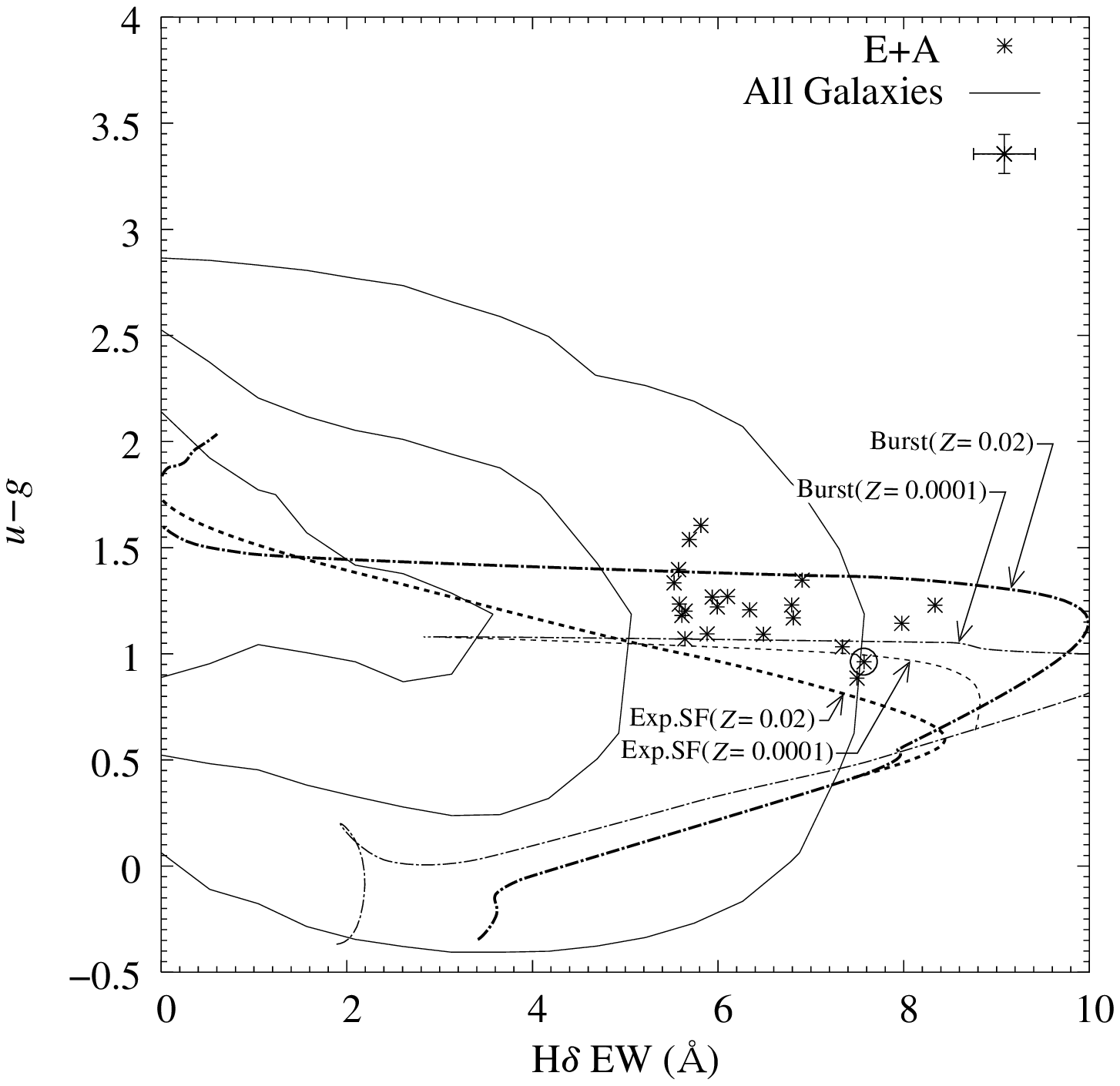}
\caption{%
The $u{\rm -}g$ colour is plotted against 
${\rm H}\delta~{\rm EWs}$ for the models with
$Z=0.0001$($0.5$\% solar) and $Z=0.02$(solar).
The dot-dashed and dotted lines are 
for the models with instantaneous burst
and exponentially decaying star formation rate.
Observational data are plotted using asterisks.
The circled asterisk is No.1 E+A galaxy.
Error bars at the top right are typical errors of the 
observational data.
The contours show the distribution of all DR1 94770 galaxies in
\citet{got03}.
}\label{fig:u-g_hd}
\end{figure}

In this subsection, we compare SED models and observational quantities
to find out the evolution scenario for E+A radial colour gradients.  
Using the GISSEL model by
\citet{bru03}, we simulated three representative star formation
histories following \citet{got03}: 
(i) The {\it Burst} model, which has an instantaneous starburst 
(with a duration of 1 Gyr) at the beginning and no star formation thereafter.  (ii) {\it Constant} star formation.  (iii) {\it Exponentially} decaying star formation (with $\tau=1$ Gyr). In all three models, we use the Salpeter initial mass function \citep{sal55}.  Figure \ref{fig:hd_time_gissel_model} plots ${\rm H}\delta~{\rm EWs}$ against time (or galaxy age) for the three models.  The dot-dashed, solid and dotted lines show the models with instantaneous burst, constant star formation and exponentially decaying star formation rate.  The burst model has a strong ${\rm H}\delta~{\rm EW}$ right after its truncation at 1 Gyr.  However, its ${\rm H}\delta~{\rm EW}$ declines rapidly, and becomes less than 3\AA~ at 1 Gyr after the truncation.  The exponentially decaying model maintains a strong ${\rm H}\delta$ for a longer time.  Its ${\rm H}\delta~{\rm EW}$ becomes 3\AA~ in 5 Gyrs.  The constantly star-forming model maintains a large ${\rm H}\delta~{\rm EW}$ ($>$6\AA) beyond 13 Gyr. 
  Although shown here for a comparison purpose, we omit the constantly star-forming model from our main discussion, since our E+A galaxies are selected to have no on-going star formation activity.

We compare the models and the data on the ${\rm H}\delta~{\rm EW}$
v.s. $D_{4000}$ plane in Figure \ref{fig:hd_d4000}.  A caveat, however,
is that models become less accurate on the plane since both 
${\rm H}\delta~{\rm EW}$ and $D_{4000}$ are more difficult quantities to
reproduce than broad band colours.  For models, we use 
${\rm H}\delta~{\rm EWs}$ given in the GISSEL model, which were measured
using the flux between 4083.50 and 4122.25\AA.  This is essentially the
same window as used to measure ${\rm H}\delta~{\rm EWs}$ from the
observational data between 4082 to 4122\AA.  For $D_{4000}$, the SED
models uses the flux ratio of the 3750-3950\AA~window to the
4050-5250\AA~window \citep{bru83}.  Observationally, $D_{4000}$ is
measured using the ratio of the flux in the 3751-3951\AA~window to that
in the 4051-4251\AA~\citep{sto02}.  We regard these two $D_{4000}$
measurements as essentially the same.

In Figure \ref{fig:hd_d4000}, 
the dot-dashed and dotted lines are for the models 
with the instantaneous burst and the exponentially decaying star
formation rate.
We subtracted 1\AA~from the model ${\rm H}\delta~{\rm EWs}$
to compensate for possible stellar absorption.  
Different line widths are for two different 
metallicities ($Z=0.001$ and $0.02$).  
\citet{got03} showed that the models might have a slight shift 
toward larger $D_{4000}$ and ${\rm H}\delta$ directions, 
compared with the distribution of all the observed galaxies.  
In the models, emission fillings are not included, and
might be a factor in gaps between observational data and the model.
However, the behavior of the models on this plane well reproduce
expected behavior of galaxies, in the sense that star-forming galaxies
evolve into large $D_{4000}$ and small ${\rm H}\delta~{\rm EWs}$.
Therefore, we regard the qualitative interpretation based on the models
as valid.  Observed data of E+A galaxies are plotted using asterisks,
and we find for the most part that our E+A galaxies are located under a
single evolution track.  But it is uncertain whether these E+A galaxies
run along the track of instantaneous burst or not, since models might
have a slight shift reported by \citet{got03}.

\citet{got03} also reported that it is more difficult to reproduce
emission lines with the current version of the model.
The model gives the amount of Ly$\alpha$ photons, 
so we could use it to estimate the amount of
Balmer or ${\rm [OII]}$ line emissions. 
However, we found that the uncertainty in this process was too large to
produce a realistic value comparable
with the observational values 
(We do not know the escaping factor of Ly$\alpha$ photons, etc.). 
We found a qualitative agreement in the long scale, but in the 
liner scale, the model ${\rm [OII]~EW}$ (or ${\rm H}\alpha~{\rm EW}$)
was far from agreement.
On the other hand, the $u{\rm -}g$ colour is preferable for examination of on-going star formation since it shows better agreement between the SED models and the observation data.  In Figure \ref{fig:u-g_hd}, we plot $u{\rm -}g$ against ${\rm H}\delta~{\rm EW}$ for the two models and the observational data.  Symbols are the same as previous figures.  The model $u{\rm -}g$ colour depends on metallicity to some extent, showing bluer $u{\rm -}g$ colour with decreasing metallicity.  Our E+A galaxies are distributed between the instantaneous burst model with $Z=0.001$ and that with solar metallicity. Neither of the exponentially decaying models can explain the observed data. This indicates that E+A galaxies are in a post-starburst phase, and cannot be explained by more normal star formation histories.

If the burst model is assumed, our young E+A galaxies with ${\rm H}\delta~{\rm EW} \sim 8{\rm \AA}$ evolve into those with ${\rm H}\delta~{\rm EW} \sim 5{\rm \AA}$ within $\sim$300 Myr by Figure \ref{fig:hd_time_gissel_model}.  Our radial colour gradient correlates with $D_{4000}$ in Figure \ref{fig:ci_d4000}, and E+A galaxies with large positive slopes of radial colour gradient tend to show strong ${\rm H}\delta~{\rm EWs}$ (Figure \ref{fig:ci_hd}).  Our E+A galaxies are placed under a single track on the $D_{4000}$ - ${\rm H}\delta~{\rm EW}$ plane; therefore, we can interpret these results as some E+A metamorphosis of large positive slopes into flat or negative slopes of radial colour gradients within the time scale. Taken together, our observational results suggest that E+A galaxies have positive colour gradients, large  ${\rm H}\delta~{\rm EW}$ and small $D_{4000}$ at the beginning, then evolve into negative colour gradients, small  ${\rm H}\delta~{\rm EW}$ and large $D_{4000}$. It is revealing that the morphological quantity,  colour gradients, correlates well with spectral properties such as ${\rm H}\delta~{\rm EW}$ and $D_{4000}$.


\begin{table*}
\begin{center}
\caption{ The list of E+A galaxies.  The numbers are labeled in ascending order with respect to $D_{4000}$.  The digits in Name represent R.A. and Dec., for example, R.A. and Dec. of Name SDSSJ010617.76+140354.00(No.16) are 01:06:17.76 and +14:03:54.00, respectively.  The unit of $a_{60}$ and line ${\rm EWs}$ are arcsec and \AA, respectively.  The morphological type $T=0,1,2,3,4,5,6$ and $-1$ represent E,S0,Sa,Sb,Sc,Sdm,Im and unclassified, respectively.  }\label{tab:e+a_list} {\tabcolsep=0.75mm
\begin{tabular}{rlrrrrrrrrrrrrrrrrrr}
\hline
\multicolumn{1}{c}{No.} &
\multicolumn{1}{c}{Name} &
\multicolumn{1}{c}{$z$} &
\multicolumn{1}{c}{$M_r$} &
\multicolumn{1}{c}{$u{\rm -}g$} &
\multicolumn{1}{c}{$g{\rm -}r$} &
\multicolumn{1}{c}{$r{\rm -}i$} &
\multicolumn{1}{c}{$a_{60}$} &
\multicolumn{1}{c}{$D_{4000}$} &
\multicolumn{1}{c}{${\rm H}\delta~{\rm EW}$} &
\multicolumn{1}{c}{${\rm H}\alpha~{\rm EW}$} &
\multicolumn{1}{c}{[OII]$~{\rm EW}$} &
\multicolumn{1}{c}{$C_e$} &
\multicolumn{1}{c}{$T$} &
\multicolumn{1}{c}{${\it CI_{g\mbox{-}r}}$} &
\multicolumn{1}{c}{${\it CI_{r\mbox{-}i}}$} \\
\hline
\hline
16 &    SDSSJ010617.76+140354.00 & 0.038 & -19.76 & 1.33 & 0.67 & 0.34 & 2.96 & 1.51 & 5.53$\pm$0.32 & 2.10$\pm$0.08 & -0.13$\pm$-0.02 & 0.28 & 1 & 0.35$\pm$0.09 & 0.12$\pm$0.05 \\
7 &     SDSSJ035652.44-061031.22 & 0.037 & -19.85 & 1.03 & 0.48 & 0.28 & 3.43 & 1.35 & 7.34$\pm$0.41 & 2.39$\pm$0.10 & -0.46$\pm$-0.05 & 0.32 & 0 & 0.64$\pm$0.21 & 0.27$\pm$0.09 \\
18 &    SDSSJ083415.60+375157.24 & 0.168 & -23.10 & 1.23 & 0.89 & 0.41 & 4.93 & 1.51 & 5.58$\pm$0.23 & 0.74$\pm$0.02 & -0.22$\pm$-0.03 & 0.32 & -1 & -0.14$\pm$0.34 & 0.02$\pm$0.05 \\
3 &     SDSSJ084918.96+462252.68 & 0.041 & -20.88 & 1.23 & 0.62 & 0.33 & 3.00 & 1.29 & 8.34$\pm$0.27 & 1.91$\pm$0.05 & -1.17$\pm$-0.08 & 0.31 & 1 & 0.35$\pm$0.11 & 0.12$\pm$0.03 \\
19 &    SDSSJ093842.96+000148.68 & 0.091 & -21.84 & 1.40 & 0.76 & 0.38 & 4.07 & 1.51 & 5.57$\pm$0.47 & 1.15$\pm$0.04 & -1.62$\pm$-0.21 & 0.36 & -1 & -0.16$\pm$0.19 & 0.02$\pm$0.05 \\
12 &    SDSSJ100743.68+554934.68 & 0.045 & -20.88 & 1.27 & 0.64 & 0.34 & 3.44 & 1.44 & 5.94$\pm$0.32 & 1.57$\pm$0.05 & -2.06$\pm$-0.20 & 0.28 & 1 & 0.35$\pm$0.12 & 0.08$\pm$0.03 \\
6 &     SDSSJ111050.88+005530.98 & 0.152 & -22.22 & 1.21 & 0.81 & 0.39 & 3.34 & 1.33 & 6.34$\pm$0.32 & 1.28$\pm$0.05 & -0.94$\pm$-0.13 & 0.33 & 1 & -0.00$\pm$0.18 & 0.03$\pm$0.04 \\
4 &     SDSSJ111108.16+004048.66 & 0.184 & -22.32 & 0.89 & 0.68 & 0.36 & 2.95 & 1.30 & 7.50$\pm$0.47 & 1.60$\pm$0.08 & -0.57$\pm$-0.08 & 0.36 & 1 & 0.31$\pm$0.53 & 0.10$\pm$0.06 \\
11 &    SDSSJ115837.68-021710.97 & 0.088 & -22.11 & 1.27 & 0.69 & 0.36 & 3.35 & 1.43 & 6.10$\pm$0.28 & 1.75$\pm$0.07 & 0.69$\pm$0.09 & 0.34 & -1 & 0.16$\pm$0.09 & 0.03$\pm$0.01 \\
2 &     SDSSJ120418.96-001855.84 & 0.094 & -22.74 & 1.14 & 0.62 & 0.35 & 4.10 & 1.27 & 7.98$\pm$0.27 & 2.11$\pm$0.06 & -1.18$\pm$-0.07 & 0.31 & -1 & 0.35$\pm$0.14 & 0.08$\pm$0.03 \\
13 &    SDSSJ120523.28+643029.52 & 0.082 & -22.78 & 1.23 & 0.66 & 0.36 & 4.64 & 1.45 & 6.79$\pm$0.29 & 1.78$\pm$0.05 & 0.91$\pm$0.09 & 0.31 & -1 & 0.30$\pm$0.16 & 0.06$\pm$0.03 \\
22 &    SDSSJ122256.16-011248.06 & 0.146 & -21.74 & 1.09 & 0.72 & 0.37 & 4.69 & 1.67 & 5.88$\pm$1.42 & 1.19$\pm$0.05 & 0.61$\pm$0.40 & 0.45 & 2 & -0.32$\pm$0.45 & -0.12$\pm$0.09 \\
20 &    SDSSJ132315.60+630726.40 & 0.175 & -21.94 & 1.35 & 0.97 & 0.43 & 3.00 & 1.52 & 6.90$\pm$0.93 & 1.11$\pm$0.07 & 1.56$\pm$0.36 & 0.35 & 1 & -0.30$\pm$0.33 & -0.04$\pm$0.05 \\
17 &    SDSSJ133350.64-001617.71 & 0.176 & -22.35 & 1.18 & 0.88 & 0.41 & 2.94 & 1.51 & 5.61$\pm$0.49 & 2.21$\pm$0.14 & 0.71$\pm$0.15 & 0.31 & 1 & 0.13$\pm$0.27 & 0.02$\pm$0.04 \\
9 &     SDSSJ140801.68+514225.92 & 0.160 & -22.34 & 1.17 & 0.83 & 0.39 & 2.88 & 1.38 & 6.81$\pm$0.37 & -0.12$\pm$-0.00 & 0.51$\pm$0.07 & 0.33 & 1 & 0.34$\pm$0.23 & 0.09$\pm$0.03 \\
5 &     SDSSJ141003.60+603229.76 & 0.171 & -22.76 & 1.07 & 0.78 & 0.39 & 2.83 & 1.32 & 5.64$\pm$0.25 & 0.67$\pm$0.05 & -1.08$\pm$-0.07 & 0.31 & -1 & 0.33$\pm$0.21 & 0.07$\pm$0.03 \\
21 &    SDSSJ141419.20-031111.51 & 0.047 & -21.37 & 1.60 & 0.79 & 0.38 & 5.53 & 1.53 & 5.81$\pm$0.34 & 0.33$\pm$0.01 & -1.11$\pm$-0.18 & 0.34 & -1 & -0.24$\pm$0.15 & -0.07$\pm$0.05 \\
1 &     SDSSJ161330.24+510335.64 & 0.034 & -20.06 & 0.96 & 0.44 & 0.25 & 9.98 & 1.09 & 7.57$\pm$0.29 & 0.54$\pm$0.02 & 1.10$\pm$0.21 & 0.36 & -1 & -0.17$\pm$0.67 & -0.00$\pm$0.20 \\
15 &    SDSSJ162702.64+432833.96 & 0.046 & -21.75 & 1.54 & 0.77 & 0.37 & 5.38 & 1.50 & 5.69$\pm$0.22 & 0.67$\pm$0.01 & -1.41$\pm$-0.12 & 0.32 & -1 & 0.08$\pm$0.12 & 0.02$\pm$0.03 \\
10 &    SDSSJ164608.16+363705.16 & 0.137 & -22.45 & 1.09 & 0.70 & 0.36 & 4.15 & 1.39 & 6.49$\pm$0.46 & 1.12$\pm$0.03 & -0.35$\pm$-0.04 & 0.35 & -1 & 0.12$\pm$0.24 & 0.02$\pm$0.04 \\
8 &     SDSSJ170636.48+334720.40 & 0.124 & -22.06 & 1.22 & 0.74 & 0.37 & 3.08 & 1.37 & 5.99$\pm$0.39 & 0.34$\pm$0.02 & -0.84$\pm$-0.10 & 0.32 & -1 & 0.20$\pm$0.23 & 0.05$\pm$0.04 \\
14 &    SDSSJ211348.24-063059.65 & 0.159 & -22.11 & 1.20 & 0.84 & 0.40 & 2.90 & 1.47 & 5.64$\pm$0.51 & 0.96$\pm$0.03 & -2.32$\pm$-0.56 & 0.33 & 1 & -0.06$\pm$0.18 & -0.03$\pm$0.03 \\
\hline
\end{tabular} 
}
\end{center}
\end{table*}

\section{Discussion}
\label{section:discussion}

Based on SDSS broad-band imaging, we investigated spatial properties
such as morphologies, 2D colour properties and radial colour gradients
of E+A galaxies. In this section, we discuss physical implications and 
the relevance of our results to other E+A-related studies.

The concentration index, $C_e$, of our E+A galaxies was distributed like
early-type galaxies, 
and our visual inspection of the SDSS images 
also showed that our E+A galaxies are predominantly early-type galaxies.
We can interpret these results as a suggestion
that all of our  E+A galaxies are bulge-dominated systems.
Although E+As with disk-like morphologies have sometimes been found in previous observations \citep{cou94,cou98,dre94,dre99,oem97,sma97,cha01}, it is not surprising if previous samples were contaminated by H$\alpha$-emitting galaxies.  Perhaps since we selected 22 E+A galaxies with strict spectroscopic criteria, `without ${\rm H}\alpha$ nor [OII] emission', our 22 E+A galaxies are all bulge-dominated.  Our results are also consistent with those by \citet{qui04} and \citet{bla04}.  

 These findings have important implications for the physical origin of E+A galaxies when compared with theoretical simulations.
 It has long been known that an elliptical galaxy can be a final product of major merging (e.g., Barnes \& Hernquist 1992), quite possibly accompanied by a tidal tail feature in its early stage.
 In a more detailed major merger model,  
\citet{bek98} showed that ancestral equal-mass galaxies are completely destroyed so
as to form a spheroid that looks more like an elliptical galaxy, and
merging of two unequal-mass spirals is finally transformed into one S0
galaxy that has a flattened oblate spheroid.
We noticed that at least half of our E+A galaxies show tidal features or
disturbed morphologies.  Thus not only disturbed morphologies but also
E/S0 morphologies of our unalloyed E+A sample indicate possibilities of
a merger origin of E+A galaxies, and we can expedite the following
discussion smoothly.

We found irregular `Colour Morphologies' -- asymmetrical and clumpy
patterns -- at the centre of the $g{\rm -}r$ and $r{\rm -}i$ 2D colourmaps of
E+A galaxies  without $K$-correction.  The HST observations of the five
bluest E+A galaxies with $z \sim 0.1$ by \citet{yan04} show the presence
of asymmetric components in the residual images obtained by subtraction of 
the smooth and symmetric model images from the data.  The physical scale
of the asymmetric components of the HST observation is comparable to
that of our asymmetrical/clumpy pattern in the 2D colourmap.  We expect
that these irregularities of residual images and our 2D colourmap 
basically have the same origin, and one possible interpretation is that they
are fluctuations created during the merger/interaction event, which also
caused the central post-starburst phase. 

Our results showed that more than half of our E+A galaxies exhibit the positive radial $g{\rm -}r$ and $r{\rm -}i$ colour gradients.  
The results of $g{\rm -}r$ radial colour gradients are in good agreement
with a long-slit spectroscopic observation of 21 E+A galaxies by
\citet{nor01} who found that young stellar populations of E+A galaxies are
more centrally concentrated than the older populations.  Although
\citet{bar01} reported that E+A galaxies on average tend to have
slightly bluer radial gradients toward the centre than the normal
early-type galaxies, these E+A galaxies are detected in rich clusters
and their spectroscopic criteria 
($3 < \langle{\rm H}\delta\gamma\beta~{\rm EW}\rangle$ 
and $-5 < {\rm [OII]}~{\rm EW}$) are less strict than ours.  It is
perhaps difficult to compare our E+A galaxies with those selected with
different criteria in different environments. However, we can find
the same tendency for a substantial number of E+A galaxies to have a
positive slope of radial colour gradient and for our strict
spectroscopic criteria to produce clearer results of a K-S test 
(with a more than 
99.99\% significance level).  Theoretically, models of interacting
gas-rich spirals suggest that the most common result of such tidal
perturbations is indeed to drive a large fraction of interstellar
material close to the centre of each galaxy; the shocked gas robbed of its
angular momentum flows inward \citep{bar92}.  
\citet{mih92} showed not only that
the vast majority of the star formation arises in 
central regions of the merging galaxies,
but also that flybys which involve prograde disks
cause significant centralized starburst.
Especially, \citet{mih94} numerically simulated galaxy
mergers and predicted colour gradients in such systems. In their Fig.7,
they show that the colour gradient of merger remnants changes the slope
of colour gradient from positive to negative during 1-5 Gyr after the
burst in a way qualitatively similar to our Fig.\ref{fig:e+a_all}. The
agreement adds more credibility to the notion of the merger/interaction origin of E+A
galaxies, and further justifies our interpretation of change of the
slope as an evolutionary sequence.
In summary, our findings of positive $g{\rm -}r$ radial colour gradients
of E+A galaxies are consistent with the merger/interaction origin of E+A
galaxies.


We found a tight correlation between radial colour gradients and 4000\AA~break.
We also found that E+A galaxies follow a single evolution track on the
$D_{4000}$ - ${\rm H}\delta~{\rm EW}$ plane (%
in Figure \ref{fig:hd_d4000}).
Taken all together, these results can be interpreted in an evolutionary
sequence where E+A galaxies change their slope from positive to
negative, and they also change irregular into moderate of 2D colourmap
during the time scale of $\sim$300 Myr. It is revealing that E+A
galaxies show morphological metamorphosis (change in colour gradients
and 2D colourmap) synchronously as their stellar population ages as
indicated by $D_{4000}$ and  ${\rm H}\delta~{\rm EW}$. This may be the
general picture of how E+A galaxies evolve after the truncation of starburst possibly caused by the merger/interaction.
It is a concern that our radial colour gradients do not show a tight
correlation with ${\rm H}\delta~{\rm EW}$, but the tendency of E+A
galaxies with large positive slopes of radial colour gradient to show
strong ${\rm H}\delta~{\rm EWs}$. However, this may be affected by our
cut-off of the inner region ($a/a_{60}<0.35$) to avoid seeing effects when measuring radial colour gradients.  Observation with high-resolution imaging might remove these effects, and produce a more accurate picture.  


\section{Conclusions}
\label{section:conclusions}

We have investigated morphologies, 2D colour property and
radial colour gradients of 22 {\it true} E+A galaxies with 
$5.5{\rm \AA} < H\delta~{\rm EW} < 8.5{\rm \AA}$
using the SDSS DR2 imaging data,
in order to reveal evolution of E+A galaxies in terms of internal galaxy structures.
We present this summary of our results:
%

\begin{itemize}
 \item
      Concentration index, $C_e$, and our visual inspection of the SDSS
      images suggested that our E+A galaxies are predominantly
      bulge-dominated systems, and at least half of our E+A galaxies 
      exhibit tidal features or disturbed morphologies.
 \item
      We found irregular `Colour Morphologies' -- asymmetrical and clumpy patterns --
      at the centre of $g{\rm -}r$ and $r{\rm -}i$ 2D colourmaps of the E+A galaxies.
 \item
      We found that a substantial number of E+A galaxies
      have positive slopes (bluer towards the centre) of radial $g{\rm -}r$ and $r{\rm -}i$ colour gradients.
 \item
      Kolomogorov-Smirnov two-sample tests
      showed that $g{\rm -}r$ and $r{\rm -}i$ colour gradient distributions of E+A 
      galaxies are different from those of early-type galaxies
      with a more than 99.99\% significance level.
 \item
      Our E+A sample showed
      tight correlation between radial colour gradients and colours,
      and
      between radial colour gradients and 4000\AA~break.
      We also found the tendency for
      E+A galaxies with large positive slopes of radial colour gradient
      to show strong ${\rm H}\delta~{\rm EWs}$.
 \item
      The comparison between the GISSEL model
      \citep{bru03}
      and E+A's observational quantities, ${\rm H}\delta~{\rm EW}$,
      $D_{4000}$ and $u{\rm -}g$ colour, indicated that almost
      all our E+A galaxies are located along a single evolution track.
      Therefore, these results are interpreted as indicating E+A
      radial colour gradients evolve from positive into negative slopes,
      coherently with ${\rm H}\delta~{\rm EW}$ (8-5\AA) within $\sim$300 Myr. 
\end{itemize}

Our findings support the notion that E+A galaxies are post-starburst
galaxies caused by merger/interaction,
having undergone a centralized starburst.  In addition, we presented a
constraint on E+A's spatial evolution properties with a physical time
scale.  Detailed comparison with future theoretical simulations would
advance our knowledge of the origin of E+A galaxies further.

\section*{Acknowledgments}

We thank Shin-ichi Ichikawa for useful discussions.
We are grateful to Ani Thakar, Naoki Yasuda and Masafumi Yagi for gracious help
with technical issues.
We thank Simona Mei for useful suggestions for proofs.
We thank Linux, XFree86, and other UNIX-related communities
for the development of various useful software.
This research has made use of the Plamo Linux.

\label{lastpage}


\begin{thebibliography}{DUM}

 \bibitem[Abadi, Moore \& Bower(1999)]{aba99}
	Abadi, M. G., Moore, B., \& Bower, R. G.
	1999, \mnras, 308, 947

 \bibitem[Abazajian et al.(2003)]{aba03}
	Abazajian, K., et al. 2003, \aj, 126, 2081

 \bibitem[Abazajian et al.(2004)]{aba04}
	Abazajian, K., et al. 2004, \aj, 128, 502

 \bibitem[Abraham et al.(1996)]{abr96}
	Abraham, R. G., et al. 1996, \apj, 471, 694

 \bibitem[Bartholomew, Rose \& Gaba(2001)]{bar01}
	Bartholomew, L. J., Rose, J. A., \& Gaba, A. E.
	2001, \aj, 122, 2913

 \bibitem[Barnes \& Hernquist(1992)]{bar92}
	Barnes, J. E., \& Hernquist L.
	1992, \araa, 30, 705

 \bibitem[Barger et al.(1996)]{bar96}
	Barger, A. J., Aragon-Salamanca, A., Ellis, R. S.,
	Couch, W. J., Smail, I., \&
	Sharples, R. M.
	1996, \mnras, 279, 1

 \bibitem[Bruzual(1983)]{bru83}
	Bruzual, A. G.
	1983, \apj, 273, 105


 \bibitem[Bruzual \& Charlot(2003)]{bru03}
	Bruzual, A. G., \& Charlot, S.
	2003, \mnras, 344, 1000

 \bibitem[Bekki(1998)]{bek98}
	Bekki, K.
	1998, \apj, 502, L133

 \bibitem[Bekki, Shioya \& Couch(2001)]{bek01}
	Bekki, K., Shioya, Y., \& Couch, W. J.
	2001, \apj, 547, L17

 \bibitem[Belloni et al.(1995)]{bel95}
	Belloni, P., Bruzual, A. G., Thimm, G. J., \&
	Roser, H.-J.
	1995, A\&A, 297, 61

 \bibitem[Blake et al.(2004)]{bla04}
	Blake, C., et al. 2004
	\mnras, 355, 713

 \bibitem[Blanton et al.(2003)]{bla03}
	Blanton, M. R., et al. 2003,  
	\aj, 125, 2348

 \bibitem[Broadhurst, Ellis \& Shanks(1988)]{bro88}
	Broadhurst, T. J., Ellis, R. S., \& Shanks, T.
	1988, \mnras, 235, 827

 \bibitem[Caldwell et al.(1993)]{cal93}
	Caldwell, N., Rose, J. A., Sharples, R. M.,
	Elllis, R. S., \& Bower, R. G.
	1993, \aj, 106, 473


 \bibitem[Caldwell \& Rose(1997)]{cal97}
	Caldwell, N., \& Rose, J. A.
	1997, \aj, 113, 492

 \bibitem[Castander et al.(2001)]{cas01}
	Castander, F. J. et al.
	2001, \aj, 121, 2331

 \bibitem[Chang et al.(2001)]{cha01}
	Chang, T., van Gorkom, J. H., Zabludoff, A. I.,
	Zaritsky, D., \& Mihos, J. C.
	2001, \aj, 121, 1965

 \bibitem[Condon(1992)]{con92}
	Condon, J. J.
	1992, ARA\&A, 30, 575

 \bibitem[Couch \& Sharples(1987)]{cou87}
	Couch, W. J., \& Shareples, R. M.
	1987, \mnras, 229, 423

 \bibitem[Couch et al.(1994)]{cou94}
	Couch, W. J., Ellis, R. S., Sharples, R. M., \& Smail, I.
	1994, \apj, 430, 121

 \bibitem[Couch et al.(1998)]{cou98}
	Couch, W. J., Barger, A. J., Smail, I., Ellis, R. S., \&
	Sharples, R. M.
	1998, \apj, 497, 188

 \bibitem[Davis, Sadler \& Peletier(1992)]{dav92}
	Davis, R. L., Sadler, E. M., \& Peletier R. F.
	1993, \mnras, 262, 650

 \bibitem[de Vaucouleurs et al.(1991)]{vau91} 
        de Vaucouleurs, G., de Vaucouleurs, A.,
        Corwin, H., Buta, R., Paturel, G.,
        \& Fouqu\'e, P. 1991,
        Third Reference Catalogue of Bright Galaxies
        (New York: Springer) (RC3)

 \bibitem[Dressler \& Gunn(1983)]{dre83}
	Dressler, A., \& Gunn, J. E.
	1983, \apj, 270, 7 

 \bibitem[Dressler \& Gunn(1992)]{dre92}
	Dressler, A., \& Gunn, J. E.
	1992, \apjs, 78, 1 

 \bibitem[Dressler et al.(1994)]{dre94}
	Dressler, A., Oemler, A. J., Sparks, W. B., \& Lucas, R. A.
	1994, \apj, 435, L23

 \bibitem[Dressler et al.(1999)]{dre99}
	Dressler, A., Smail, I., Poggianti, B. M., Butcher, H.,
	Couch, W. J., Ellis, R. S., \& Oemler, A. J.
	1999, \apjs, 78, 1

 \bibitem[Farouki \& Shapiro(1980)]{far80}
	Farouki, R., \& Shapiro, S. L.
	1980, \apj, 241, 928

 \bibitem[Fabricant, McClintock \& Bautz(1991)]{fab91}
	Fabricant, D. G., McClintock, J. E., \& Bautz, M. W.
	1991, \apj, 381, 33

 \bibitem[Fan et al.(2003)]{fan03}
	Fan, X. et al.
	2003, \aj, 125, 1649

 \bibitem[Fisher et al.(1998)]{fis98}
	Fisher, D., Fabricant, D., Franx, M., \&
	van Dokkum, P.
	1998, \apj, 498, 195

 \bibitem[Franx(1993)]{fra93}
	Franx, M. 1993, \apj, 407, L5

 \bibitem[Fujita \& Nagashima(1999)]{fuj99}
	Fujita, Y., \& Nagashima, M. 1999, \apj, 516, 619

 \bibitem[Fujita(2004)]{fuj04}
	Fujita, Y. 2004, \pasj, 56, 29

 \bibitem[Fujita \& Goto(2004)]{fg04}
	Fujita, Y., \& Goto, T.
	2004, \pasj, 56, 621

 \bibitem[Fukugita et al.(1996)]{fuk96}  
        Fukugita, M., Ichikawa, T., Gunn, J. E., Doi, M.,
        Shimasaku, K., \& Schneider, D. P. 
        1996, \aj, 111, 1748

 \bibitem[Goto(2003)]{got03}
	Goto, T., 2003, PhD Thesis, 
	The University of Tokyo, astro-ph/0310196

 \bibitem[Goto et al.(2003a)]{got03a}
	Goto, T. et al. 2003a, \pasj, 55, 739

 \bibitem[Goto et al.(2003b)]{got03b}
	Goto, T., Okamura, S.,
	Sekiguchi, M., et al. 2003b, \pasj, 55, 757

 \bibitem[Goto et al.(2003c)]{got03c}
	Goto, T. et al. 2003c, \pasj, 55, 771

 \bibitem[Goto et al.(2003d)]{got03d}
	Goto, T. et al. 2003d, \pasj, submitted

 \bibitem[Goto et al.(2003e)]{got03e}
	Goto, T., Yamauchi, C., Fujita, Y., Okamura, S.,
	Sekiguchi, M., Smail, I., Bernardi, M., \&
	Gomez, P. L.
	2003e, \mnras, 346, 601

 \bibitem[Goto(2004)]{got04a}
	Goto, T.
	2004, \aap, 427, 125

 \bibitem[Goto(2005)]{got05}
 	Goto, T.
 	2005, \mnras, 357, 937

 \bibitem[Gunn \& Gott(1972)]{gun72}
	Gunn, J. E., \& Gott, J. R. I.
	1972, \apj, 176, 1

 \bibitem[Gunn et al.(1998)]{gun98} 
	Gunn, J. E., et al. 1998, \aj, 116, 3040

 \bibitem[Hogg et al.(2001)]{hog01} 
        Hogg, D. W., Schlegel, D. J., \&
        Finkbeiner, D. P., 
        \& Gunn, J. E. 2001, \aj, 122, 2129

 \bibitem[Hopkins et al.(2003)]{hop03}
	Hopkins, A. M., et al. 
	2003, \apj, 599, 971

 \bibitem[Kennicutt(1992a)]{ken92a}
	Kennicutt, R. C.
	1992a, \apjs, 79, 255

 \bibitem[Kennicutt(1992b)]{ken92b}
	Kennicutt, R. C.
	1992b, \apj, 388, 310

 \bibitem[Kennicutt(1998)]{ken98}
	Kennicutt, R. C. 
	1998, ARA\&A, 36 189

 \bibitem[Kent(1981)]{ken81}
	Kent, S. M.
	1981, \apj, 245, 805



 \bibitem[Lavery \& Henry(1986)]{lav86}
	Lavery, R. J. \& Henry, J. P.
	1986, \apj, 304, L5

 \bibitem[Lavery \& Henry(1988)]{lav88}
	Lavery, R. J. \& Henry, J. P.
	1988, \apj, 330, 596

 \bibitem[Liu \& Kennicutt(1995a)]{liu95a}
	Liu, C. T., \& Kennicutt, R. C.
	1995a, \apjs, 100, 325

 \bibitem[Liu \& Kennicutt(1995b)]{liu95b}
	Liu, C. T., \& Kennicutt, R. C.
	1995b, \apj, 450, 547

 \bibitem[Lupton et al.(2001)]{lup01}
        Lupton, R. H., Gunn, J. E., Ivezic, Z., Knapp, G. R., Kent, S.,
        \& Yasuda, N.
        2001, 
        Astronomical Data Analysis Software and Systems X, 
        ASP Conference Proceedings, 
        238, 269

 \bibitem[MacLaren, Ellis \& Couch(1988)]{mac88}
	MacLarn, I., Ellis, R. S., \& Couch, W. J.
	1988, \mnras, 230, 249

 \bibitem[Miller \& Owen(2001)]{mil01}
	Miller, N. A., \& Owen, F. N.
	2001, \apj, 554, L25

 \bibitem[Mihos, Richstone \& Bothun(1992)]{mih92}
	Mihos, J. C., Richstone, D. O., \&
	Bothun, G. D. 
	1992, \apj, 400, 153

 \bibitem[Mihos \& Heanquist(1994)]{mih94}
	Mihos, J. C., \& Heanquist, L.
	1994, \apj, 427, 112


 \bibitem[Morris et al.(1998)]{mor98}
	Morris, S. J., Hutchings, J. B., Carlberg, R. G.,
	Yee, H. K. C., Ellingson, E., Balogh, M. L.,
	Abraham, R. G., \& Smecker-Hane, T. A.
	1998, \apj, 507, 84

 \bibitem[Newberry, Boroson \& Kirshner(1990)]{new90}
	Newberry, M. V., Boroson, T. A., \& Kirshner, R. P.
	1990, \apj, 350, 585

 \bibitem[Nikolic, Cullen \& Alexander(2004)]{nik04}
	Nikolic, B., Cullen, H., \& Alexander, P.
	2004, \mnras, 355, 874

 \bibitem[Norton et al.(2001)]{nor01}
	Norton, S. A., Gebhardt, K., Zabludoff, A. I., \&
	Zaritsky, D.
	2001, \apj, 557, 150

 \bibitem[Oegerle, Hill \& Hoessel(1991)]{oeg91}
	Oegerle, W. R., Hill, J. M., \& Hoessel, J. G.
	1991, \apj, 381, L9

 \bibitem[Oemler, Dressler \& Butcher(1997)]{oem97}
	Oemler, A. J., Dressler, A., \& Butcher, H. R.
	1997, \apj, 474, 561

 \bibitem[Owen et al.(1999)]{owe99}
	Owen, F. N., Ledlow, M. J., Keel, W. C., \& Morrison, G. E.
	1999, \aj, 118, 633

 \bibitem[Pier et al.(2003)]{pie03}  
        Pier, J. R., Munn, J. A., Hindsley, R. B.,
        Hennessy, G. S., Kent, S. M.,
        Lupton, R. H., \& Ivezic, Z.
        2003, \aj, 125, 1559

 \bibitem[Poggianti \& Wu(2000)]{pog00}
	Poggianti, B. M., \& Wu, H.
	2000, \apj, 529, 157

 \bibitem[Quilis, Moore \& Bower(2000)]{qui00}
	Quilis, V., Moore, B., \& Bower, R.
	2000, Sci, 288, 1617

 \bibitem[Quintero(2004)]{qui04}
	Quintero, A. D., et al. 2004, \apj, 602, 190

 \bibitem[Rose et al.(2001)]{ros01}
	Rose, J. A., Gaba, A. E., Caldwell, N., \&
	Chaboyer, B. 2001, \aj, 121, 793

 \bibitem[Salpeter(1955)]{sal55}
	Salpeter, E. E.
	1955, \apj, 121, 161

 \bibitem[Schlegel, Finkbeiner \& Davis(1998)]{sch98}
	Schlegel, D. J.,
	Finkbeiner, D. P., \&
	Davis, M.
	1998, \apj, 500, 525

 \bibitem[Schweizer(1982)]{sch82}
	Schweizer, F. 1982, \apj, 252, 455

 \bibitem[Schweizer(1996)]{sch96}
	Schweizer, F. 1996, \aj, 111, 109

 \bibitem[Sharples et al.(1985)]{sha85}
	Sharples, R. M., Ellis, R. S., Couch, W. J., \&
	Gray, P. M.
	1985, \mnras, 212, 687

 \bibitem[Shimasaku et al.(2001)]{shi01}
        Shimasaku, K.
        et al. 2001, \aj, 122, 1238

 \bibitem[Smail et al.(1997)]{sma97}
	Smail, I., Dressler, A., Couch, W. J.,
	Ellis, R. S., Oemler, A. J., Butcher, H.,
	\& Shaples, R. M.
	1997, \apjs, 110, 213

 \bibitem[Smail et al.(1999)]{sma99}
	Smail, I., Morrison, G., Gray, M. E., Owen, F. N.,
	Ivison, R. J., Kneib, J.-P., \& Ellis, R. S.
	1999, \apj, 525, 609

 \bibitem[Smith et al.(2002)]{smi02}  
        Smith, J. A., et al. 2002, \aj, 123, 2121

 \bibitem[Spitzer \& Baade(1951)]{spi51}
	Spitzer, L. J., \& Baade, W.
	1951, \apj, 113, 413

 \bibitem[Stoughton et al.(2002)]{sto02} 
        Stoughton, C., et al. 2002, \aj, 123, 485

 \bibitem[Strateva et al.(2001)]{str01} 
        Strateva, I.,
        et al. 2001, \aj, 122, 1861

 \bibitem[Strauss et al.(2002)]{str02}  
	Strauss, M. A. et al. 2002, \aj, 124, 1810

 \bibitem[Tanaka et al.(2004)]{tan04}
	Tanaka, M., Goto, T., Shimasaku, K., Okamura, S.,
	Shimasaku, K., Brinkmann, J.
	2004, \apj, 128, 2677

 \bibitem[Yamauchi \& Goto(2004)]{yam04}
	Yamauchi, C., \& Goto, T.
	2004, \mnras, 352, 815

 \bibitem[Yamauchi et al.(2005)]{yam04s}
	Yamauchi, C., Ichikawa, S., Doi, M., Yasuda, N.,
	Yagi M., Fukugita, M., Okamura, S., Nakamura, O.,
	Sekiguchi, M., \& Goto, T.,
	\aj, submitted.

 \bibitem[Yang et al.(2004)]{yan04}
	Yang, Y., Zabludoff, D., Zaritsky, D.,
	Lauer, T., \& Mihos, J. C.
	2004, \apj, 607, 258

 \bibitem[York et al.(2000)]{yor00} 
	York, D. G., et al. 2000, \aj, 120, 1579
 

\end{thebibliography}
\end{document}